\documentclass[twocolumn,trackchanges]{aastex701}
\usepackage{natbib}
\usepackage{graphicx}
\usepackage{url}
\usepackage{ulem}
\usepackage{xcolor}
\newcommand{\cmt}{\,cm$^{-3}$}

\newcommand{\kms}{\,km\,s$^{-1}$}

\newcommand{\myr}{\,$M_{\odot}\,{\rm yr}^{-1}$}
\newcommand{\es}{$\rm\,erg\,s^{-1}$}
\newcommand{\ecs}{$\rm\,erg\,s^{-1}\,cm^{-2}$}
\newcommand{\ecsa}{$\rm\,erg\,s^{-1}\,cm^{-2}\,\AA^{-1}$}
\newcommand{\ha}{H$\alpha$}
\newcommand{\hb}{H$\beta$}
\newcommand{\hg}{H$\gamma$}
\newcommand{\hi}{H\,{\scriptsize I}}
\newcommand{\hii}{H\,{\scriptsize II}}
\newcommand{\hei}{He\,{\scriptsize I}}
\newcommand{\heii}{He\,{\scriptsize II}}
\newcommand{\ro}{\,$R_{\odot}$}
\newcommand{\mo}{\,$M_{\odot}$}

\begin{document}

\title{The first hours and days of the 2021 explosion of 
       the recurrent symbiotic nova RS Ophiuchii}
\author[orcid=0000-0002-8312-3326,gname=Augustin,sname=Skopal]{Augustin Skopal}
\affiliation{Astronomical Institute, Slovak Academy of Sciences,
             059\,60 Tatransk\'a Lomnica, Slovakia}
\email[show]{skopal@ta3.sk}

\author[gname=Martin,sname=Vra\v{s}\v{t}\'ak]{Martin Vra\v{s}\v{t}\'ak}
\affiliation{Private observatory Liptovsk\'{a} \v{S}tiavnica,
             K\v{l}u\v{c}iny 457/74, Slovakia}
\email{mvrastak@live.com}

\author[orcid=0000-0002-9107-9791,gname=Francois,sname=Teyssier]{Francois Teyssier}
\affiliation{ARAS Eruptive Stars Group, F-76100 Rouen, France}
\email{francoismathieu.teyssier@gmail.com}

\author[gname=Mitsugu,sname=Fujii]{Mitsugu Fujii}
\affiliation{Fujii Kurosaki Observatory, 4500 Kurosaki, Tamashima,
             Kurashiki,Okayama 713-8126, Japan}
\email{fkobs@sky.plala.or.jp}

\author[orcid=0000-0001-8718-6765,gname=Sergei,sname=Shugarov]{Sergei Shugarov}
\affiliation{Astronomical Institute, Slovak Academy of Sciences,
             059\,60 Tatransk\'a Lomnica, Slovakia}
\email{shugarov@ta3.sk}

\author[orcid=0000-0002-6819-2331,gname=Miroslav,sname=\v{S}lechta]{Miroslav \v{S}lechta}
\affiliation{Astronomical Institute, Czech Academy of Sciences, 
             251~65, Ondřejov, Czech Republic}
\email{miroslav.slechta@asu.cas.cz}

\author[orcid=0000-0002-4387-6358,gname=Marek,sname=Wolf]{Marek Wolf}
\affiliation{Astronomical Institute of Charles University,
            V Hole\v{s}ovi\v{c}k\'{a}ch 2, 180~00, Praha 8, Czech Republic}
\email{MarekWolf@seznam.cz}

\begin{abstract}
The accretion of matter on a massive white dwarf (WD) can lead to 
repeated nuclear explosions on its surface over a timescale of 
years to decades. The seventh explosion of the recurrent symbiotic 
nova RS~Ophiuchi (RS~Oph) was recorded on August 8, 2021. In this 
paper, we examine its early evolution, from 9 hours before its 
optical maximum until day~42. 
We achieved our goal by modeling the spectral energy distribution 
(SED) using optical spectroscopy and simultaneous 
$BVR_{\rm C}I_{\rm C}$ photometry, supplemented by $JHKL$ 
photometry and ultraviolet spectroscopy from previous explosions 
in 2006 and 1985. 
Our SED models revealed an early stage of development of the ejecta 
bipolar structure, consisting of a flared, density-enhanced 
equatorial disk and low-density regions in bipolar directions. 
The comparability of the internal shocks' luminosity in 
the equatorial outflow, inferred from our model parameters, 
with the luminosity of the warm WD pseudophotosphere 
during its presence in the spectrum (until $\sim$day 42) 
confirmed that a significant part of its radiation originates 
from reprocessed shock emission. 
We explain the formation and evolution of the bipolar ejecta 
structure during RS~Oph explosions by the rotation of the accreting 
WD. Such an ejecta structure provides a natural framework for 
the generation of strong internal shocks and thus $\gamma$-ray 
emission inside the ejecta. 
\end{abstract}
%
%
%
\section{Introduction}
\label{s:intro}
A nonmagnetic white dwarf (WD) accreting from a main-sequence 
star in cataclysmic variables (CVs) or from an evolved giant in 
symbiotic stars (SySts) can accumulate hydrogen-rich matter 
until its pressure and temperature at the WD surface reach 
critical values at which thermonuclear runaway (TNR) is 
triggered, causing an optical brightening up to $\sim$15\,mag. 
We observe the classical nova (CN; in CVs) or symbiotic nova 
(SyN; in SySts) explosion 
\citep[e.g.,][for a review]{2008clno.book.....B,
                            2016PASP..128e1001S,
                            2025CoSka..55c..47M}. 
If mass is accreted onto a massive WD ($>$1\mo), an explosive TNR 
under degenerate conditions due to strong surface gravity will 
ignite, and form a very fast nova. If in this case, accretion 
proceeds at a high rate of several times (10$^{-8} - 10^{-7}$)\myr, 
the TNR will repeat on a timescale of human life. 
These events are called recurrent (symbiotic) novae 
\citep[e.g.,][]{2005ApJ...623..398Y}. 

RS~Ophiuchi (RS~Oph) is a recurrent SyN whose explosions are 
characterized with an optical brightening by $\sim$7\,mag and 
an average recurrence period of $\sim$15 years 
\citep[][]{2010ApJS..187..275S}. The most recent explosion was 
reported on August 8, 2021 
(see vsnet-alert 26131\footnote{\url{http://ooruri.kusastro.kyoto-u.ac.jp/mailarchive/vsnet-alert/26131}.}). 
The binary of RS~Oph consists of a massive WD accreting from the 
wind of a K7\,III -- M2\,III red giant (RG) in a 453.6-d orbit 
\citep[][]{1999A&AS..137..473M,
           2008A&A...485..541P,
           2009A&A...497..815B,
           2018MNRAS.480.1363Z}. 
The mass ratio of the binary components, 
$M_{\rm RG}/M_{\rm WD} = 0.56$, and the assumed WD mass of 
1.2--1.4\mo, correspond to the orbital inclination of 
49${\degr}$--$52{\degr}$ \citep[][]{2009A&A...497..815B}. 

The attractiveness of the recurrent novae phenomenon has led 
to the use of available observational techniques to obtain 
observations in the widest possible energy range, as well as 
spatial imaging of the explosion remnants. Radio observations 
of the 1985 explosion, thus first detected the non-spherical 
shape of the nova ejecta \citep[][]{1989MNRAS.237...81T}, which 
was later confirmed and refined for the 2006 
\citep[e.g.,][]{2006Natur.442..279O,2008ApJ...688..559R} and 
the 2021 eruption \citep[][]{2022A&A...666L...6M,
                             2023MNRAS.523..132D,
                             2024A&A...692A.107L}. 
The latter studies on the 2021 outburst showed that the bipolar 
structure of the radio emission is remarkably similar to that 
from the 2006 outburst, with a density enhancement in the orbital 
plane that obscures the receding emission lobe 
\citep[][]{2022A&A...666L...6M}. 
In particular, for the ejecta of the 2006 outburst, the bipolar 
shaping was also indicated by the interferometric technique in 
the near- and mid-IR 
\citep[e.g.,][]{2007A&A...464..119C,2008ApJ...677.1253B}, by 
the $HST$ imaging in the optical \citep[][]{2007ApJ...665L..63B}, 
and by the $Chandra$ X-ray Observatory 
\citep[][]{2009ApJ...707.1168L}. Later, in 2009 and 2011, 
the $Chandra$ X-ray images detected the east--west extended 
emission with opening angles of $\approx$70${\degr}$, consistent 
with a picture of sharply slowing equatorial ﬂow but fast bipolar 
outﬂow \citep[][]{2022ApJ...926..100M}. 
Such shaping was recently supported by the discovery of 
a prolonged ($\sim$16$\times$5\,pc) super-remnant cavity 
surrounding RS~Oph formed by sweeping up local interstellar 
medium by the nova eruptions over its lifetime 
\citep[][]{2024MNRAS.529L.175H}. 
Also, the spectropolarimetric monitoring of the 2021 eruption 
revealed the presence of dust between days 2 and 9, the spatial 
distribution of which was asymmetric, with components both aligned 
with and perpendicular to the orbital plane of the binary 
\citep[][]{2023A&A...679A.150N}. 
Since the early dust was present at about the same time when both 
the optical brightness and $\gamma$-ray emission detected by 
Fermi-LAT peaked \citep[see Fig.~2 of][]{2022ApJ...935...44C}, 
the spectropolarimetric measurements supported the link between 
shocks and dust in RS~Oph \citep[see][]{2023A&A...679A.150N}. 

The 2021 outburst of RS~Oph was intensively observed in the 
$\gamma$-ray energies, using Fermi-LAT 
\citep[0.1 -- 23\,GeV,][]{2022ApJ...935...44C,2022PhRvD.106j3011Z}, 
H.E.S.S \citep[100\,GeV -- 2\,TeV,][]{2022Sci...376...77H}, MAGIC 
\citep[60 -- 250\,GeV,][]{2022NatAs...6..689A}, and LST-1 
\citep[$\sim$(40 -- 700)\,GeV,][]{2025A&A...695A.152A}. 
To explain the very-high-energy (VHE; 100\,GeV$ < E < 100$\,TeV) 
and high-energy (HE; 100\,MeV$ < E < 100$\,GeV) $\gamma$-ray emission, 
a hadronic scenario of particle acceleration in the nova shock was 
preferred over an alternative leptonic scenario in these studies. 
However, modeling the HE--VHE spectral energy distribution (SED) 
during 4 days after the outburst, \cite{2023ApJ...951...62D} 
showed that the VHE $\gamma$-ray data can be explained by a hadronic 
component, whereas the HE data satisfy rather a leptonic component. 
The SED model was consistent with radio low-frequency data. 
\cite{2023ApJ...947...70D} demonstrated that, contrary to previous 
studies, a single shock cannot simultaneously explain the observed 
$\gamma$-ray spectrum of RS~Oph. Instead, they proposed a model 
involving multiple shocks that reproduces the GeV to TeV spectrum 
and its temporal evolution. The model is supported by the presence 
of multiple distinct velocity components in the optical spectrum 
during the first few days of the explosion 
\citep[see][]{2021arXiv210901101M,2023A&A...671A..49T}. 
By modeling X-ray observations of RS~Oph obtained with NICER 
(0.2--12\,keV), \cite{2024ApJ...960..125I} found evidence of the 
interaction of the ejecta with dense equatorial matter during 
the initial phase, and with less dense matter in the polar regions 
in the later phases of the outburst, which is consistent with 
the shaping of the ejecta suggested by radio and $\gamma$-ray 
observations. 
Similar to other powerful $\gamma$-ray novae, RS~Oph shows 
a correlation between optical and $\gamma$-ray light curves 
\citep[LCs;][]{2022ApJ...935...44C}, suggesting that a significant 
portion of the energy of the internal shocks is converted into 
optical light around its maximum 
\citep[see][]{2014MNRAS.442..713M,
              2017NatAs...1..697L,
              2020NatAs...4..776A}.

Multifrequency observations of the last three RS~Oph explosions 
(1985, 2006, and 2021) revealed their striking similarity. For 
the 2006 and 1985 eruptions, \cite{1987rorn.conf....1R}, 
\cite{2009MNRAS.399..357B}, and \cite{2007PThPS.169..187N} 
pointed out the similarity of the LC profiles in the 
optical, near-IR, and supersoft X-ray regions. 
The similarity of the 2021 eruption to the previous one in 2006 
has also been noted in several studies, mostly as a side result 
\citep[e.g.,][]{2022NatAs...6..689A,
                2022A&A...666L...6M,
                2023A&A...674A.139A}. 
However, the supersoft X-ray source (SSS) emission 
of the 2021 explosion 
was found to be weaker and switched off earlier than in the 2006 
explosion \citep[][]{2022MNRAS.514.1557P,2023ApJ...955...37O}. 
Analyzing the X-ray grating spectra of both the 2006 and 2021 
outbursts, \cite{2023A&A...670A.131N} explained 
the lower SSS emission observed in 2021 by a higher absorption 
from cold (neutral) and hot (ionised) material in the line of 
sight. 

In this study, we analyze the optical continuum emitted 
by the material ejected during the 2021 RS~Oph explosion, with 
the primary goal of determining the evolution of its physical 
parameters and fundamental structure from 14 hours after the start 
of brightening to the supersoft X-ray source phase ($\approx$day 40). 
We also analyze the evolution of the \ha\ line, which helps us 
understand the properties of the ejecta, and, in 
particular, the surrounding environment it penetrates. 

Our analysis and its results are described in Sect.~\ref{s:results}, 
their discussion and connections with the results obtained from other 
spectral domains are given in Sect.~\ref{s:discuss}, whereas 
Sect.~\ref{s:summ} provides a summary of our results. 
%
%
%
\begin{table*}
\caption{
Log of observatories and instruments used to obtain amateur 
spectroscopic observations. 
}
\begin{center}
\begin{tabular}{cccccccc}
\hline
\hline
\noalign{\smallskip}
Observatory               &
Telescope$^{\dagger}$     & 
Spectrograph              & 
Camera                    &
Resolution$^{\ddagger}$   &
Observer                  &
Reference                 & 
Label$^{\ast}$            \\
\noalign{\smallskip}
\hline
\noalign{\smallskip}
%
OSJ-CA$^{(a)}$& 37  & Alpy\,600     & ATIK\,414EX  & 1,000 & Desrosiers
              &  2  & (i)  \\
FKO$^{(b)}$   & 40  & FBSPEC-IV     & ASI6200MM-PRO&  500 & Fujii 
              &1,4  & (ii) \\
WCO-UK$^{(c)}$& 28  & LISA          & SXVR-H694    & 1,000 & Boyd 
              &1,2  & (iv) \\
KOL-SK$^{(d)}$& 28  & LISA          & ATIK\,460EX  &  680 & Dubovsk\'y 
              &3,2  & (iii)\\
THO-UK$^{(e)}$& 28  & Alpy\,600     & ATIK\,428    & 1,000 & Leadbeater 
              &  2  & (vi)\\
LSO$^{(f)}$   & 36  & UVEX          & ATIK\,414EX  & 870  & Vra\v{s}\v{t}\'{a}k 
              &  2  & (vii)\\
LSO$^{(f)}$   & 36  & FEST          & QHY 294M-PRO & 13,500& Vra\v{s}\v{t}\'{a}k 
              &  2  & (vii)\\
SMM-SP$^{(g)}$& 40  & B60050-VI     & ATIK\,460EX  & 9,500 & Teyssier 
              &1,2  & (xiii)\\
SMM-SP$^{(g)}$& 40  & B60050-VI     & ASI\,2600MM  & 8,500 & Guarro
              &1,2  & (x) \\
DUR-FR$^{(h)}$& 51  & echelle       & ATIK\,460EX  &11,000 & Charbonnel 
              &2,4  & (xi) \\
ARM-AU$^{(i)}$& 25  & LISA          & ATIK\,314L   &16,000 & Bohlsen 
              &2,4  & (xiv)\\
OCT-FR$^{(j)}$& 18  & Lhires3\,2400 & ATIK\,460EX  &18,700 & Boussin 
              &  2  & (xv) \\
AQL-IT$^{(k)}$& 22  & Lhires3\,2400 & ST\,8300     &13,100 & Sollecchia 
              &  2  & (xx)\\
AQL-IT$^{(k)}$& 22  & Lhires\,600   & SXVR-H694    & 3,900 & Berardi 
              &  2  & (xviii)\\
UNM-US$^{(l)}$& 35  & Lhires3\,2400 & ATIK\,460EX  &20,000 & Doctor 
              &  2  & (xix)\\
HOR-UK$^{(m)}$& 35  & Lhires3\,2400 & ATIK\,460EX  &18,500 & Martin
              &  2  & (xxi)\\
ANT-FR$^{(n)}$& 20  & SOLEX500      & ASI\,183MM   & 1,300 & Buil 
              &  2  & (v)  \\
ANT-FR$^{(n)}$& 20  & SOLEX2400     & ASI\,183MM   &16,700 & Buil 
              &  2  & (xvi)\\
ORL-FR$^{(o)}$& 25  & Alpy\,600     & ATIK\,414EX  &  600 & Lecocq 
              &  2  & (viii)\\
OBE-FR$^{(p)}$& 31  & echelle       & ATIK\,460EX  &11,000 & Thizy
              &  2  & (ix) \\
OTO-FR$^{(q)}$& 40  & echelle       & ATIK\,460EX  &11,000 & Garde 
              &1,2  & (xii) \\
MAR-AU$^{(r)}$&100  & eShel         & ATIK\,460EX  &10,000 & Eldridge 
              &  2  & (xvii)\\
\noalign{\smallskip}
\hline
\end{tabular}
\end{center}
{\bf Notes.} 
$^{(a)}$\,Mont St-Joseph Observatory, 
$^{(b)}$\,Fujii Kurosaki Observatory, 
$^{(c)}$\,West Challow Observatory, 
$^{(d)}$\,Astronomical Observatory at the Kolonica Saddle, 
$^{(e)}$\,Three Hills Observatory, 
$^{(f)}$\,Liptovsk\'{a} \v{S}tiavnica Observatory, 
$^{(g)}$\,Santa Maria de Montmagastrell Observatory, 
$^{(h)}$\,Durtal Observatory, 
$^{(i)}$\,Mirranook Observatory, 
$^{(j)}$\,Observatory de l'Eridan, 
$^{(k)}$\,L'Aquila Bellavista Observatory, 
$^{(l)}$\,private station in Las Cruces, New Mexico, 
$^{(m)}$\,Huggins Observatory, Rayleigh, 
$^{(n)}$\,Antibes,
$^{(o)}$\,Orlienas,
$^{(p)}$\,Observatoire de la Belle Etoile,
$^{(q)}$\,Observatoire de la Tourbi\`{e}re, 
$^{(r)}$\,Mardella. \\
$^{\dagger}$\,Diameter of the primary mirror in cm, 
$^{\ddagger}$\,average resolution, 
$^{\ast}$\,label of the observatory in 
           Tables~\ref{tab:low}, \ref{tab:medium}, 
           and \ref{tab:hapar}. \\
{\bf References.} 
1 -- \cite{2017A&A...604A..48S}, 
2 -- \cite{2019CoSka..49..217T}, 
3 -- \cite{2014CoSka..43..429K}, 
4 -- \cite{2014A&A...569A.112S}. 
\label{tab:spec}
\end{table*}
%
%
\section{Observations}
\label{s:obs}
\subsection{Photometry}
\label{ss:phot}
For the purpose of this work, we used multicolor 
$BVR_{\rm C}I_{\rm C}$ photometry 
collected by members of the American Association of Variable Star 
Observers (AAVSO)\footnote{\url{https://www.aavso.org/databases}}, 
by members of the Variable Star Observers League in Japan (VSOLJ), 
Itoh, Kiyota, Moriyama, Sano, and Sato as collected by Hiroyuki 
Maehara\footnote{\url{
http://kws.cetus-net.org/~maehara/LCG.html}}, 
and those published by \cite{2022NatAs...6..689A}. 

Further $UBVR_{\rm C}I_{\rm C}$ photometry was carried out 
using the \textsl{FLI ML3041} CCD camera (2048$\times$2048 px, 
pixel size: 15\,$\mu$m $\times $15\,$\mu$m, scale: 0.4 arcsec/px, 
FoV: $14^{\prime}\times 14^{\prime}$), mounted at the Cassegrain 
focus of the 60 cm, f/12.5 telescope, operated by the Astronomical 
Institute of the Slovak Academy of Sciences. The data were reduced 
using the \textsl{IRAF} software package\footnote{\url{
http://iraf.noao.edu}} as described by \cite{2005CoSka..35...35P}. 
Corresponding magnitudes (Table~\ref{tab:ubvri}) were obtained 
using the comparison stars listed in Table~1 of 
\cite{2019CoSka..49...19S}. 

Multicolor photometry was primarily used to scale the relative 
flux units of the low-resolution spectra (Table~\ref{tab:low}) 
to absolute units and to verify the calibration of the 
high-resolution spectra (Table~\ref{tab:uves}), which were used 
for modeling the SED (see Appendix~\ref{app:calib}). 
%

%
\subsection{Spectroscopy}
\label{ss:spec}
In this work, we primarily use ground-based optical spectroscopy. 
Most of the spectra were obtained by amateurs within the initiative 
{\it Astronomical Ring for Access to Spectroscopy} (ARAS), which 
promotes cooperation between professional and amateur astronomers 
in the field of spectroscopy\footnote{
\url{https://aras-database.github.io/database/}}.  
The spectra containing the Balmer jump, obtained by one of us 
(MV), were crucial for our work. 
The amateurs spectra were taken at 18 different observatories 
and/or private stations. Basic information about their spectral 
acquisition equipment is introduced in Table~\ref{tab:spec}. 
The list of spectra with low ($R = 500 - 1\,300$) and medium 
resolution ($R = 8,500 - 20,000$) is in Table~\ref{tab:low} 
and \ref{tab:medium}. 

At the Ond\v{r}ejov Observatory, medium-resolution spectra 
($R\sim13,000$) covering the red spectral region were 
secured at the coud\'{e} focus of the 2.0\,m reflector 
\citep[see][]{2002PAICz..90.....S}. 
Standard initial reduction (bias subtraction, flat-fielding, 
creation of 1D spectra, and wavelength calibration) was 
performed using modified \textsl{IRAF} packages. 
The times and spectral ranges of individual observations 
are listed in Table~\ref{tab:medium}, where they are 
labeled by 'Ondrejov'. 

Furthermore, we used high-resolution spectra downloaded from 
the European Southern Observatory (ESO) Science Archive Facility 
with DOI: https://doi.org/10.18727/archive/50.
The spectra were obtained using the high-resolution echelle 
spectrograph UVES \citep[see][]{2000SPIE.4008..534D} attached 
to the Nasmyth B focus of the UT2 Kueyen telescope at the ESO 
Paranal Observatory \citep[see][]{2023MNRAS.518.2614M}. 
They cover the spectrum from $\sim$304 to $\sim$946\,nm and have 
a resolution power from $\sim$58\,640 to $\sim$107\,200. The list 
of the used UVES spectra is given in Table~\ref{tab:uves}. 
They are labeled as 'Paranal' in Table~\ref{tab:hapar}. 

The low-resolution spectra (Table~\ref{tab:low}), preferentially 
those covering the Balmer discontinuity, as well as 
the high-resolution UVES spectra, after an appropriate 
reduction of their resolution, were primarily used to model 
their SED (Sect.~\ref{ss:sed}). Their flux calibration was 
verified with the aid of the (near-)simultaneous 
$(U)BVR_{\rm C}I_{\rm C}$ photometry 
(see Appendix~\ref{app:calib}). 
Medium- and high-resolution spectra served to analyze variations 
in the line profiles and fluxes 
(Figs.~\ref{fig:ha}, and \ref{fig:brandnar}, 
\ref{fig:brandnar}, and \ref{fig:hahb}, Table~\ref{tab:hapar}). 
Absolute fluxes of their used parts were obtained with the aid 
of the two neighboring SED models (see Fig.~\ref{fig:cont}). 

The strong similarity of the RS~Oph outbursts 
(see Sect.~\ref{s:intro}) allows us to consider observations 
obtained at the same time after the optical maximum of different 
outbursts as simultaneous. 
Therefore, we also used ultraviolet (UV) spectra made 
with the {\it International Ultraviolet Explorer} ($IUE$) taken 
during the 1985 outburst and the near-infrared (near-IR) $JHKL$ 
photometry obtained during the 1985 and 2006 outbursts, published 
by \cite{1988MNRAS.234..755E}, \cite{2006IAUC.8683....3W}, and 
\cite{2009MNRAS.399..357B}. 
These observations confirmed the strong similarity between RS~Oph 
outbursts in the UV to near-IR and, conversely, verified 
the correctness of our optical SED models (see Sect.~\ref{sss:parhot}, 
Fig.~\ref{fig:seduvir}). 

Observations were dereddened with $E_{\rm B-V}$ = 0.69 
\citep[][]{2018MNRAS.480.1363Z} using the extinction curve of 
\cite{1989ApJ...345..245C} and the distance-dependent parameters 
were scaled to a canonical value of 1.6\,kpc 
\citep[][]{1986ApJ...305L..71H}. 
%
%
%
\section{Analysis and results}
\label{s:results}
\subsection{Times of maxima and onsets of the last 3 explosions}
\label{ss:maxima}
\begin{figure}[t!]
\begin{center}
\resizebox{8.5cm}{!}{\includegraphics[angle=-90]{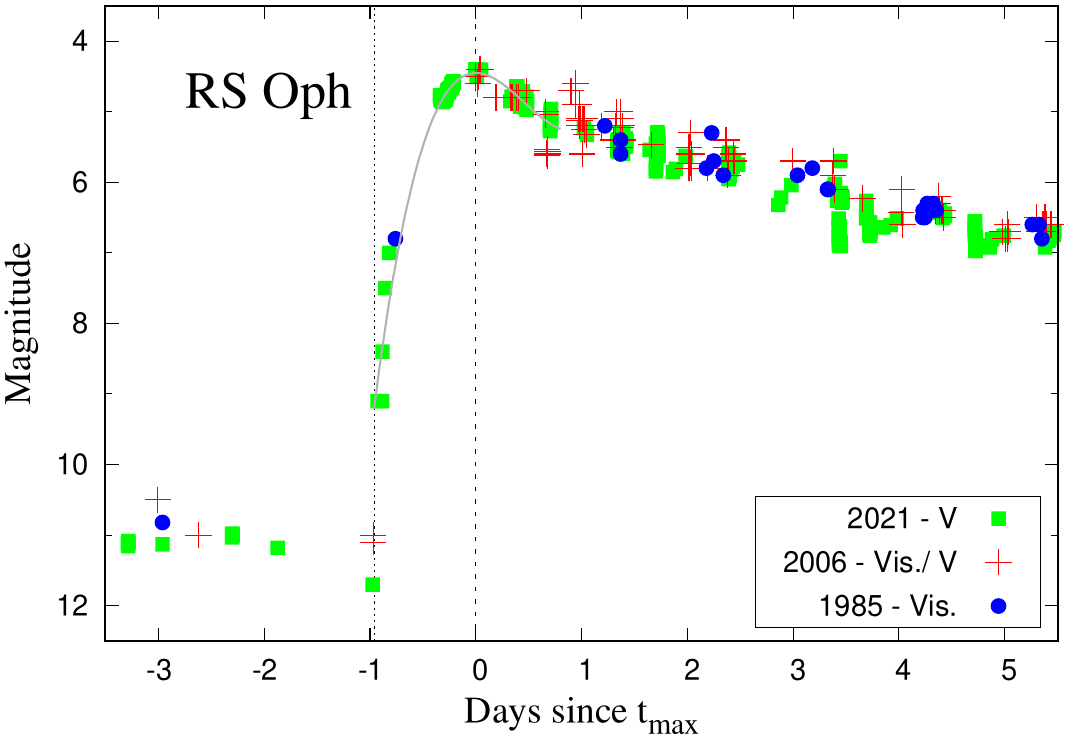}}
\end{center}
\caption{
Comparison of $V$ and visual LCs of RS~Oph around 
the maxima of its explosions in 2021, 2006, and 1985. 
The gray curve is a third-degree polynomial that fits 
the $V$ magnitudes around the 2021 maximum. 
The vertical dotted and dashed lines indicate the time 
of the brightness onset and its maximum, respectively 
(Table~\ref{tab:maxima}, Sect.~\ref{ss:maxima}). 
}
\label{fig:maxima}
\end{figure}

Figure~\ref{fig:maxima} compares the $V$ and visual LCs around 
the brightness maximum of the RS~Oph explosion in 1985, 2006, 
and 2021. The time of the maximum brightness, $t_{\rm max}$, 
could only be determined for the 2021 explosion, which was 
observed both before and after the maximum. 
By fitting 131 $V$ magnitudes around the maximum with 
a polynomial of the third degree, we determined $t_{\rm max}$ 
of the 2021 eruption at JD\,2,459,435.96$\pm$0.06 
(August 9.46$\pm$0.06, 2021). 
The last observation in quiescence (11.7, no filter) was 
made on August 8.49, 2021, by the VSOLJ observer Tadashi 
Kojima (vsnet-alert 26136). 
This date and the first date, when RS~Oph was indicated in 
outburst (August 8.53, 2021; $V$ = 9.10$\pm$0.032), allowed us 
to estimate the time of the start of the optical brightening, 
$t_{\rm 0}$, at JD\,2,459,435.01$\pm$0.02 
(August 8.51$\pm$0.02, 2021), 0.95$\pm$0.08 days before 
the time $t_{\rm max}$. 

The 2006 explosion was captured right at its visual maximum 
of 4.5\,mag by \cite{2006IAUC.8671....2N} on February 12.83, 
2006. A previous observation on February 11.86, indicated a 
quiescent phase of RS~Oph at visual magnitude 11.0 (see AAVSO 
database), 0.97 days before the peak brightness. The onset 
and the peak brightness of the 2006 explosion coincide with 
the timing of the 2021 explosion within its uncertainties 
(see Fig.~\ref{fig:maxima}). Therefore, for 
the 2006 explosion, we adopted the same uncertainties in 
$t_{\rm 0}$ and $t_{\rm max}$ times. 

The first visual magnitudes of the 1985 explosion, 6.8 and 5.2, 
were estimated by \cite{1985IAUC.4030....2M} in January 1985 
26.47 and 28.45, respectively. Based on the similarity 
of the LCs of all three explosions, shifting the 1985 LC to match 
the LC profiles of the 2006 and 2021 explosions showed that 
the first magnitude is located before the maximum, while 
the second one is placed after it 
(see Fig.~\ref{fig:maxima})\footnote{We omitted the estimate 
from January 27.20 (6.2\,mag) because it was more than 1.5\,mag 
below the predicted values.}. This allowed us to estimate 
the $t_{\rm max}$ time of the 1985 explosion at 
January 27.23$\pm0.12$ (JD\,2,446,092.73$\pm0.12$), 
where the uncertainty is the sum of 
the uncertainties in the $t_{\rm max}$ times of the explosions 
in 2006 and 2021, i.e., 0.12\,d. For time $t_{\rm 0}$ we adopted 
the same error as we estimated for the 2006 and 2021 eruptions. 
Table~\ref{tab:maxima} summarizes $t_{\rm max}$ and $t_{\rm 0}$ 
times for the 1985, 2006, and 2021 RS~Oph explosions. 

Finally, the time elapsed after time $t_{\rm max}$ 
is called the nova age, unless otherwise stated. 
%
%
%
%
%
\subsection{Modeling the SED}
\label{ss:sed}
In this section, using the method of disentangling composite 
spectra of symbiotic stars \citep[see][]{2005A&A...440..995S}, 
we determine the basic physical parameters of the WD 
pseudophotosphere (WDP) along the outburst, from our first 
observations on day $-$0.366 to the beginning of 
the SSS phase ($\sim$day 40). The presence of emission lines 
in the spectrum from the very beginning of the explosion, 
especially prominent \hi\ lines of the Balmer series 
\citep[e.g.,][]{2021ATel14838....1T,
                2021arXiv210901101M}, 
and the Balmer jump in emission 
(see Appendix~\ref{app:cont}) reflect the presence 
of a nebular continuum in the spectrum. Therefore, we assume 
that the spectrum of the ejecta consists of two main components 
of radiation: The stellar component(s) from the WDP(s), and 
the nebular continuum emitted by the hydrogen plasma. 

In the early stages of nova evolution, the radiation 
from the WDP often dominates the optical and resembles the spectra 
of A to F-type stars \citep[e.g.,][]{2008clno.book...16W}. 
Here we refer to it as {\it a warm} WDP. 
In later phases, usually when the nebular radiation starts 
to be more pronounced in the spectrum, the radiation from 
the warm WDP is better approximated by a blackbody radiation 
\citep[see Fig.~\ref{fig:sedopt} here or Fig.~13 of][]
{2025MNRAS.540.3549P}. 

A relative strong nebular component of radiation in the spectrum 
signals the presence of a hot ionizing source in 
the system -- {\it a hot} WDP, because the warm WDP is unable 
to generate the observed amount of nebular emission 
(see Sect.~\ref{ss:ionejecta}). 

\begin{figure}[t]
\begin{center}
\resizebox{\hsize}{!}{\includegraphics[angle=-90]{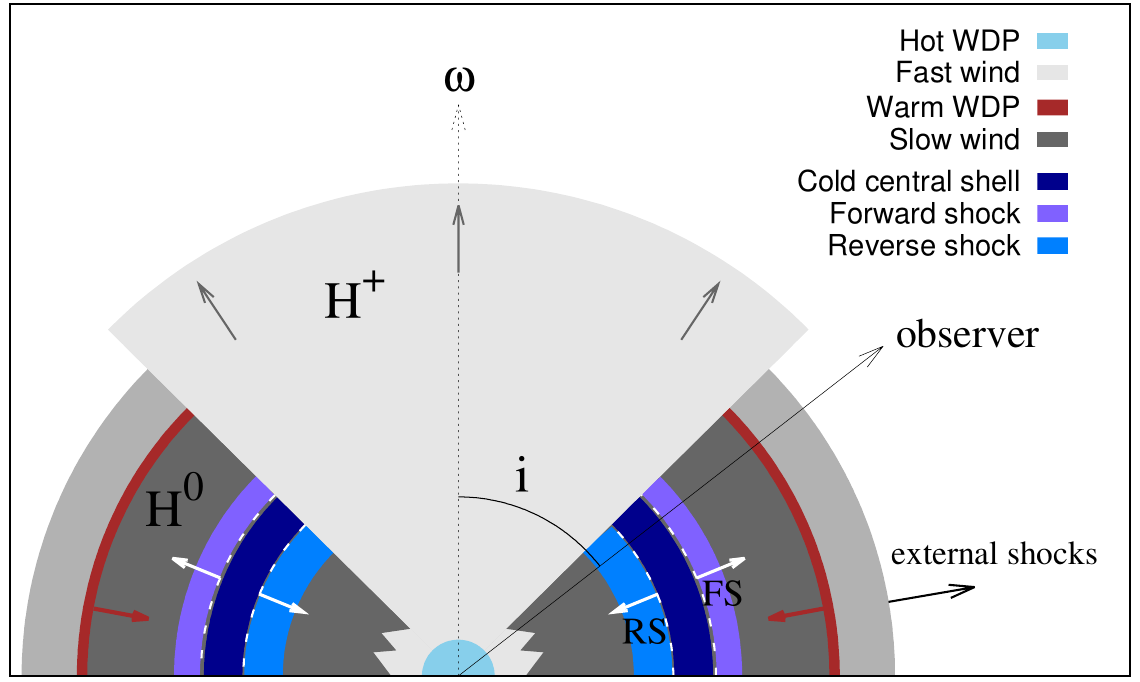}}
\end{center}
\caption{
Sketch for the RS~Oph ejecta inferred from SED models (side view, 
Sect.~\ref{ss:ionejecta}). 
Due to the WD rotation, the wind is compressed and slowed down 
towards the equator at the expense of the polar directions, where 
we indicate a fast ionized wind (Sect.~\ref{ss:cause}). 
The radius of the optically thick/thin compressed wind interface 
(i.e., the warm WDP, brown line) starts to shrink after 
the maximum due to its expansion 
and the lowering of $\dot M_{\rm WD}$ (brown arrow, see 
Table~\ref{tab:par}). The earliest optically thin wind above 
the warm WDP (medium gray band) shocks the surrounding much less 
dense and relatively stable giant wind. 
Structure of the internal shocks in the slow dense outflow was 
adapted according to Fig.~1 of \cite{2015MNRAS.450.2739M}. 
White arrows indicate the propagation of forward (FS) and 
reverse (RS) shocks. H$^0$ and H$^{+}$ denote the neutral and 
ionized parts of the ejecta, and $i$ is the orbital inclination. 
The axis of rotation ($\omega$) is shown by the dotted arrow. 
          }
\label{fig:sketch}
\end{figure}

The radiation of the warm and hot WDP is described and modeled in 
Sects.~\ref{sss:parwarm} and \ref{sss:parhot}. Their geometrical 
structure within the ejecta is suggested in 
Sect.~\ref{ss:ionejecta} and sketched in Fig.~\ref{fig:sketch}. 
%
%
%
\begin{table}[t!]
\caption{
Times of brightening onset, $t_{\rm 0}$, and maximum, $t_{\rm max}$, 
for the RS~Oph explosions in 2021, 2006, and 1985 
(see Sect.~\ref{ss:maxima}). 
}
\begin{center}
\begin{tabular}{ccc}
\hline
\hline
\noalign{\smallskip}
Outburst & $t_{\rm 0}$    & $t_{\rm max}$               \\
         & Date (UT)      & Date (UT)                   \\
         & JD$_0$\,2,4...     & JD$_{\rm max}$\,2,4...  \\
         & Orbital phase$^{(a)}$  & Orbital phase       \\
\noalign{\smallskip}
\hline
\noalign{\smallskip}
 2021 & August 8.51$\pm$0.02      & August 9.46$\pm$0.06     \\
      & 59,435.01$\pm$0.02        & 59,435.96$\pm$0.06       \\
      &   0.727                   &    0.729                 \\
 2006 & February 11.86$\pm$0.02   & February 12.83$\pm$0.06  \\
      & 53,778.36$\pm$0.02        & 53,779.33$\pm$0.06       \\
      &   0.257                   &    0.259                 \\
 1985 & January 26.27$\pm$0.02    & January 27.23$\pm$0.12   \\
      & 46,091.77$\pm$0.02        & 46,092.73$\pm$0.12       \\
      &   0.311                   &    0.313                 \\
\noalign{\smallskip}
\hline
\end{tabular}
\end{center}
{\bf Notes.} 
$^{(a)}$\,According to the ephemeris of the inferior conjunction 
of the RG, 
$T_{\rm conj.} = {\rm JD}\,2,445,043.54 + 453.6\times E$ 
\citep[][]{2009A&A...497..815B}. 
%
\label{tab:maxima}
\end{table}
%
%
%
%
%
\subsubsection{Parameters of the warm WD pseudophotosphere and the nebula}
\label{sss:parwarm}
According to the predicted components of the radiation from 
the RS~Oph ejecta, similar to the case of CN V339~Del SED 
modeling \citep[see Sect.~3.1. of][]{2014A&A...569A.112S}, 
the reddening-free continuum of the RS~Oph ejecta measured 
at Earth, can be written in the form, 
%
\begin{equation}
 F(\lambda) = 
   \theta_{\rm WD}^2\,\mathcal{F}_{\lambda}(T_{\rm eff})\,
   +\,k_{\rm N}\times\varepsilon_{\lambda}({\rm H},T_{\rm e}),
\label{eq:fl1}
\end{equation}
or
\begin{equation}
 F(\lambda) = 
   \theta_{\rm WD}^2 \pi B_{\lambda}(T_{\rm BB})\,
   +\,k_{\rm N}\times\varepsilon_{\lambda}({\rm H},T_{\rm e}), 
\label{eq:fl2}
\end{equation}
where the first term at the right represents the stellar component 
of radiation from the warm WDP compared to a synthetic 
spectrum $\mathcal{F}_{\lambda}(T_{\rm eff})$ (Eq.~(\ref{eq:fl1}))
or to a blackbody radiation (Eq.~(\ref{eq:fl2})), calculated for 
the temperature $T_{\rm eff}$ or $T_{\rm BB}$, respectively. 
Their scaling to the measured spectrum, the angular radius 
of the WDP, $\theta_{\rm WD} = R_{\rm WD}^{\rm eff}/d$ is given by 
its effective radius $R_{\rm WD}^{\rm eff}$ (which is 
the radius of a sphere with the same luminosity) and 
the distance $d$. 
The second term at the right is the nebular continuum expressed 
by its volume emission coefficient, 
$\varepsilon_{\lambda}({\rm H},T_{\rm e})$ 
(${\rm erg\,cm^3\,s^{-1}\,\AA^{-1}}$), scaled with the factor 
$k_{\rm N} = EM/4\pi d^2$, where $EM$ (\cmt) is the emission 
measure given by the proton and electron concentration within 
a volume of the ionized hydrogen. The electron temperature 
$T_{\rm e}$ and thus $\varepsilon_{\lambda}({\rm H},T_{\rm e})$ 
are assumed to be constant throughout 
the nebula\footnote{This assumption is justified by 
the fact that elastic collisions between free electrons occur on 
a significantly shorter timescale than their inelastic collisions 
with other atoms or ions in the plasma, which leads to the rapid 
thermalization of electrons and, hence, a Maxwellian velocity 
distribution \citep[][]{1947ApJ...105..131B}.}. 
For the sake of simplicity, we considered only contributions 
from f--b and f--f transitions in hydrogen plasma. 

In modeling the SED, we compared a grid of models given by 
Eq.~(\ref{eq:fl1}) or (\ref{eq:fl2}) with the observed spectrum. 
The grid of synthetic spectra was prepared using the library of 
\cite{2005A&A...442.1127M} for $T_{\rm eff} = 5,000 - 15,000$\,K 
with a step of 250\,K, and fixed other atmospheric parameters 
($\log(g) = 2.5$, [M/H] = 0, [$\alpha$/{\rm Fe}] = 0, 
$v_{\rm rot} = 20$\,\kms, and microturbulent velocity = 
2\,\kms)\footnote{The fixed parameters mainly affect the spectral 
line profile, while $T_{\rm eff}$ mainly determines the continuum 
profile.}. 
For the spectra covering the Balmer discontinuity, we estimated 
the continuum fluxes by eye, adopted their uncertainty at 5\%, and 
selected the model corresponding to a minimum of the reduced $\chi^2$ 
function. 
In the remaining cases, when the Balmer jump region is not 
available, we first estimated $T_{\rm e}$ with the aid of models 
of the neighboring spectra covering this region\footnote{Note 
that $T_{\rm e}$ determines the height of the Balmer jump and 
the slope of the nebular continuum.}. 
Second, we adjusted the other parameters ($k_{\rm N}$, 
$\theta_{\rm WD}$, $T_{\rm eff}$/$T_{\rm BB}$) and, based on 
our experience, we selected the one whose model matched 
the observed continuum. 
In this way we determined the model parameters, $\theta_{\rm WD}$, 
$T_{\rm WD}$ (i.e., $T_{\rm eff}$ or $T_{\rm BB}$), $k_{\rm N}$, 
and $T_{\rm e}$ that correspond to the luminosity of the warm WDP, 
$L_{\rm WD}^{\rm warm} = 4\pi d^2 \theta_{\rm WD}^2 \sigma T_{\rm WD}^4$, 
$R_{\rm WD}^{\rm eff} = d\,\theta_{\rm WD}$, and 
$EM = 4\pi d^2 k_{\rm N}$ of the nebular continuum 
(see Table~\ref{tab:par}, Fig.~\ref{fig:sedopt}). 
Finally, the radiation from the giant was adopted according to 
\cite{2015NewA...36..128S}. However, its contribution in 
the spectrum was negligible, at least, until $\sim$day 30 
(see Fig.~\ref{fig:seduvir}). 

We note that the warm WDP dominated the spectrum only around 
the brightness maximum. From day $-$0.366 to day $+$0.130, 
its effective radius increased from $\sim$114 to $\sim$290\ro, 
and cooled from $\sim$11,000 to $\sim$7,250\, K, respectively. 
Before the maximum, the nebular spectrum was indicated only by 
hydrogen emission lines. Its continuum could not be extracted 
because it was relatively faint, and our spectrum did not contain 
the Balmer jump region. The nebular continuum was first 
unambiguously determined by the UVES spectrum on day 0.617 
(see Fig.~\ref{fig:sedopt}\,d). 
%
%
%
\begin{table*}[t]
\caption{Physical parameters of the nova RS~Oph 
from the first hours of its life.
        }
\begin{center}
\begin{tabular}{cccccccccc}
\hline
\hline
\noalign{\smallskip}
                                                   &
\multicolumn{3}{c}{Warm WD pseudophotosphere}      &
\multicolumn{3}{c}{Hot WD pseudophotosphere}       &
\multicolumn{2}{c}{Nebula}                         &
                                                  \\
                                                   &
\multicolumn{3}{l}{---------------------------------------------} &
\multicolumn{3}{l}{------------------------------------}          &
\multicolumn{2}{l}{--------------------------}                    &
                                                                 \\
Age$^{\rm a}$                                 & 
$T_{\rm eff}/T_{\rm BB}$                      & 
$R_{\rm WD}^{\rm eff}$                        & 
$L_{\rm WD}^{\rm warm}$                       & 
$T_{\rm BB}^{\rm ion}$                        & 
$R_{\rm WD}^{\rm eff}$                        &
$L_{\rm WD}^{\rm hot}$                        & 
$T_{\rm e}$                                   & 
$EM$                                          & 
$\dot M_{\rm WD}$                             \\
JD-JD$_{\rm max}$                             &
(kK)                                          &
($R_{\odot}$)                                 & 
($10^{37}{\rm erg\,s^{-1}}$)                  & 
(kK)                                          & 
($R_{\odot}$)                                 & 
($10^{39}{\rm erg\,s^{-1}}$)                  &
(kK)                                          &
($10^{62}\,{\rm cm^{-3}}$)                    & 
($10^{-4}$\myr)                               \\
%
\noalign{\smallskip}
\hline
\noalign{\smallskip}
-0.366 & 11.0$\pm 0.5^{\rm b}$    & 114$\pm 6$  & 65$\pm 6$ 
       &   --                     &     --      &   --         
       &   --                     &     --                     
       &   --   \\  
 0.130 & 7.25$\pm 0.25^{\rm b}$   & 290$\pm 14$ & 80$\pm 8$ 
       &   73$^{\rm d}$           & 3.7         & $>$1.4
       &  $\approx$15             & $\gtrapprox$1.5
       &  $\sim$2.29  \\
 0.348 & 7.15$\pm 0.15^{\rm b}$   & 258$\pm 13$ & 60$\pm 7$
       &   73$^{\rm d}$           &  4.8        & $>$2.3
       & $\sim$16                 & $\sim$2.6 
       &  $\sim$3.47  \\
 0.478 & 7.00$\pm 0.20^{\rm b}$   & 251$\pm 13$ & 52$\pm 6$
       &   73$^{\rm d}$           &  5.0        & $>$2.4
       & $\sim$18                 & $\sim$3.1 
       &  $\sim$3.85  \\
 0.617 & 7.00$\pm 0.20^{\rm b}$   & 201$\pm 11$ & 34$\pm 5$
       &   73$^{\rm d}$           & 6.3         & $>$3.9
       & 18.0$\pm 2$              & 4.90$\pm 0.30$  
       &  5.50  \\
 1.006 & 7.00$\pm 0.20^{\rm b}$   & 179$\pm 9$  & 27$\pm 4$
       &   73$^{\rm d}$           &  6.1        & $>$3.6
       & 19.0$\pm 2$              & 4.90$\pm 0.35$  
       & 5.36  \\
 1.409 & 7.00$\pm 0.20^{\rm b}$   & 164$\pm 9$  & 22$\pm 3$
       &   73$^{\rm d}$           &  5.2        & $>$2.7
       & 19.0$\pm 2$              & 3.67$\pm 0.27$
       &  4.47  \\
 1.567 & 7.00$\pm 0.20^{\rm b}$   & 161$\pm 9$  & 22$\pm 3$
       &    73$^{\rm d}$          & 5.0         & $>$2.4
       & 19.5$\pm 2$              & 3.32$\pm 0.24$
       &  4.15 \\
 2.391 & 7.20$\pm 0.20^{\rm c}$   & 144$\pm 8$  & 19$\pm 2$
       & 73$^{\rm d}$             & 4.8         & $>$2.3
       & 19.0$\pm 2$              & 3.10$\pm 0.21$
       &  3.99  \\
 3.395 & 7.50$\pm 0.30^{\rm c}$   & 105$\pm 5$  & 12$\pm 2$
       & 73$^{\rm d}$             & 4.6         & $>$2.1
       & 16.0$\pm 2$              & 2.40$\pm 0.20$
       &  3.31  \\
 3.759 & 5.20$\pm 0.30^{\rm c}$   &190$\pm 10$   & 9.1$\pm 1.0$
       & 73$^{\rm d}$             &  3.7         & $>$1.4
       & 26.0$\pm 2$              & 2.42$\pm 0.20$
       &  3.03  \\
 4.751 & 4.60$\pm 0.20^{\rm c}$   &206$\pm 10$   & 6.6$\pm 1.0$
       & 73$^{\rm d}$             &  3.6         & $>$1.3 
       & 26.0$\pm 2$              & 2.25$\pm 0.20$
       &  2.88  \\
 5.416 & 4.40$\pm 0.20^{\rm c}$   & 196$\pm 10$  & 5.0$\pm 0.7$
       & $\sim$86$^{\rm e}$       & $\sim$2.5    & $\sim$1.2 
       & 25.0$\pm 2$              & 1.96$\pm 0.20$
       &  2.13  \\
 5.759 & 4.10$\pm 0.30^{\rm c}$   & 223$\pm 13$  & 4.8$\pm 0.7$
       & $\sim$89$^{\rm e}$       & $\sim$2.2    & $\sim$1.1 
       & 26.0$\pm 2$              & 1.84$\pm 0.20$
       &  2.02  \\
 7.343 & 3.70$\pm 0.20^{\rm c}$    & 230$\pm 12$  & 3.4$\pm 0.6$
       & $\sim$84$^{\rm e}$        & $\sim$2.4    & $\sim$0.99 
       & 18.0$\pm 2$               & 1.23$\pm 0.15$
       &  1.61  \\
 9.510 & 3.90$\pm 0.30^{\rm c}$    & 183$\pm 12$  & 2.7$\pm 0.5$
       & $\sim$80$^{\rm e}$        & $\sim$2.2    & $\sim$0.67
       & 19.5$\pm 2$               & 0.92$\pm 0.11$
       &  1.41  \\
11.048 & 3.70$\pm 0.20^{\rm c}$    & 179$\pm 10$  & 2.1$\pm 0.4$
       & $\sim$84$^{\rm e}$        & $\sim$1.9    & $\sim$0.62
       & 22.0$\pm 2$               & 0.89$\pm 0.09$
       &  1.18  \\
11.559 & 3.80$\pm 0.20^{\rm c}$    & 160$\pm 10$  & 1.8$\pm 0.4$
       & $\sim$92$^{\rm e}$        & $\sim$1.6    & $\sim$0.65
       & 20.0$\pm 2$               & 0.89$\pm 0.09$
       &  1.16  \\
12.421 & 3.80$\pm 0.20^{\rm c}$    & 183$\pm 10$  & 2.4$\pm 0.4$
       & $\sim$72$^{\rm e}$        & $\sim$2.2    & $\sim$0.45 
       & 22.0$\pm 2$               & 0.68$\pm 0.09$
       &  1.11  \\
15.998 & 3.80$\pm 0.20^{\rm c}$    & 139$\pm 8$   & 1.4$\pm 0.2$
       & $\sim$90$^{\rm f}$        & $\sim$1.6    & $\sim$0.57
       & 20.0$\pm 2$               & 0.58$\pm 0.08$
       &  0.863  \\
19.989 & 3.60$\pm 0.20^{\rm c}$    & 121$\pm 7$   & 0.85$\pm 0.2$
       & $\sim$117$^{\rm f}$       & $\sim$1.1    & $\sim$0.80
       & 19.0$\pm 2$               & 0.40$\pm 0.05$
       &  0.587  \\
21.583 & 3.80$\pm 0.20^{\rm c}$ & 105$\pm 6$   & 0.80$\pm 0.2$
       & $\sim$134$^{\rm f}$    & $\sim$0.94   & $\sim$0.98
       & 21.0$\pm 2$            & 0.41$\pm 0.05$
       &  0.563  \\
25.356 & $\sim$3.00$^{\rm c}$   & $\sim$85     & $\sim$0.20
       & $\sim$144$^{\rm f}$    & $\sim$0.90   & $\sim$1.2
       & 18.0$\pm 2$            & 0.34$\pm 0.04$
       &  0.498  \\
26.559 & 3.20$\pm 0.18^{\rm c}$ & 113$\pm 6$   & 0.46$\pm 0.1$
       & $\sim$153$^{\rm f}$    & $\sim$0.75   & $\sim$1.1
       & 23.0$\pm 2$            & 0.28$\pm 0.04$
       &  0.425  \\
31.349 &   --                   &  --          &  -- 
       & $\sim$157$^{\rm f}$    & $\sim$0.63   & $\sim$0.84
       & 20.0$\pm 2$            & 0.29$\pm 0.03$
       &  0.378  \\
32.549 & 3.60$\pm 0.20^{\rm c}$ & 59$\pm 4$    & 0.20$\pm 0.05$
       & $\sim$153$^{\rm f}$    & $\sim$0.56      & $\sim$0.59
       & 25.0$\pm 2$            & 0.21$\pm 0.03$
       &  0.319  \\
33.361 &  --                    &  --          &  -- 
       & $\sim$155$^{\rm f}$    & $\sim$0.63   & $\sim$0.79
       & 20.0$\pm 2$            & 0.21$\pm 0.03$
       & 0.329  \\
37.532 & 3.60$\pm 0.20^{\rm c}$ & 40$\pm 4$    & 0.09$\pm 0.03$
       & $\sim$155$^{\rm f}$    & $\sim$0.39   & $\sim$0.30
       & 25.0$\pm 2$            & 0.14$\pm 0.02$
       &  0.221  \\
42.566 & 3.60$\pm 0.20^{\rm c}$ & 39$\pm 4$    & 0.09$\pm 0.03$
       & $\sim$162$^{\rm f}$    & $\sim$0.37   & $\sim$0.32
       & 27.0$\pm 2$            & 0.11$\pm 0.015$
       &  0.180  \\
\noalign{\smallskip}
\hline
\end{tabular}
\end{center}
{\bf Notes:}\\
$^{\rm a}$\,JD$_{\rm max}$ = 2\,459\,435.96 (2021 August 9.46) 
            is the date of the optical maximum in $V$ 
            (Table~\ref{tab:maxima}). 
$^{\rm b}$\,$T_{\rm eff}$, 
$^{\rm c}$\,$T_{\rm BB}$, 
$^{\rm d}$\,adopted value, 
$^{\rm e}$\,$Zanstra$-temperature, 
$^{\rm f}$\,from \heii\,$\lambda$4686/\hb\ flux ratio 
            (see Sect.~\ref{sss:parhot} for determining 
            the temperature $T_{\rm BB}^{\rm ion}$). 
\label{tab:par}
\end{table*}
%
%
\subsubsection{Parameters of the hot WD pseudophotosphere}
\label{sss:parhot}
\begin{figure*}[t!]
\begin{center}
\resizebox{18cm}{!}{\includegraphics[angle=-90]{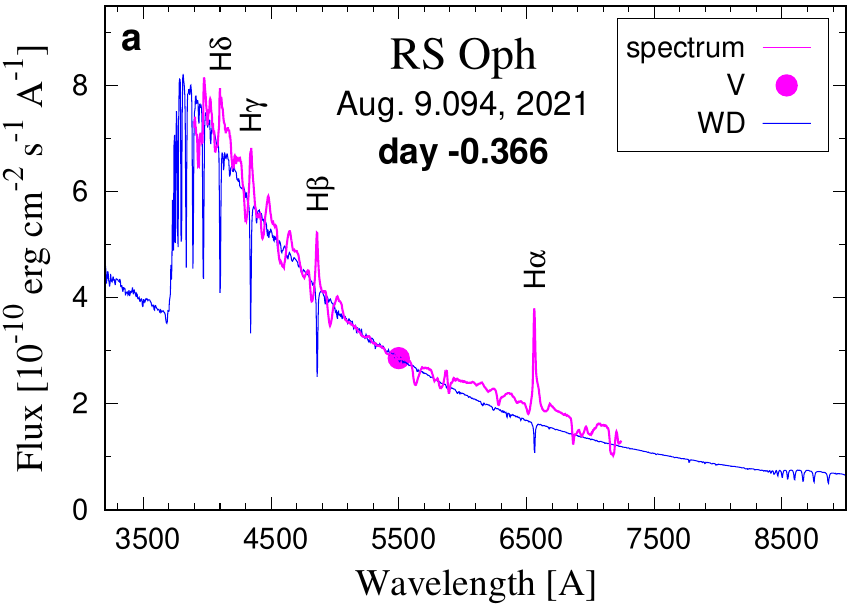}
                    \includegraphics[angle=-90]{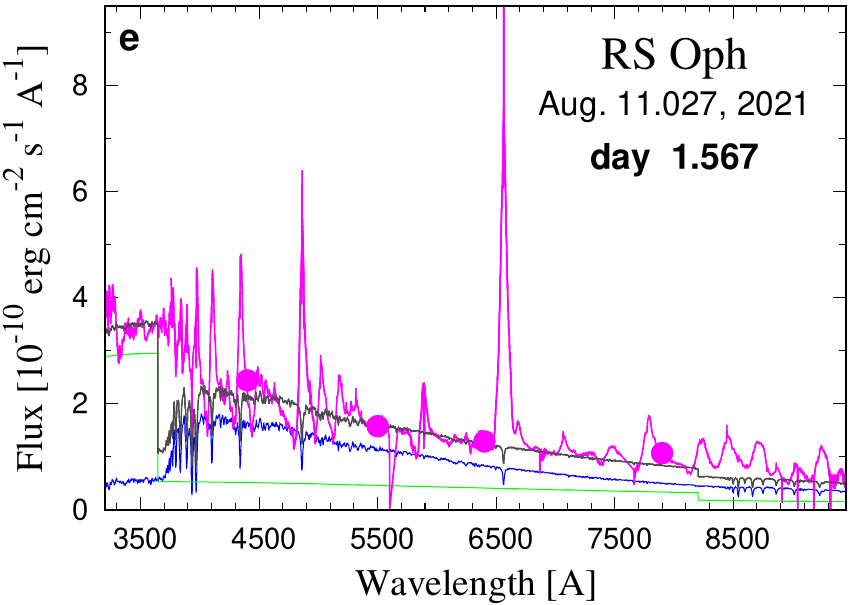}
                    \includegraphics[angle=-90]{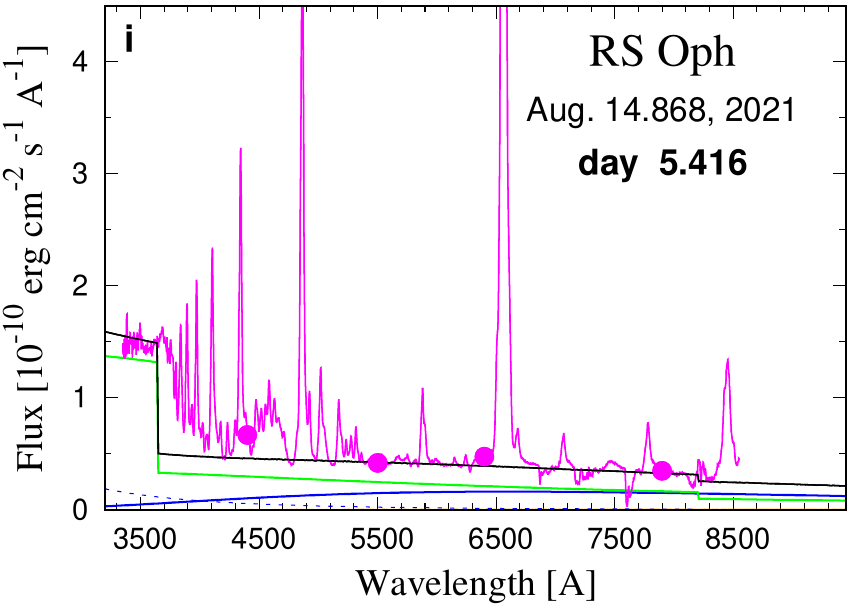}}
\resizebox{18cm}{!}{\includegraphics[angle=-90]{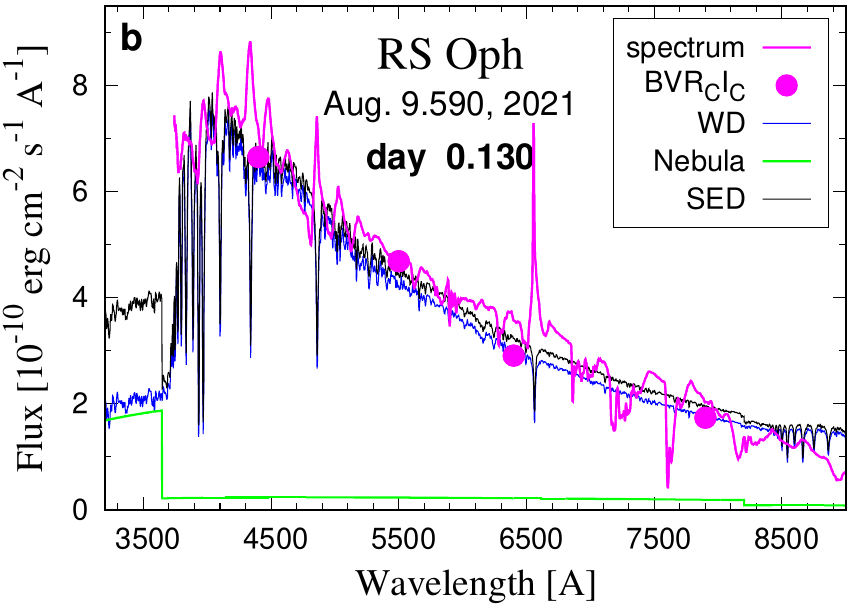}
                    \includegraphics[angle=-90]{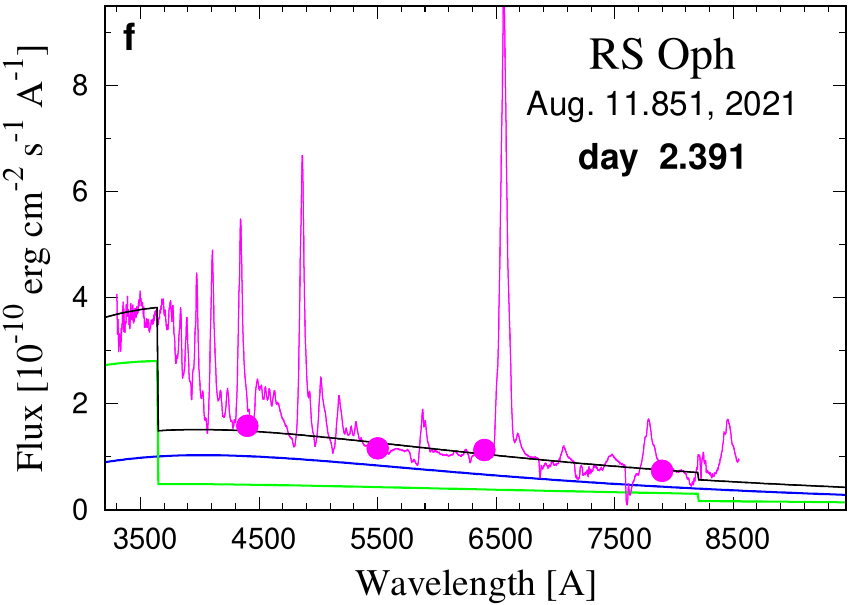}
                    \includegraphics[angle=-90]{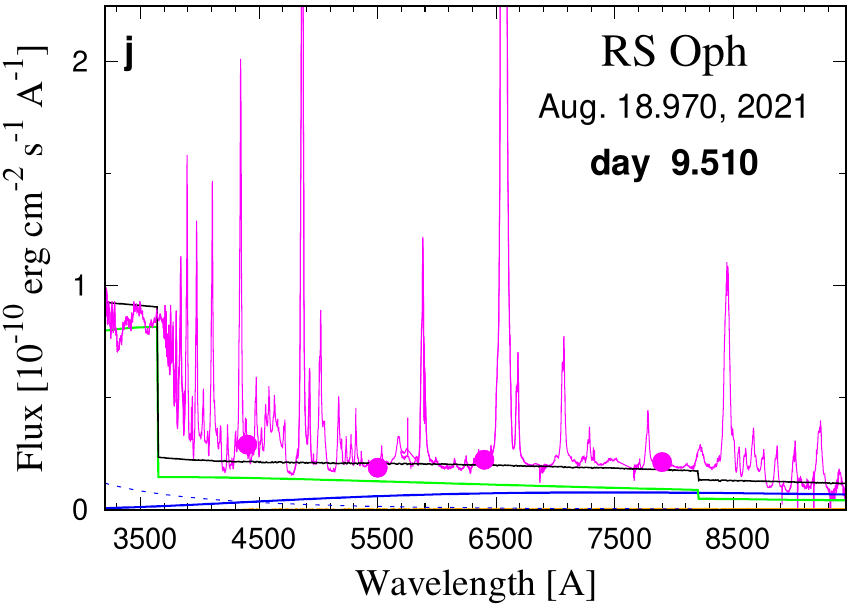}}
\resizebox{18cm}{!}{\includegraphics[angle=-90]{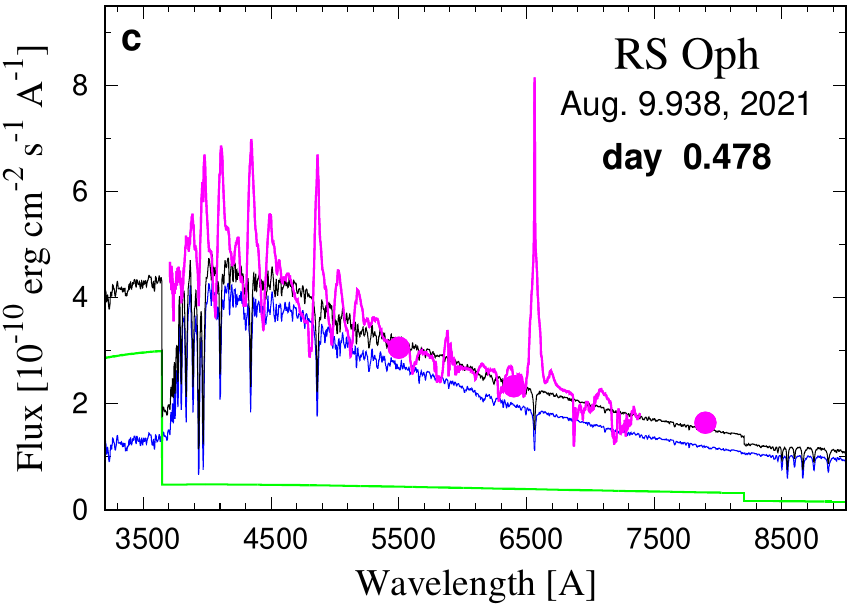}
                    \includegraphics[angle=-90]{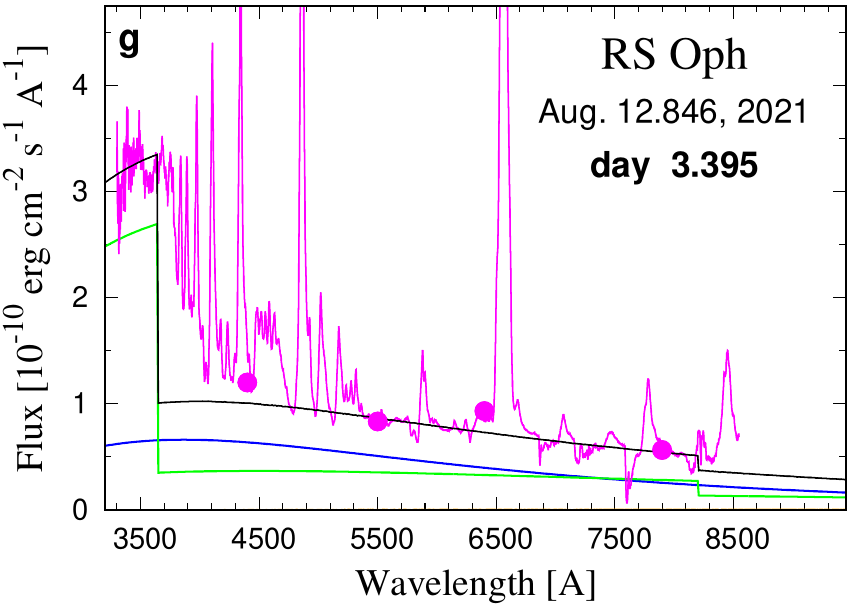}
                    \includegraphics[angle=-90]{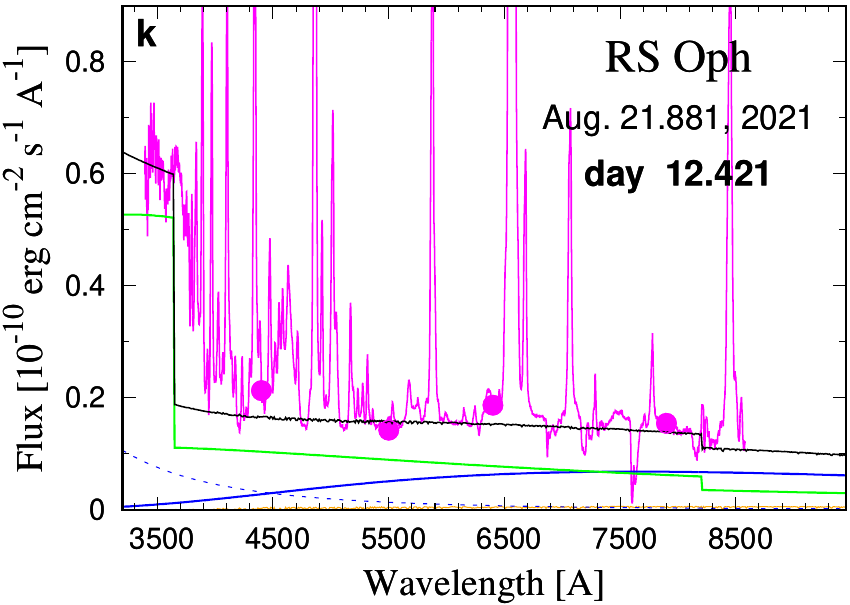}}
\resizebox{18cm}{!}{\includegraphics[angle=-90]{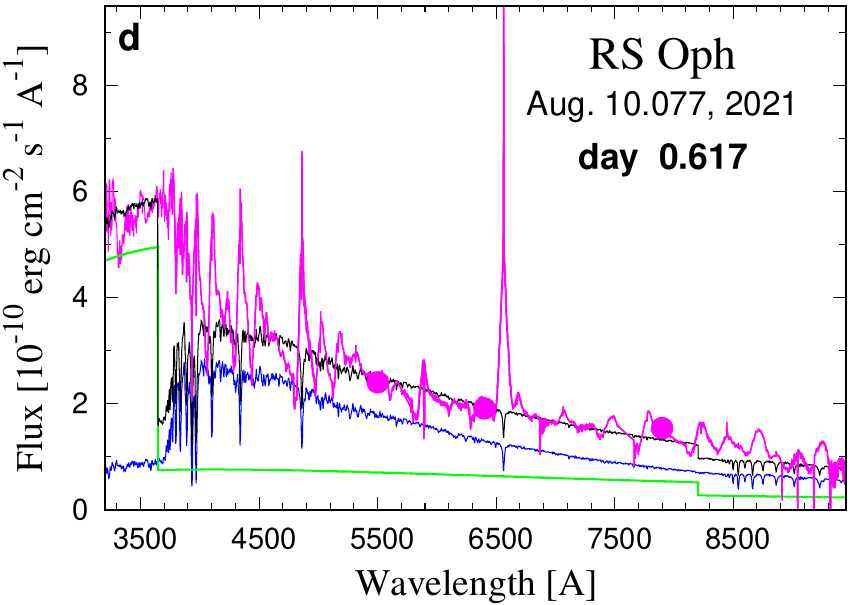}
                    \includegraphics[angle=-90]{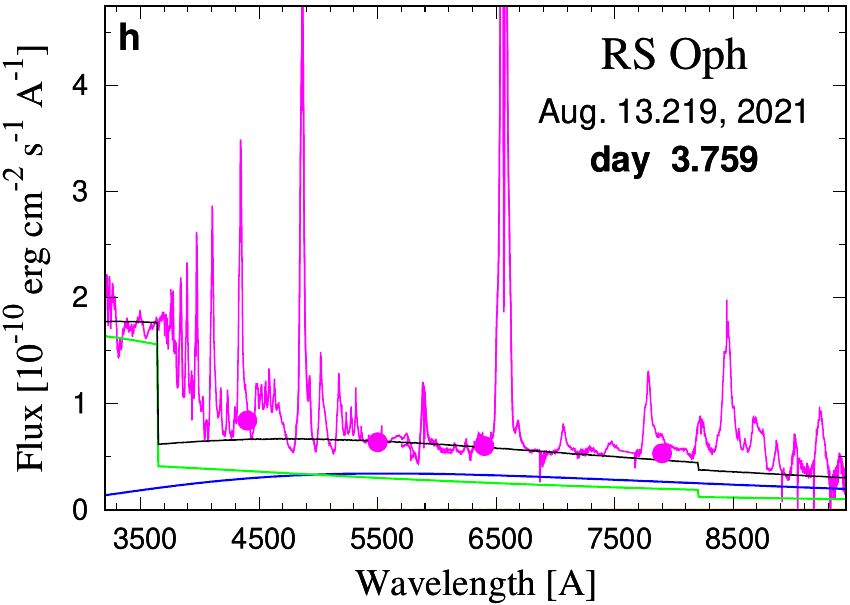}
                    \includegraphics[angle=-90]{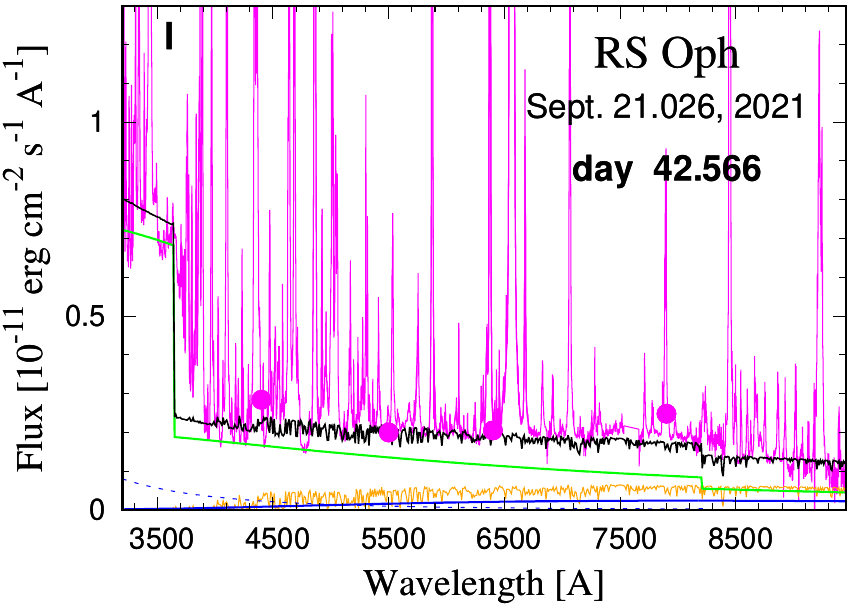}}
\end{center}
\caption{
Examples of the observed (in magenta) and modeled (black lines) 
SEDs of RS~Oph at selected dates from the onset of its 2021 
outburst. The meaning of lines and symbols is as shown in the keys 
(top left panels). The solid blue line represents the radiation 
of the warm WDP, while the dotted blue line (from day 5.416) 
represents the radiation of the hot WDP (see Sect.~\ref{ss:sed}). 
The spectrum of the giant (in orange, panel {\bf l}) was taken 
from \cite{2015NewA...36..128S}, recalculated for 
$E_{\rm B-V} = 0.69$. 
          }
\label{fig:sedopt}
\end{figure*}

\begin{figure*}[t!]
\begin{center}
\resizebox{18cm}{!}{\includegraphics[angle=-90]{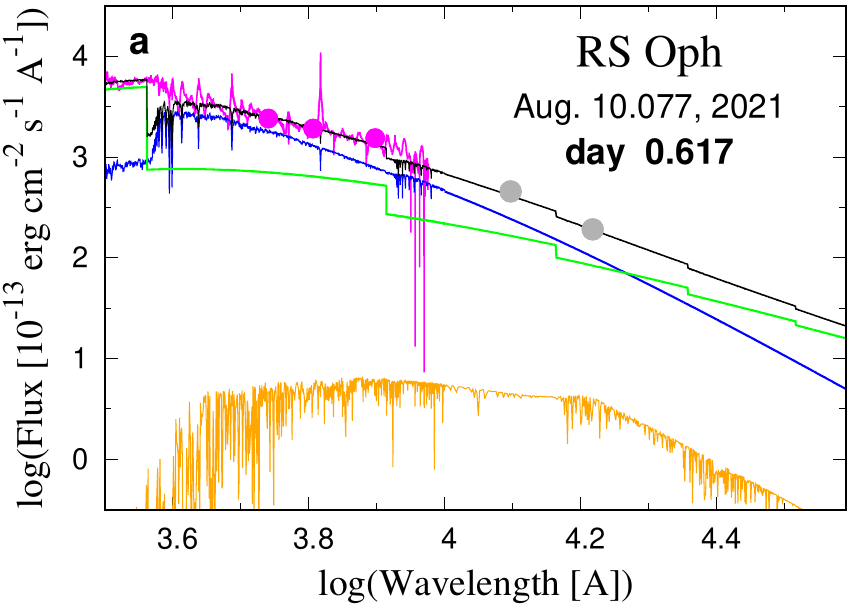}
                    \includegraphics[angle=-90]{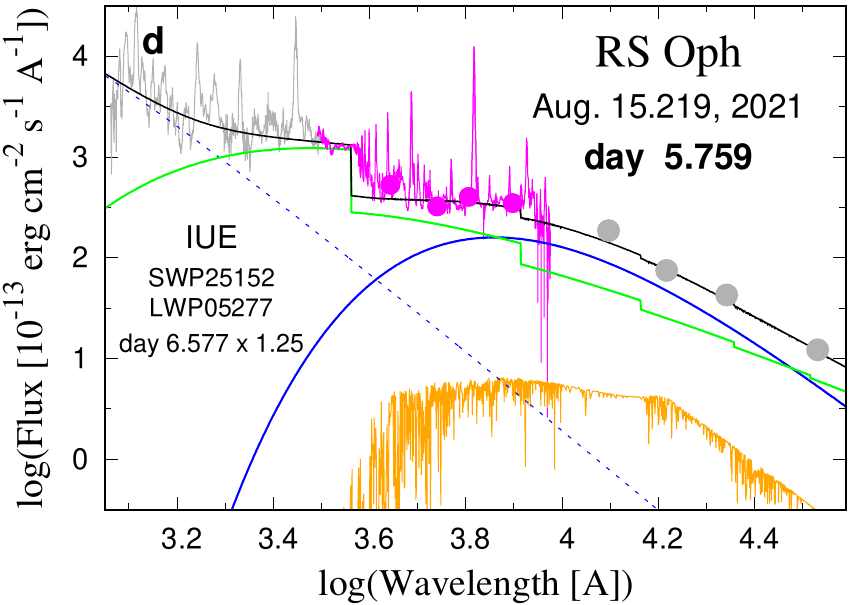}
                    \includegraphics[angle=-90]{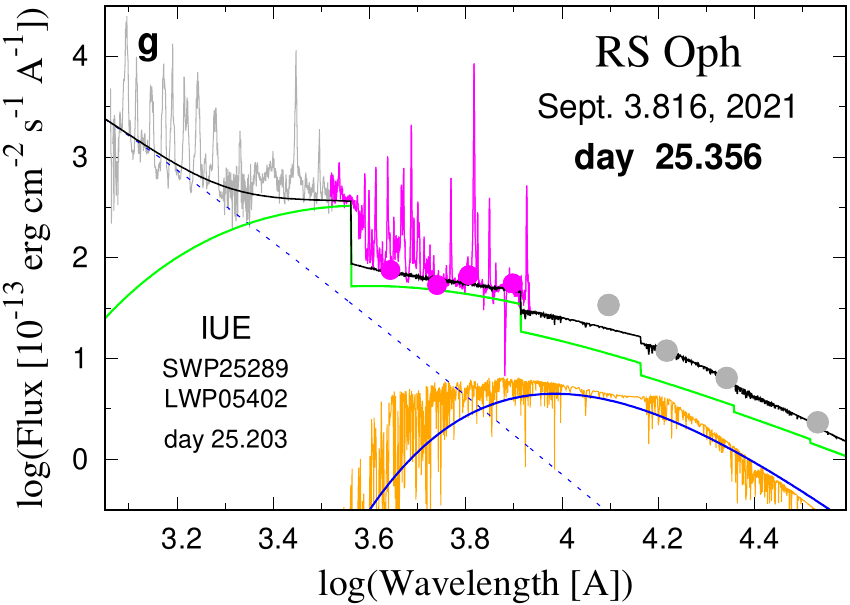}}
\resizebox{18cm}{!}{\includegraphics[angle=-90]{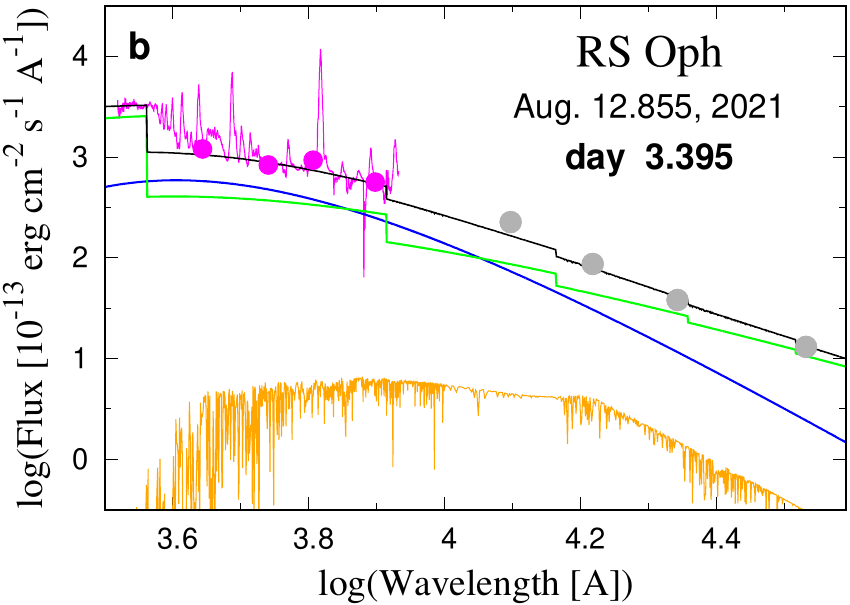}
                    \includegraphics[angle=-90]{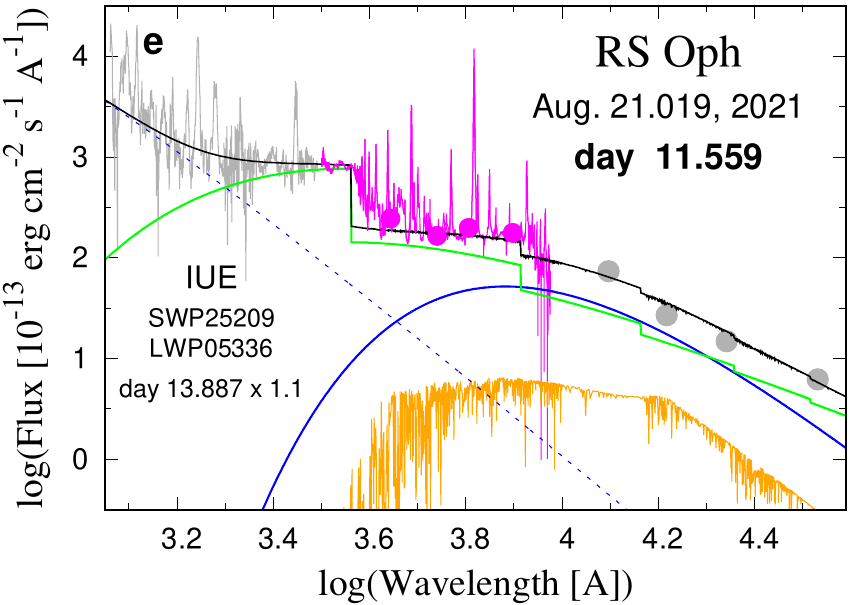}
                    \includegraphics[angle=-90]{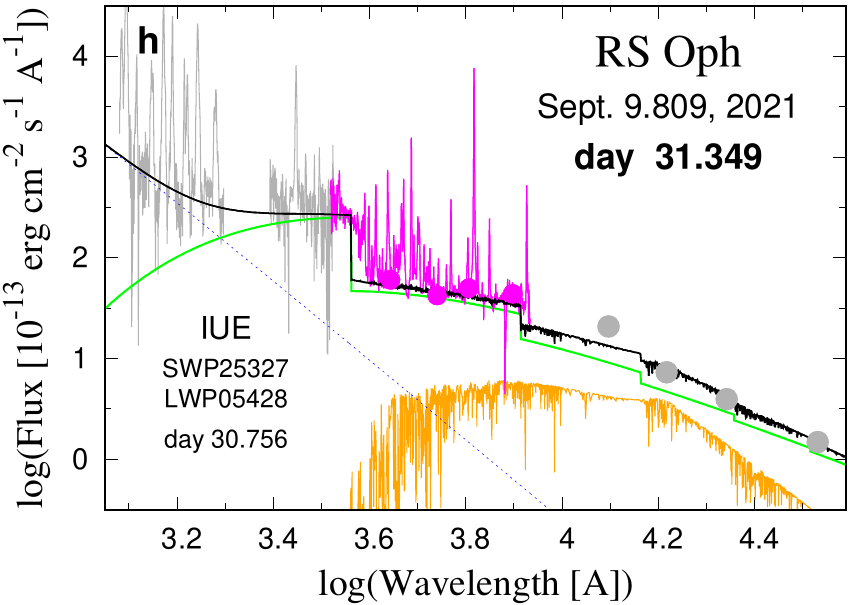}}
\resizebox{18cm}{!}{\includegraphics[angle=-90]{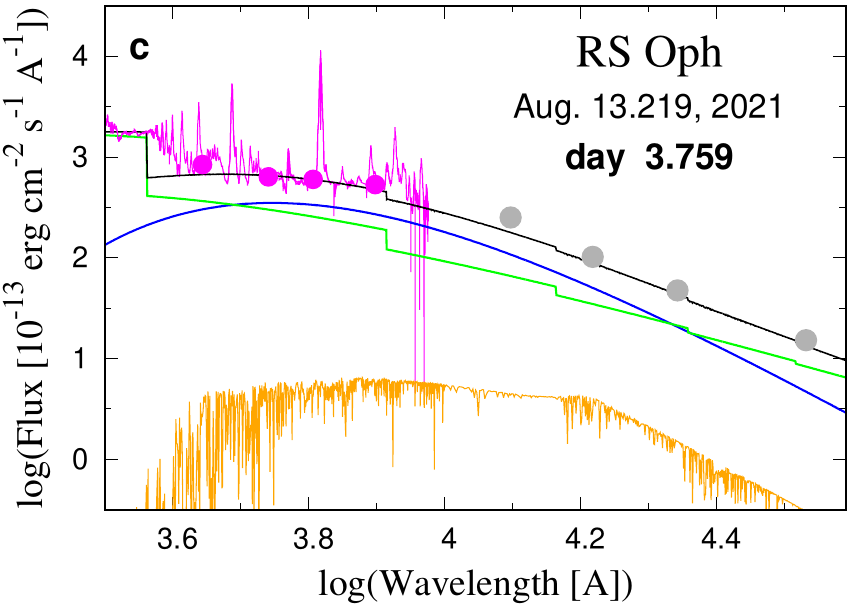}
                    \includegraphics[angle=-90]{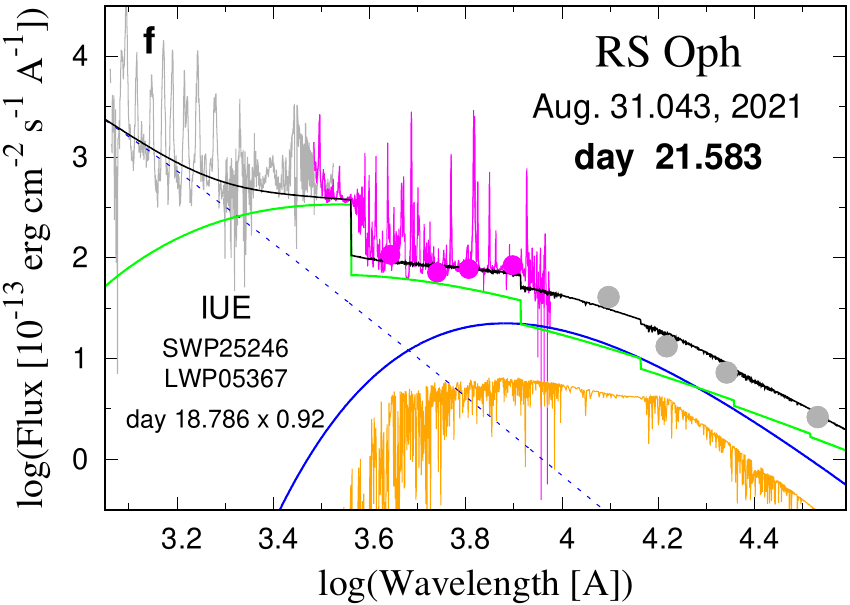}
                    \includegraphics[angle=-90]{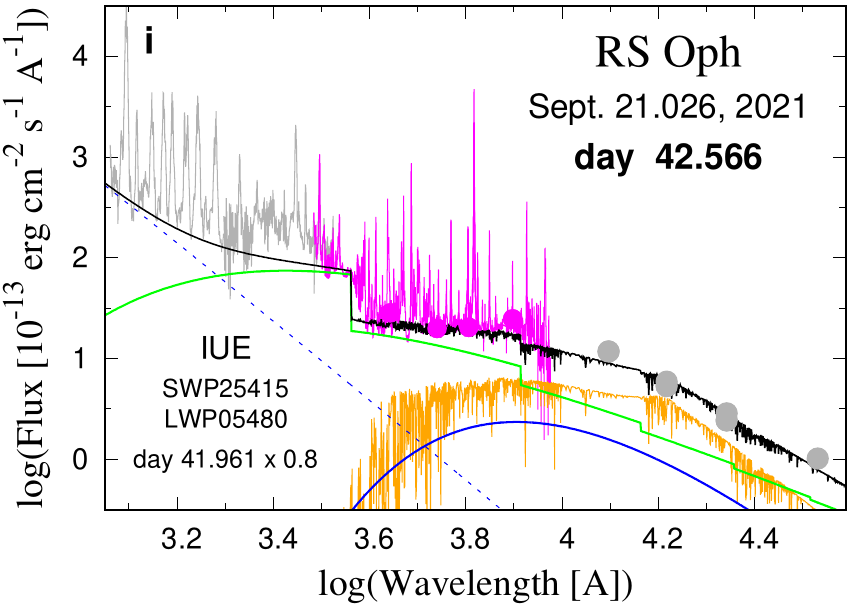}}
\end{center}
\caption{
Examples of SED models in the UV/optical/near-IR domain. Given 
the similarity of the RS~Oph eruptions, our observations were 
supplemented with (near-)simultaneous observations of the 
previous two eruptions (in gray). Denotation of lines and points 
is the same as in Fig.~\ref{fig:sedopt}. 
          }
\label{fig:seduvir}
\end{figure*}

The original radiation of the hot WDP was first measured with $IUE$ 
on day 6.6 after the peak of the 1985 outburst. This component 
dominates the far-UV (see Fig.~\ref{fig:seduvir}\,d -- i), but it 
is very weak in the optical -- seen as a faint Rayleigh-Jeans tail 
of a hot blackbody radiation (see Fig.~\ref{fig:sedopt}\,i -- l). 
We considered its contribution from day 5.416, when this 
component began to be identifiable in the vicinity of 
the Balmer jump (see Fig.~\ref{fig:sedopt}\,i). We adjusted its 
radiation using the far-UV fluxes of the nearest $IUE$ spectrum 
scaled to the short-wavelength part of the modeled optical 
spectrum. It slightly reduced the height of the Balmer jump, 
which required a slightly higher $T_{\rm e}$. 

Assuming that the nebular continuum grows from complete absorption 
of hydrogen ionizing photons emitted by the hot WDP, we can determine 
its fundamental parameters, $L_{\rm WD}^{\rm hot}$, 
$R_{\rm WD}^{\rm eff}$, and temperature, $T_{\rm BB}^{\rm ion}$. 
From day 5.416, when the flux of the hot WDP can be estimated from 
the far-UV, $T_{\rm BB}^{\rm ion}$ can be approximated by 
the temperature of a blackbody radiation that is just capable of 
producing the observed $EM$ when scaled to far-UV fluxes -- the 
so-called $Zanstra$--temperature. 
According to \cite{2015NewA...36..128S}, $T_{\rm BB}^{\rm ion}$ 
can be obtained by solving the equation 
%
%
\begin{equation}
  \frac{k_{\rm N}}{F_{\rm WD}(\lambda)}\alpha_{\rm B}
       ({\rm H},T_{\rm e})\pi B_{\lambda}(T_{\rm BB}^{\rm ion}) 
       - f(T_{\rm BB}^{\rm ion}) = 0 ,
\label{eq:tion}
\end{equation}
where the flux of the ionizing source, $F_{\rm WD}(\lambda)$, is 
selected at a wavelength $\lambda$, at which it dominates the 
spectrum (in the far-UV), $\alpha_{\rm B}({\rm H},T_{\rm e})$ is 
the total recombination coefficient to all but the ground state 
of hydrogen (cm$^{3}$\,s$^{-1}$), and the function 
$f(T_{\rm BB}^{\rm ion})$ determines the flux of ionizing photons 
emitted by 1\,cm$^2$ area of the blackbody source 
(cm$^{-2}$\,s$^{-1}$). 
Solution of Eq.~(\ref{eq:tion}) for the values of $k_{\rm N}$, 
$F_{\rm WD}(\lambda)$, $\lambda$ and 
$\alpha_{\rm B}({\rm H},T_{\rm e})$ provides 
$T_{\rm BB}^{\rm ion}$. 
Having $T_{\rm BB}^{\rm ion}$, we can estimate the 
$L_{\rm WD}^{\rm hot}$ of the hot WDP as 
\citep[see e.g.,][]{2017A&A...604A..48S}, 
\begin{equation}
  L_{\rm WD} = \alpha_{\rm B}({\rm H},T_{\rm e})\,\textsl{EM}
               \frac{\sigma {T_{\rm BB}^{\rm ion}}^4}
                    {f(T_{\rm BB}^{\rm ion})}, 
\label{eq:lwdem}
\end{equation}
and $R_{\rm WD}^{\rm eff}$ is given by the Stefan--Boltzmann law. 

From the beginning of the eruption to day 4.751, when the hot WDP 
was not directly identifiable in the spectrum, we adopted 
$T_{\rm BB}^{\rm ion} = 73,000$\,K, at which the source generates 
maximum flux of hydrogen ionizing photons for a given luminosity. 
In other words, this temperature determines the minimum of the 
function (\ref{eq:lwdem}) that, when scaled with the measured 
$EM$, provides the lower limit of $L_{\rm WD}^{\rm hot}$ 
\citep[see Fig.~A.1. of][]{2015NewA...36..128S}. 

From $\sim$day~16, when the emission of the recombination 
\heii\,$\lambda$4686 line appeared in the spectrum, 
we estimated $T_{\rm BB}^{\rm ion}$ from the 
\heii\,$\lambda$4686/\hb\ flux ratio 
\citep[e.g.,][]{1997pdpn.book.....G}. We used the approach of 
\cite{2020A&A...636A..77S} for the average $T_{\rm e} = 22,000$\,K 
during this period (see their Eq.~(6)). 
%

\begin{figure}
\begin{center}
\resizebox{8.5cm}{!}{\includegraphics[angle=-90]{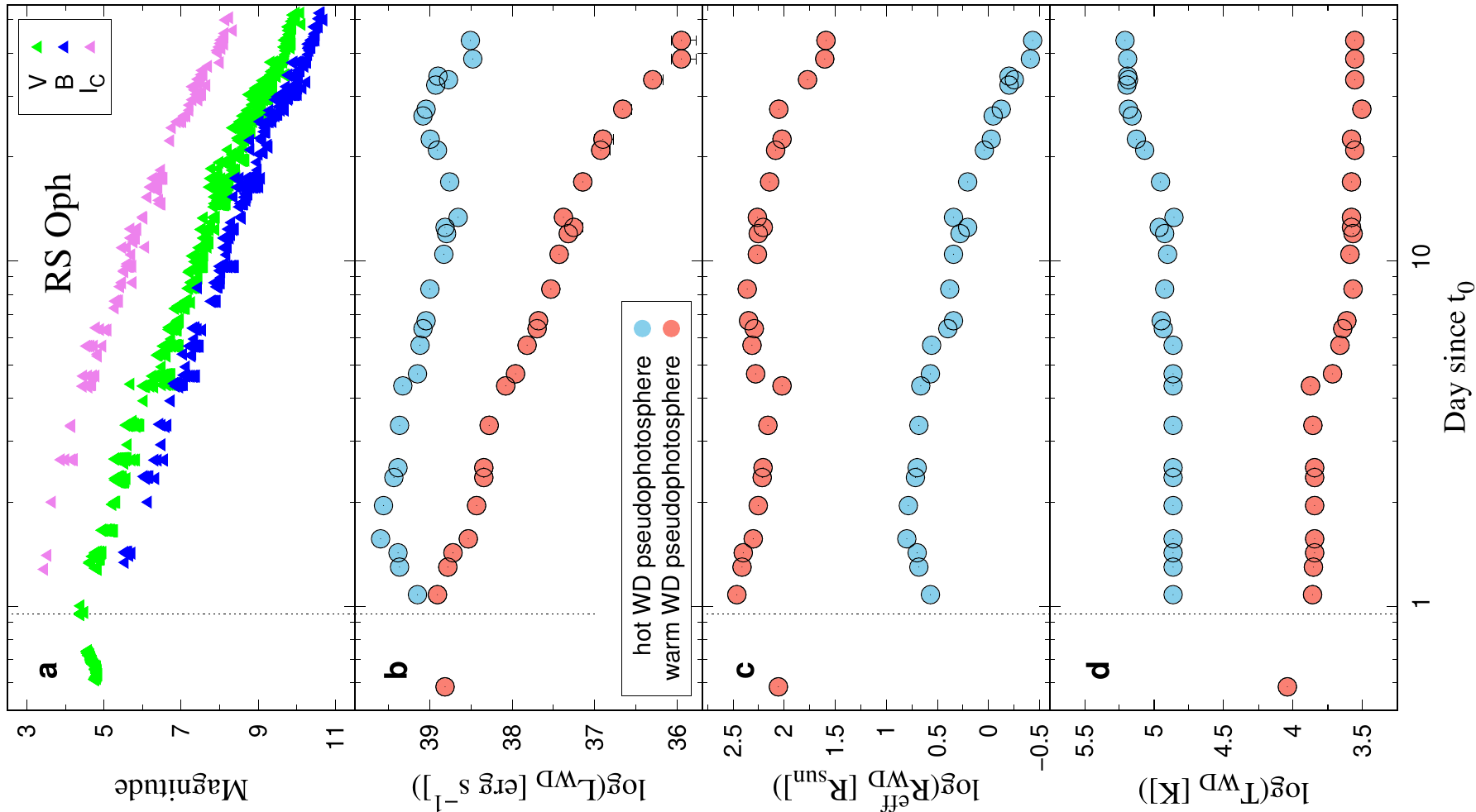}}
\end{center}
\caption{
The $B$, $V$, and $I_{\rm C}$ LCs (panel ({\bf a})), 
and evolution in $L_{\rm WD}$ ({\bf b}), $R_{\rm WD}^{\rm eff}$ 
({\bf c}), and $T_{\rm WD}$ ({\bf d}) parameters from the onset 
of the RS~Oph explosion, $t_0$. Parameter values are in 
Table~\ref{tab:par}. The dotted line indicates the time of 
the optical maximum. 
          }
\label{fig:lrtcont}
\end{figure}

Parameters of the hot WDP, $T_{\rm BB}^{\rm ion}$, 
$R_{\rm WD}^{\rm eff}$, and $L_{\rm WD}^{\rm hot}$ are introduced 
in Table~\ref{tab:par}. Their temporal evolution is depicted 
in Fig.~\ref{fig:lrtcont} and examples of SED models are 
shown in Figs.~\ref{fig:sedopt} and \ref{fig:seduvir}. 
\subsubsection{Plausibility of the hot WDP parameters}
\label{sss:plausibility}
The parameters $L_{\rm WD}^{\rm hot}$, $R_{\rm WD}^{\rm eff}$ and 
$T_{\rm BB}^{\rm ion}$ of the hot WDP, determined in the manner 
described in Sect.~\ref{sss:parhot}, represent only estimates 
of their actual values. 
The assumption of ionization equilibrium is probably fulfilled 
only in the initial phase of the outburst, when the maximum $EM$ 
is observed. However, the lack of knowledge of the exact value 
of $T_{\rm BB}^{\rm ion}$ during this period allows only a lower 
limit of the $L_{\rm WD}^{\rm hot}$ to be determined. 
In the later phase of nova evolution ($\gtrsim$days~5), 
$T_{\rm BB}^{\rm ion}$ can be determined more accurately, 
as $Zanstra$-temperature. 
However, ionization equilibrium may not be achieved, as 
indicated by a marked decrease in $EM$. In this case, a part 
of ionizing photons is not absorbed in the nebula, leading 
to a decrease in $EM$ that corresponds to a lower 
$L_{\rm WD}^{\rm hot}$ than the real one 
(see Eq.~(\ref{eq:lwdem})). 
%

Consistent with the bipolar structure of the ejecta (see 
Sect.~\ref{ss:ionejecta}), the hot WDP will probably not be 
spherically symmetric with a non-spherical temperature 
distribution -- hotter regions with smaller radii are located 
at/around its poles and cooler regions with larger radii are 
located toward the WD equator \citep[see Sect. 4.2.1 of][ for 
comparison with CN V339~Del]{2019ApJ...878...28S}. 
However, viewing the hot WDP from a single polar angle (the orbital 
inclination), we can determine only one $T_{\rm BB}^{\rm ion}$ 
for which we determine the scaling (i.e., $R_{\rm WD}^{\rm eff}$) 
and $L_{\rm WD}^{\rm hot}$ assuming spherical symmetry, which is 
not true. This is probably the main source of systematic errors 
in determining the parameters of the hot WDP. 
%
%
%
%
%
%
%
%
%
%
%
%
\begin{figure}
\begin{center}
\resizebox{8.5cm}{!}{\includegraphics[angle=-90]{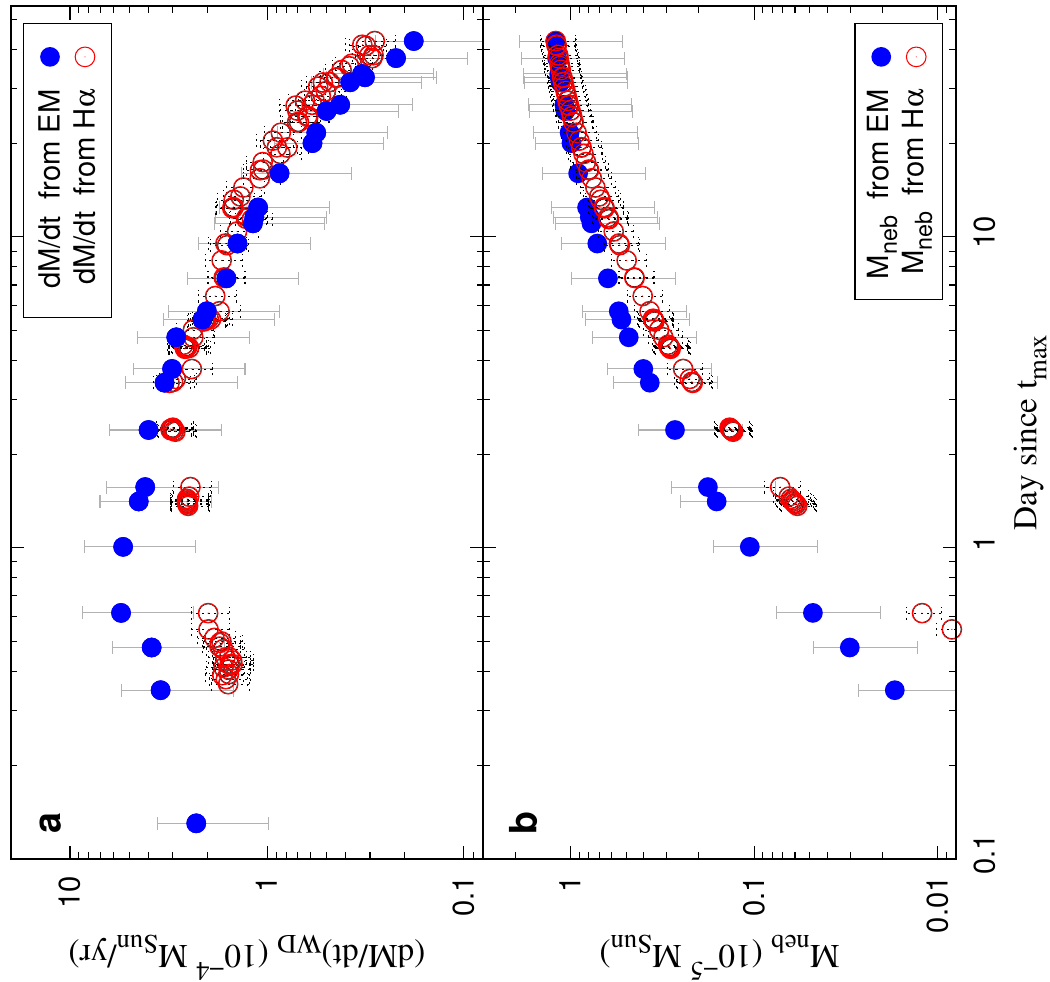}}
\end{center}
\caption{
Mass-loss rate by the ionized part of the ejecta, $\dot M_{\rm WD}$ 
(panel ({\bf a})), and its corresponding mass, $M_{\rm neb}$ ({\bf b}). 
See Sects.~\ref{ss:mdot} and \ref{ss:mass}. 
          }
\label{fig:dotM}
\end{figure}
%
%
\subsection{Mass-loss rate by the ionized ejecta}
\label{ss:mdot}
At the very beginning of a nova outburst, the optically thick wind 
occurs deep inside the WD photosphere, extinguishing the X-ray 
flash by strong absorption 
\citep[e.g.,][]{1994ApJ...437..802K,2022ApJ...935L..15K}. 
From a certain point in a nova evolution, the wind is ionized by 
the radiation from the hot WDP, converting it into the nebular 
radiation via f--f and f--b transitions. 
Knowing the result of these processes, i.e., $EM$ of the nebular 
continuum and/or the flux in the line (e.g., in \ha), we can 
determine the mass and mass-loss rate of the ionized ejecta as 
a function of the nova age. 
In determining these parameters, we will assume a biconical 
structure of the ejecta, in which the ionized part is distributed 
symmetrically in the polar regions, determined by the opening 
angle from the polar axis (see Sect.~\ref{ss:ionejecta}, 
Fig.~\ref{fig:sketch}). 

In our case, the nebular component of radiation was 
measurable already from the maximum of the optical brightness 
(day 0.130, Table~\ref{tab:par}, Fig.~\ref{fig:sedopt}). 
According to the relationship between $EM$ and ${\dot{M}}_{\rm WD}$ 
\citep[see Eq.~(C.5) of][]{2017A&A...604A..48S}, derived for 
a spherically symmetric, $\beta$-law wind of 
\cite{1999isw..book.....L}, the mass-loss rate can be expressed 
as 
\begin{equation}
\dot M_{\rm WD}({\rm EM}) = \mu m_{\rm H} v_{\infty}\left[f\,4\pi\,
        EM\,b\,R_0\,(1-2\beta)\,B^{-1}\right]^{1/2}, 
\label{eq:dotM}
\end{equation}
where
\begin{equation}
 B =  \left[\left(1-\frac{bR_0}{R_{\rm neb}}\right)^{1-2\beta} - 
      \left(1-\frac{bR_0}{R_{\rm WD}^{\rm eff}}\right)^{1-2\beta}\right]. 
      \nonumber
\label{eq:bracket}
\end{equation}
The parameter $f$ is the fractional solid angle that 
bounds the ionized part of the ejecta around the polar axis, 
$\mu$ is the mean molecular weight, and $m_{\rm H}$ is the mass 
of the hydrogen atom. 
The wind starts at the radial distance, $R_0$, from the WD center 
with the initial velocity $a$, and it is accelerated with the factor 
$\beta$ to the terminal velocity, $v_{\infty}$. It becomes optically 
thin at the radius of the hot WDP, $R_{\rm WD}^{\rm eff}$, and it 
reaches the outer radius of the ionized ejecta, $R_{\rm neb}$, at 
the given time of the observation. 
The parameter $b = 1-(a/v_{\infty})^{1/\beta}$. 

In determining $\dot M_{\rm WD}$ from the very beginning of 
the 2021 RS~Oph eruption, we used parameters of the SED models, 
$EM$ and $R_{\rm WD}^{\rm eff}$ (Table~\ref{tab:par}), $v_{\infty}$ 
from the extension of \ha\ wings (Table~\ref{tab:hapar}), 
$\beta = 0.56$ as suggested by fitting the broad \ha\ profile just 
after the 2006 eruption \citep[see][]{2008ASPC..401..227S}, and 
$a = 50$\,\kms\ \citep[][]{2017A&A...604A..48S}\footnote{Due to 
the very fast wind acceleration, the high value of $v_{\infty}$, 
and its uncertain beginning inside the photosphere, the value 
of $a$ between $\sim$10 and $\sim$100\kms\ affects the resulting 
$\dot M_{\rm WD}$ within a few per-cents only}. 
The outer radius $R_{\rm neb} = R_{290}\,v_{\infty}\,\Delta t$, 
where, $R_{290}$ is the radius of the warm WDP on day 0.130, 
when the nebular component was first detected, and $\Delta t$ 
is the time between day 0.130 and a later observation. 
Since the actual beginning of the wind is not further specified, 
we adopted two limiting values for $R_0$: 
$R_0 = R_{\rm WD}^{\rm eff}$ that corresponds to the lower limit 
of $\dot M_{\rm WD}$ and $R_0$ equals the WD core radius of 
0.004\ro\ 
\citep[for a WD mass of $\sim$1.3\mo, see][]{1972ApJ...175..417N}, 
which provides the upper limit of $\dot M_{\rm WD}$. 
Table~\ref{tab:par} presents the average of these limits, while 
Fig.~\ref{fig:dotM}\,a also shows its maximum deviations. As 
the limits of $R_0$ are not physically possible, the actual 
uncertainties of $\dot M_{\rm WD}$ are smaller than the figure 
shows. 

As the nebular continuum and emission lines (e.g., of hydrogen) 
arise in the same ionized volume, the value of $\dot M_{\rm WD}$ 
can also be determined from the measured ﬂux of the \ha\ broad 
component, $F_{\alpha}$, produced by the ionized part of the ejecta. 
Its luminosity, $L_{\alpha} = 4\pi d^2 F_{\alpha}$, is related 
to the line emissivity, $\varepsilon_{\alpha} n_{\rm e}n_{\rm p}$ 
(erg\,cm$^{-3}$\,s$^{-1}$), by 
%
\begin{equation}
 L_{\alpha} = 
  \varepsilon_{\alpha}\int_{V}\! n_{\rm e}n_{\rm p}\,{\rm d}V = 
  \varepsilon_{\alpha} EM, 
\label{eq:lha}
\end{equation}
where $\varepsilon_{\alpha}$ is the volume emission coefficient 
in \ha, $n_{\rm e}$ and $n_{\rm p}$ are concentrations of electrons 
and protons, and $V$ is the volume of the ionized part of the wind. 
Accordingly, using the same assumptions as for the continuum, 
Eq.~(\ref{eq:dotM}) can be rewritten as 
\begin{equation}
\dot M_{\rm WD}({\rm H\alpha}) = 
     \mu m_{\rm H} v_{\infty}\left[f\,\frac{4\pi}{\varepsilon_{\alpha}}\,
     L_{\alpha}\,b\,R_0\,(1-2\beta)\,B^{-1}\right]^{1/2}. 
\label{eq:dotMa}
\end{equation}
However, in a line transition, the wind becomes optically thin 
at distances $>R_{\rm WD}^{\rm eff}$. \cite{1988ApJ...326..356L} 
found that winds of O stars become optically thin in the \ha\ 
line from a distance of $\sim$1.5 times the star’s radius. 
Also, modeling of the broad \ha\ wings in the spectra of 
symbiotic stars suggested the optically thin conditions from 
about 1.2 to 1.5\,$R_{\rm WD}^{\rm eff}$ 
\citep[][]{2006A&A...457.1003S}. 
In the expression $B$ of Eq.~(\ref{eq:dotM}), we assume 
the optically thin wind in the \ha\ line from a radius of 
1.2\,$R_{\rm WD}^{\rm eff}$. 
Using our values of $F_{\alpha}$ (Table~\ref{tab:hapar}), radii 
of the hot WDP in Table~\ref{tab:par} interpolated to dates of 
high- and medium-resolution spectra containing the \ha\ line, and 
$\varepsilon_{\alpha} = 1.83\times 10^{-25}\,\rm erg\,s^{-1}\,cm^3$ 
\citep[][]{1989agna.book.....O} for our average $T_{\rm e}$ of 
20,000\,K (see Table~\ref{tab:par}), we obtained 
the corresponding $\dot M_{\rm WD}$ (Fig.~\ref{fig:dotM}\,a, 
Table~\ref{tab:hapar}). 

The values of $\dot M_{\rm WD}$ determined in this way suffer 
from systematic errors. For example, it is the unknown value 
of the factor $f$ at a given observation time. 
For simplicity, we have used $f = 1$. 
The uncertain determination of the hot WDP parameters (see 
Sect.~\ref{sss:plausibility}) represents another source of errors 
that cannot be specified. For example, a small jump in 
$\dot M_{\rm WD}$ from the \ha\ flux since $\sim$day~15 
(Fig.~\ref{fig:dotM}\,a) could 
be related to the increase in $T_{\rm BB}^{\rm ion}$ when 
the \heii\,$\lambda$4686 emission appeared in the spectrum. 
However, its reality is not supported by the $\dot M_{\rm WD}$ 
values determined from $EM$. 
%
%
\subsection{Mass released by the ionized ejecta}
\label{ss:mass}
The mass of the ionized part of the ejecta can be estimated 
by integrating its outflow rate from the first detection of 
the nebular continuum on day 0.13 to our last SED model on 
day 42.6, i.e. 
\begin{equation}
  M_{\rm neb} = \int_{0.13}^{42.6}\! \dot M_{\rm WD}\,{\rm d}t .
\label{eq:mass}
\end{equation}
Figure~\ref{fig:dotM}\,b shows the increase in $M_{\rm neb}$ 
from the peak brightness to day 43, when its mean value was 
$1.2\times 10^{-5}$\mo. 
This value can be considered final, as the decrease in $EM$ as 
well as \ha\ flux after $\approx$day~33 (see Tables~\ref{tab:par} 
and \ref{tab:hapar}) indicates a decrease in the wind mass-loss 
rate, which is consistent with the onset of the SSS phase in 
the nova evolution \citep[see][]{2023A&A...670A.131N}. According 
to theoretical considerations by \cite{2010ApJ...709..680H}, 
the emergence of supersoft X-rays indicates that the optically 
thick wind has stopped or substantially weakened and therefore 
the total mass of the ejecta remains constant in time. 
However, our value of $M_{\rm neb}$ represents rather a lower 
limit, because a part of the ejected mass at the very beginning 
of the eruption was not ionized. 

Assuming $T_{\rm e} = 10^4$\,K, \cite{2023MNRAS.518.2614M} 
determined the ejected mass to be
$2.9\times 10^{-5}\,(d/2.4\,{\rm kpc})$\mo\ from the \hb\ flux 
measured at the beginning of the nebular phase. This value is 
$\sim$1.6 times higher than our value and represents an upper 
limit, since the \hb\ emission includes a contribution from 
the ionized wind from the giant, measured during quiescence. 
Therefore, \cite{2023MNRAS.518.2614M} also used a flux of 
\heii\,$\lambda$4686, which corresponded to an ejected mass of 
$1.1\times 10^{-5}\,(d/2.4\,{\rm kpc})$\mo\, which is 
$\sim$1.6 times lower than our value, 
$1.2\times 10^{-5}\,(d/1.6\,{\rm kpc})$\mo. 
The authors noted that the uncertainty in $T_{\rm e}$, 2,000\,K, 
corresponds to changes in $M_{\rm neb}$ by a factor of $\sim$2, 
being higher at higher $T_{\rm e}$, and vice versa. 
For comparison, $T_{\rm e}\equiv 10^4$\,K 
\citep[i.e., $\varepsilon_{\alpha} = 
3.56\times 10^{-25}\,\rm erg\,s^{-1}\,cm^3$,][]{1989agna.book.....O}
would reduce our value of $M_{\rm neb}$ to 
$8.6\times 10^{-6}\,(d/1.6\,{\rm kpc})$\mo. 
%
%
\subsection{Evolution of the \ha\ line}
\label{ss:ha}
\cite{2021arXiv210901101M,2022arXiv220301378M} showed 
the detailed evolution of the \ha\ line profile from the maximum 
of the RS~Oph outburst to its nebular phase. An example of such
evolution in our spectra is shown in Fig.~\ref{fig:hahb} 
(Appendix~\ref{app:hahb}), while Fig.~\ref{fig:brandnar} compares 
the \ha\ line profile at the beginning (day 1.6) and at the end 
(day 37.5) of optical brightening. 
Its typical feature is a broad emission component, triangular 
in profile, with FWZI of $\sim$8,000\kms, and a blueward-shifted 
absorption component at $\sim$-4,500\kms, which was indicated 
until day 3--4. 
Narrow absorption and emission components are superimposed on 
the top of the broad component at $\sim$-73 and $\sim$-12\kms, 
respectively, well measurable until day 5--6, but still 
distinguishable until the end of our observations. Their early 
evolution is shown in Fig.~\ref{fig:brandnar}. 
%
%
%
\begin{figure*}
\begin{center}
\resizebox{17cm}{!}{\includegraphics[angle=-90]{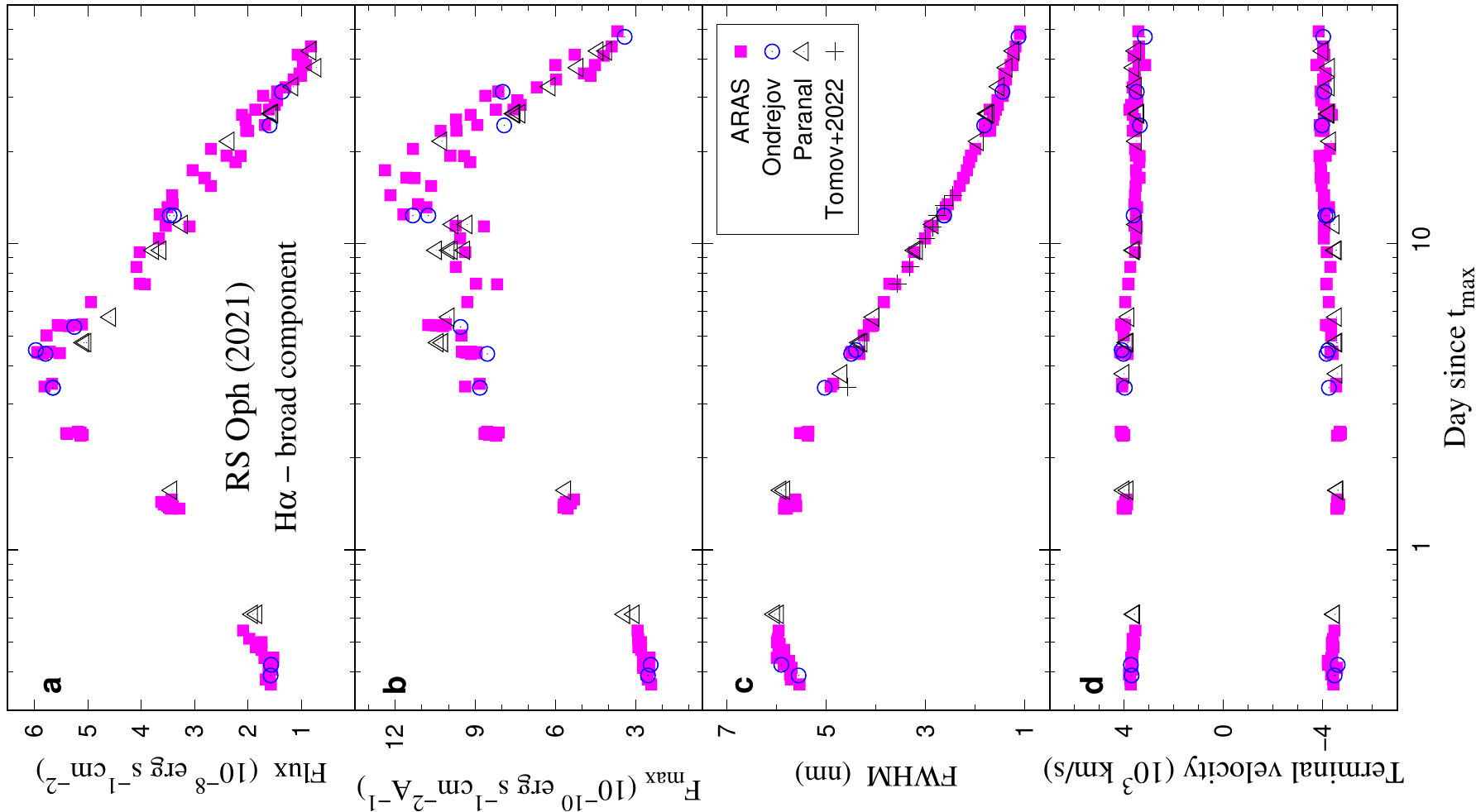}
                    \includegraphics[angle=-90]{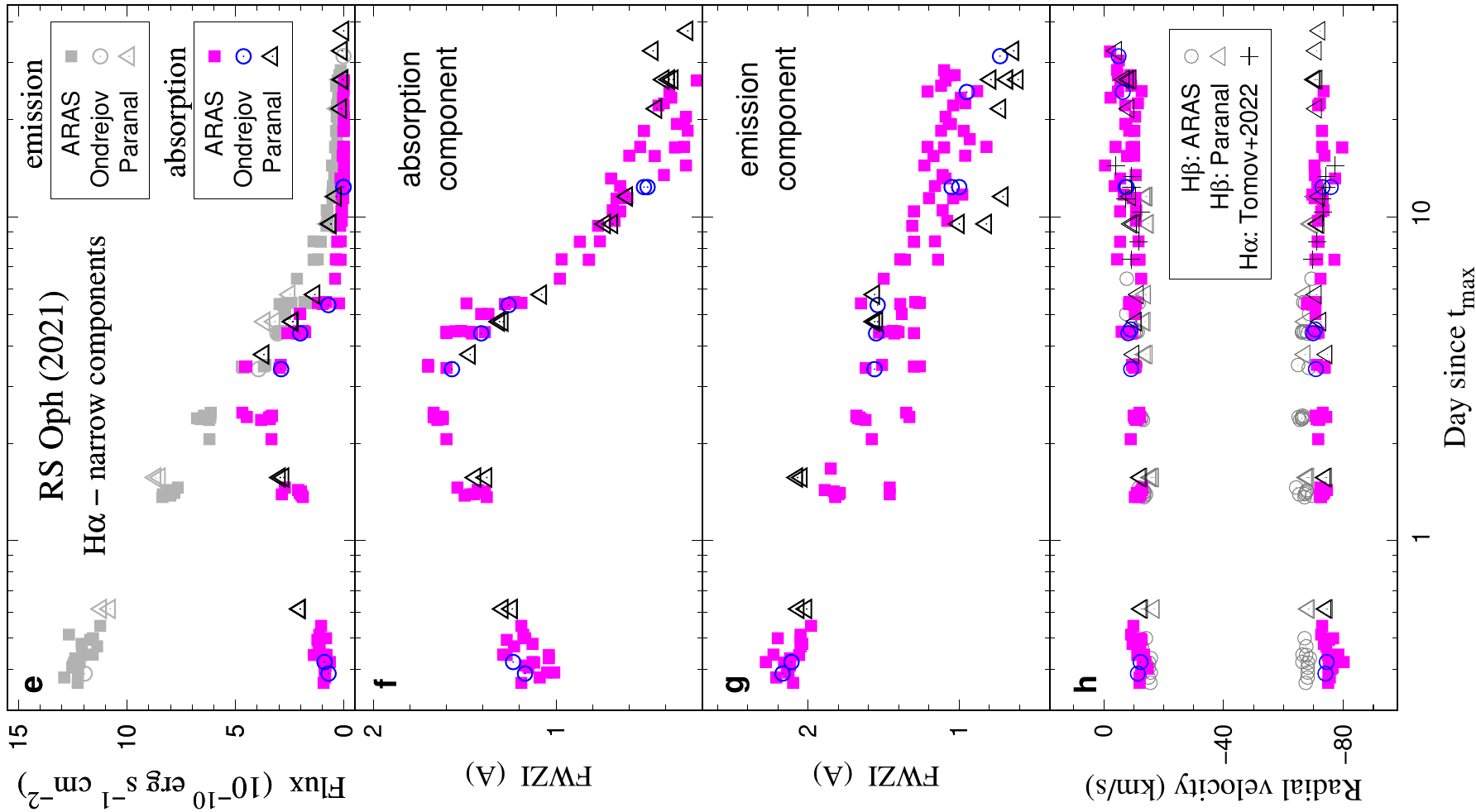}}
\end{center}
\caption{
Evolution of the \ha\ line parameters. 
Left: The broad component: 
The flux (panel ({\bf a})), its maximum height, $F_{\rm max}$, 
above the continuum ({\bf b}), FWHM ({\bf c}), and terminal 
velocities of the wings ({\bf d}). The description is in 
Sect.~\ref{sss:habroad} and the data in Table~\ref{tab:hapar}. 
Right: The narrow components: 
The fluxes ({\bf e}), FWZI of the absorption ({\bf f}) and 
emission ({\bf g}) component, and their radial velocities 
({\bf h}). Their description and interpretation are found in 
Sects.~\ref{sss:hanarrow} and \ref{ss:environment}, respectively. 
          }
\label{fig:ha}
\end{figure*}
\subsubsection{The broad component}
\label{sss:habroad}
\begin{figure*}[t]
\begin{center}
\resizebox{\hsize}{!}{\includegraphics[angle=-90]{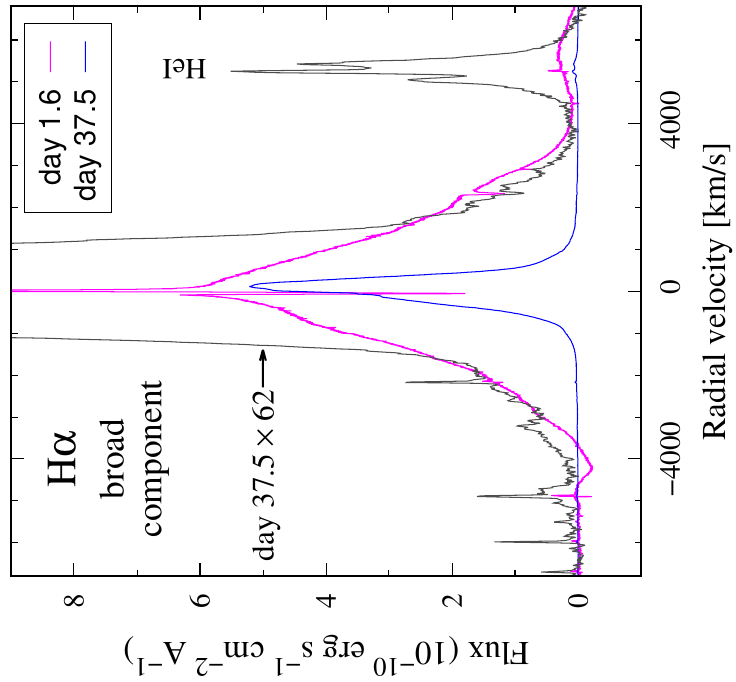}\hspace*{+5mm}
                      \includegraphics[angle=-90]{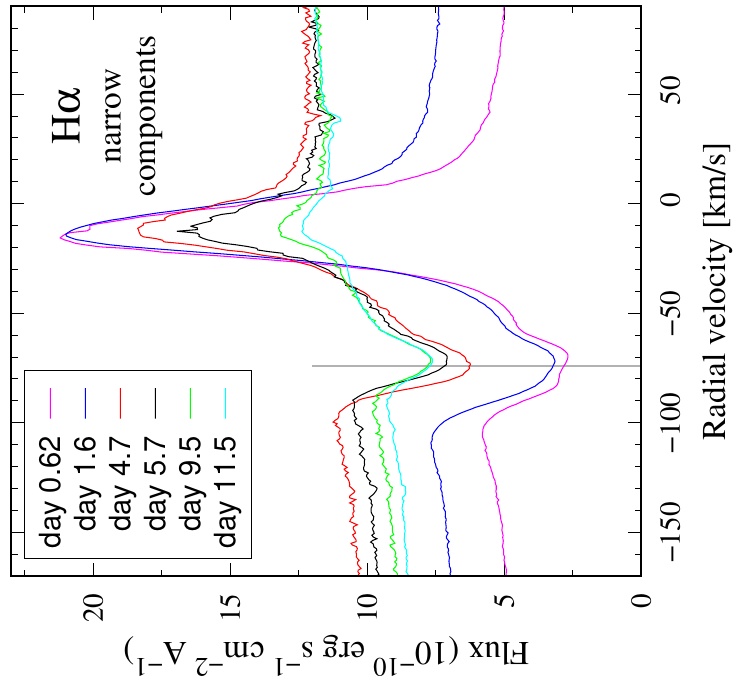}
                      \includegraphics[angle=-90]{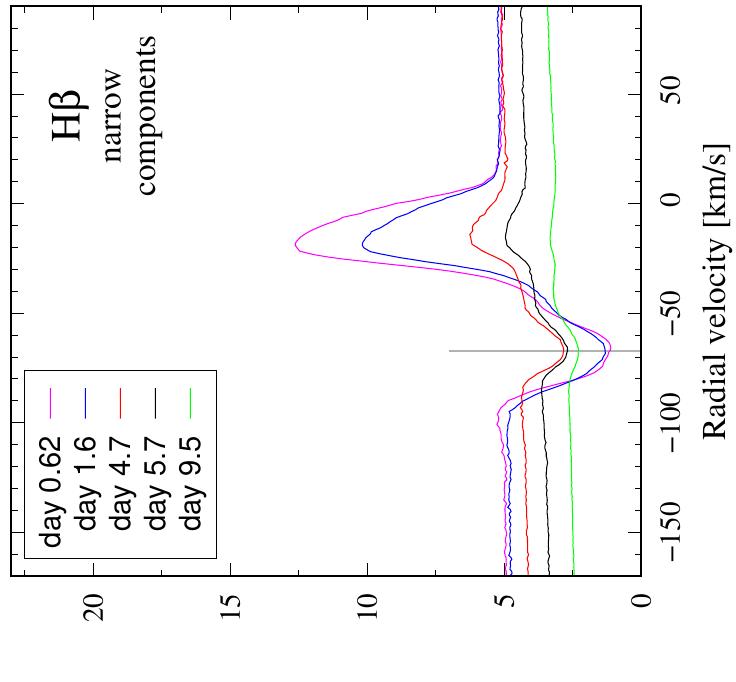}
\hspace*{+5mm}}
\end{center}
\caption{
Example of the evolution of the broad and narrow components of 
the hydrogen line profiles measured on UVES spectra. 
Left panel shows the \ha\ broad component measured on days 1.567 
and 37.532. 
This comparison shows a significant narrowing of the line emission 
core (see Fig.~\ref{fig:ha}\,c), although the expansive, weak line 
wings remain present even in the later stage of nova evolution. 
The local continuum is subtracted (see Sect~\ref{sss:habroad}). 
Rihgt panels show the evolution of the narrow components in 
the \ha\ and \hb\ profiles. Note that the transition 
from the emission to the absorption component is not smooth, so 
the resulting profile is not exactly of the P~Cygni type. 
The vertical gray lines indicate the position of the absorption 
component in the \ha\ and \hb\ profiles, which differ by 
$\sim$7\kms\ during the first days after $t_{\rm max}$ 
(see Fig.~\ref{fig:ha}\,h and Sect.~\ref{sss:hanarrow}). 
          }
\label{fig:brandnar}
\end{figure*}

The development of the \ha\ broad component parameters is 
shown in Fig.~\ref{fig:ha}. 
The flux of its emission $F_{\alpha}$ (Fig.~\ref{fig:ha}\,a) 
was gradually increasing from our first observation on day 0.36, 
reached a maximum of $\sim$5.5$\times 10^{-8}$\ecs\ between 
day 3 and 6, and then was decreasing to values of 
$\lesssim$1$\times 10^{-8}$\ecs\ on day 42. 
It is natural to assume that the broad \ha\ component is emitted 
by the ionized wind ejected from the WD during the explosion. 
The outflowing wind increases the particle density of the ionized 
medium ($n\propto \dot M_{\rm WD}$), which leads to an increase 
in its emissivity ($\propto n^2$). At the same time, the expansion 
of the ionized wind reduces its density with $v_{\infty}^{-3}$, 
which lengthens the recombination time ($\propto 1/n$), thereby 
reducing the total emission per unit time. The measured flux, 
which is given by the rate of recombinations, then results mostly 
from the rivalry of these physical processes. 
During the first $\sim$5 days, the high value of $\dot M_{\rm WD}$ 
(Fig.~\ref{fig:dotM}\,a) and a relatively small volume of 
the ionized ejecta result in a gradual increase of both 
the $F_{\alpha}$ flux and the height of the broad component above 
the continuum, $F_{\rm max}$ (see Fig.~\ref{fig:ha}\,a and b). 
Furthermore, an increase in the visible volume of the biconical 
ionized wind due to the increase in its opening angle (see 
Sect.~\ref{ss:ionejecta}) may also contribute to the increase 
of both the parameters during this early evolution of the ejecta. 
In later stages of the nova evolution, the decreasing value 
of $\dot M_{\rm WD}$ and the continued wind expansion led to 
a marked decrease in the particle density, reducing the 
contributions at all wavelengths of the line. We measured 
a decline in all parameters (see Figs.~\ref{fig:ha}\,a, 
b, c, and \ref{fig:brandnar}). 
Figures~\ref{fig:ha}\,d and \ref{fig:brandnar} show that 
the deceleration of the ejecta in the direction towards 
the observer is small (see Sect.~\ref{sss:ourshocks}). 

We estimated the uncertainties of the broad component parameters 
as follows. 
The uncertainty of its flux is primarily determined by 
the estimate of the local continuum, which is influenced 
mainly by the \hei\,$\lambda$6678 emission. We estimated 
its errors to be less than 10\%. We note that the \ha\ emission 
in the UVES spectra was saturated by day 5.76, even for 
the shortest exposure times. Its values are therefore below 
by $\gtrsim$10\%\ (see panel {\bf a}). Similar error values 
apply for the maximum height above the local continuum, while 
for the FWHM, the error estimate is roughly half. 
The terminal velocity values of the broad component are affected 
by the blend of a faint but broad Fe\,{\scriptsize II}\,$\lambda$6456 
and \hei\,$\lambda$6678 emission lines on the blue and red sides 
of the extended \ha\ wing. 
We estimated their errors to be 5-10\%. 

Finally, the high-velocity absorption component may be associated 
with the neutral wind from the giant, which is swept off and 
shocked by the ejecta, and accumulated at its front 
\citep[][ Sect.~\ref{sss:ourshocks} here]{2006Natur.442..279O}. 
This material absorbs the light from the underlying warm WDP 
in a given line transition, which is best seen in the hydrogen 
lines. 
However, its presence is limited to the first 3--4 days only 
when the warm WDP dominates the spectrum 
(see Figs.~\ref{fig:sedopt} and \ref{fig:fsfn}). 
Such behavior is related to the temporal evolution of 
the ejecta (see Sect.~\ref{ss:evolstr}). 

Further absorption/emission continues in the undisturbed wind 
from the giant, which is responsible for the formation of 
the narrow components (see Sects.~\ref{sss:hanarrow} and 
\ref{ss:environment}). 
%
%
\subsubsection{The narrow component}
\label{sss:hanarrow}
A pair of narrow absorption and emission lines, located on the top 
of the broad emission, is observed from the very beginning of 
the eruptions of SyNe \citep[][]{2025CoSka..55c..47M}. 
The case of RS~Oph was recently shown in detail by 
\cite{2021arXiv210901101M}. According to the small FWHM of 
$\approx$20\kms, their origin was ascribed to the wind from 
the giant surrounding the binary that is not perturbed by 
the ejecta \citep[e.g.,][]{2019arXiv190901389M}. 

Here, Fig.~\ref{fig:ha} (panels e--h) shows the evolution of the narrow 
components parameters in \ha\ for the recent 2021 RS~Oph explosion. 
Panel (e) shows that the flux of the emission component 
decreases rapidly from a maximum of around $1.3\times 10^{-9}$\ecs\ 
measured on our first spectra (day 0.36) to $<$10$^{-10}$\ecs\ 
on day $\gtrsim$10, while the flux absorbed by the narrow component 
gradually increases from an initial value of $\sim$10$^{-10}$\ecs\ 
to a value of $\sim$4$\times 10^{-10}$\ecs\ on $\sim$day~4, with 
a subsequent decline to values $<$10$^{-10}$\ecs\ during 
days 4--6. Thereafter, very low flux values of both components 
were observed until our last observations. 
A similar course was also observed in the evolution of the width 
of both components; here, characterized with FWZI for simplicity 
(see panels (f) and (g)). 
Finally, panel (h) shows their RVs. 
The relatively stable value of the absorption component in the 
\ha\ line, $-73\pm 3$\kms, corresponds to the terminal velocity 
of the RG wind $34\pm 3$\kms\ after subtracting the systemic 
velocity of -39\kms\ \citep[][]{2009A&A...497..815B}. 
Interestingly, the RV of the \hb\ absorption core was shifted 
by $\sim$7\kms\ compared to the values for the \ha\ core 
during the first $\sim$5--6 days (gray symbols in 
Fig.~\ref{fig:ha}\,h). 
This shift is likely due to the formation of the narrow \hb\ 
absorption deeper in the absorbing column, i.e., closer to 
the warm WDP, where smaller RVs of the RG wind are expected 
compared to larger distances above the WDP, where absorption 
in \ha\ arises \citep[see Fig.~1 of][]{2016MNRAS.457..822B}. 
Later, from $\sim$day~5.4, the RVs of \ha\ and 
\hb\ become equal, probably due to the reduced size of the 
undisturbed column of the wind located at larger distances 
from the binary (see Sect.~\ref{ss:environment}). 
As concerns the narrow emission component, it is placed at 
($-12\pm 3$)\kms\ during the first $\sim$4 days with a slow 
increasing trend to ($-8\pm 4$)\kms\ and significantly reduced 
flux during the following days (see also Fig.~\ref{fig:brandnar}). 
Further properties of the narrow components are described and 
interpreted in Sect.~\ref{ss:environment}. 

We estimated the uncertainties of the measured parameters as 
follows. 
By day $\sim$5, flux uncertainties were as small as $\sim$5\%, 
as the narrow line profiles were well-defined. Later, as these 
components weakened, their errors increased to more than $\sim$10\%. 
A similar trend in errors is estimated for the FWZI. However, in 
later stages (day $\gtrsim$9), the FWZI estimate suffers from 
significant errors ($\gtrsim$50\%) due to the superposition of 
narrow lines with the broad component, the profile of which is 
not precisely known. 
The intrinsic uncertainties in the RVs of the narrow components 
were relatively small, ranging from 1 to 2\kms, but were 
affected by systematic errors arising from inaccurate wavelength 
calibration in some amateur spectra. Also, RVs measured in 
amateur spectra do not include a correction for Earth's rotation, 
which causes an error of $<$0.5\kms. 

Finally, Fig.~\ref{fig:brandnar} shows the evolution of the profiles 
of narrow components in the \ha\ and \hb\ lines during the first 
days of the explosion in very high-resolution UVES spectra. 
As can be seen, the emission and absorption components do not 
represent exactly the components of the P-Cygni profile. The two 
components are separated, in contrast to what can be seen in 
medium-resolution spectra 
\citep[see][]{2022BlgAJ..37...24Z,2021arXiv210901101M}. 
The profile of the narrow components, which is not of the P~Cyg 
type, suggests that the wind from the giant is not spherically 
symmetric around the binary, being diluted on the polar sides 
(see Sect.~\ref{ss:environment}). 
%
%
\section{Discussion}
\label{s:discuss}
%
%
\subsection{Bipolar density structure of the ejecta}
\label{ss:ionejecta}
Our SED models showed that the spectrum of the ejecta from 
the very beginning of the RS~Oph explosion consists of a relatively 
cool stellar component (i.e., the warm WDP) and nebular component 
of radiation, the contributions of which change during the explosion 
(Sect.~\ref{sss:parwarm}, Figs.~\ref{fig:sedopt} and \ref{fig:lrtcont}). 

For the interpretation of such a two-component spectrum, it is 
essential to know whether the warm WDP is capable of 
producing the measured amount of nebular emission. 
If not, there must be a strong ionizing source inside the ejecta. 
The spectrum of RS~Oph has this feature from the very beginning of 
the explosion until the end of the detection of the cold stellar 
component. 
For example, on day 0.617, the warm WDP ($L_{\rm WD}^{\rm warm} = 
3.4\times 10^{38}$\es, $T_{\rm eff} = 7,000$\,K) generates only 
$\sim$5$\times 10^{42}$ hydrogen ionizing photons per second 
(assuming blackbody radiation), while the measured $EM$ of 
$4.9\times 10^{62}$\cmt\ is generated by 
$\alpha_{\rm B}({\rm H},T_{\rm e})\times EM = 7.8\times 10^{49}$ 
recombinations per second for the total hydrogen recombination 
coefficient for the Case~B, 
$\alpha_{\rm B}({\rm H},18,000\,K) = 1.6\times 10^{-13}$ 
$\rm cm^{3}\,s^{-1}$ \citep[e.g.,][]{1987A&A...182...51N}. 
The very low flux of ionizing photons produced by the warm WDP, 
which is several orders of magnitude lower than the required 
rate of recombinations, thus signals the presence of a strong 
ionizing source in the ejecta, which we have called the hot WDP 
(see Sect.~\ref{ss:sed}, Fig.~\ref{fig:sketch}). 
A spectrum that arises from the superposition of the stellar 
and nebular components of radiation, where the former is unable 
to give rise to the latter, is called a spectrum of the 
'two-temperature type'. It was first detected during Z~And-type 
outbursts of SySts with high orbital inclinations 
\citep[see Sect.~5.3.4 of][]{2005A&A...440..995S}, but has also 
been found for CN V339~Del 
\citep[see Sect.~4.2 of][]{2014A&A...569A.112S}. 

The two-temperature-type of spectrum can be produced by 
an optically thick equatorial disk-like structure of 
the outflowing wind (the ejecta) surrounding the burning WD, 
whose ﬂared outer rim (the warm WDP) occults the central 
ionizing source (the hot WDP) in the line of sight, while 
the optically thin parts of the ejecta above/below the disk 
structure are ionized, giving rise to the strong nebular 
emission (see Fig.~27 of \cite{2005A&A...440..995S}, or 
Fig.~4 of \cite{2015NewA...36..128S} for RS~Oph, and 
Fig.~\ref{fig:sketch} here). 
This means that the density is not uniformly distributed in 
the ejecta, suggesting that an optically thin, low-density 
parts are located in bipolar directions above the flared 
density-enhanced disk-like structure around the equator, 
which is thought to coincide with the orbital plane. 

This interpretation is consistent with the bipolar shaping of 
the RS~Oph ejecta suggested in previous studies using various 
techniques and observations from $\gamma$-ray to radio 
(see references in Sect.~\ref{s:intro}). 
The development of the two-temperature type spectrum from the 
very beginning of the RS~Oph eruption suggests an equally early 
development of the bipolar structure of the ejecta. This is 
supported, for example, by the modeling of the X-ray spectrum 
monitored by NICER \citep[see][]{2024ApJ...960..125I} and/or 
spectropolarimetric observations that revealed a similarly 
asymmetric dust structure \citep[][]{2023A&A...679A.150N}, 
in both cases, measured as early as two days after the start of 
the eruption. 
Such a structure of the nova ejecta is probably responsible 
for the generation of multiple shocks within the ejecta, which 
produce the observed complex $\gamma$-ray spectrum 
\citep[][ Sect.~\ref{ss:shocks} here]{2023ApJ...947...70D}. 
%
%
\subsection{Possible cause of the ejecta bipolar structure}
\label{ss:cause}
It is thought that the formation of a bipolar density structure 
in classical nova ejecta may be caused by the orbital motion of 
the WD companion in the dense ejecta during a lengthy (weeks to 
months) 'common-envelope' phase. The donor experiences 
frictional forces in the nova envelope and transfers orbital 
energy and angular momentum to the ejecta, producing a 'density 
contrast' in the ejected mass between the polar and equatorial 
(orbital) directions, with concentrated mass in the latter 
\citep[see][]{1990ApJ...356..250L,1997MNRAS.284..137L}. 
However, for long-period symbiotic novae, this mechanism cannot 
be effective because the timescale of the bipolar structure 
formation represents only a small fraction of the orbital period. 
Moreover, the photospheric radius of the ejecta is smaller than 
the separation between the binary components -- the 
'common-envelope' phase does not even exist. Therefore, for 
symbiotic novae, we favor rapid WD rotation as the primary 
mechanism creating the bipolar density structure of their ejecta. 

According to \cite{1993ApJ...409..429B}, the equatorial 
density-enhanced structure may be the result of the rotation of 
the star with a radiatively driven wind, which leads to 
compression of the outflowing material towards the equatorial 
regions at the expense of the polar regions due to the 
conservation of angular momentum of the wind particles 
\citep[see][ for a review]{1999isw..book.....L}. 
\cite{2012A&A...548A..21C} applied the wind compression model 
to accreting WDs in symbiotic binaries during the Z~And-type 
outbursts. In their model, the wind begins at the surface of 
the rotating WD and becomes optically thin at/above the hot WDP. 
According to the model, the wind creates a denser neutral disk-like 
zone in the equatorial plane, the optically thick outer edge 
of which represents the warm WDP, while the remaining, less dense 
parts of the wind above/below the disk are ionized by the hot WDP 
(see their Fig.~6, and compare Fig.~\ref{fig:sketch} here). 
Such a biconical ionization structure of the ejecta can then 
produce the two-temperature type of spectrum as described 
in Sect.~\ref{ss:ionejecta}. 
The model was applied to the RS~Oph eruption in 
2006 (Sect.~\ref{ss:ionejecta}) and we therefore adopt it 
for its recent explosion in 2021. 

The rotation of the accreting WD may therefore be responsible 
for the formation of the density-enhanced equatorial region and 
the density-reduced polar regions during their thermonuclear 
outbursts, when the mass-loss rate is significantly enhanced. 
Furthermore, due to centrifugal force, a rotating WD requires 
the accretion of more material to trigger a TNR, which can 
result in a higher luminosity and a larger amount of ejected 
mass than in the case of a non-rotating WD. 
%
%
\subsection{Evolution of the ejecta structure}
\label{ss:evolstr}
According to the wind compression model (Sect.~\ref{ss:cause}), 
the formation of a dense, optically thick equatorial region of 
the ejecta depends mainly on the mass-loss rate, 
${\dot{M}}_{\rm WD}$, which significantly increases during 
the outburst. In particular, the opening angle of 
the \hii/\hi\ boundary\footnote{We assume that 
the \hii/\hi\ boundary roughly corresponds to 
the optically thin/thick interface in the ejecta.} 
is proportional to ${\dot{M}}_{\rm WD}^{-2}$ 
\citep[see Fig.~1 and Eq.~(14) of][]{2012A&A...548A..21C}. 

\begin{figure}[t]
\begin{center}
\resizebox{8.5cm}{!}{\includegraphics[angle=-90]{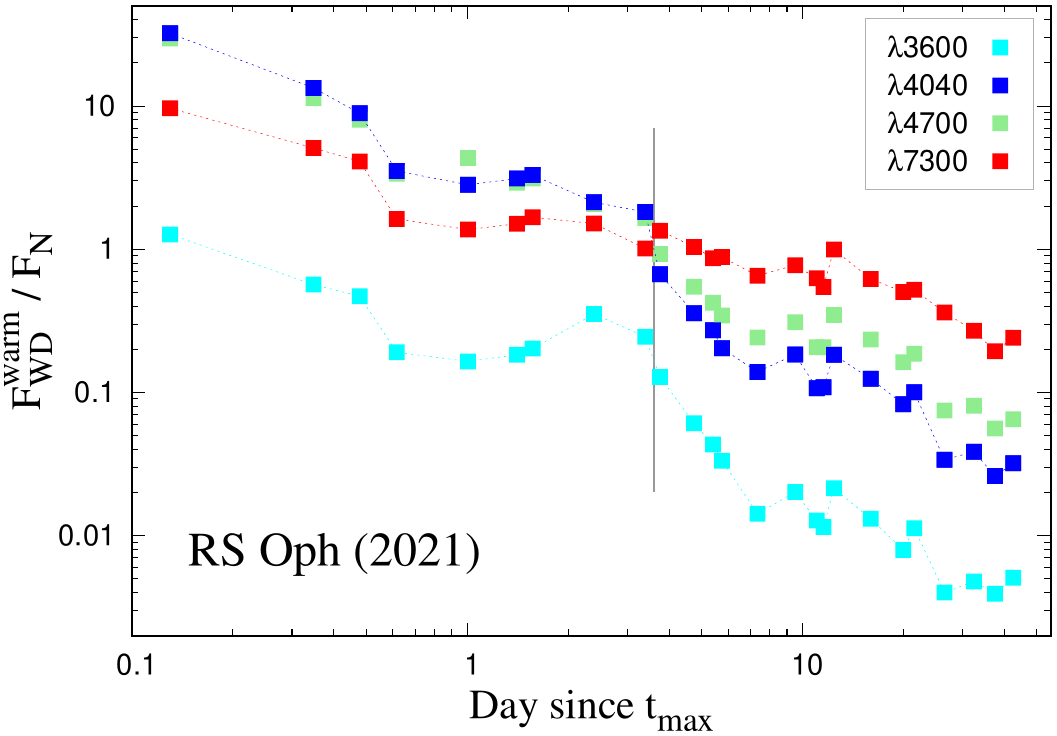}}
\end{center}
\caption{
The ratio of fluxes from the warm WDP, $F_{\rm WD}^{\rm warm}$, 
and the nebula, $F_{\rm N}$, at selected wavelengths, 3,600\,\AA, 
4,040\,\AA, 4,700\,\AA, and 7,300\,\AA, given by SED models. 
The rapid decrease of this ratio indicates a rapid opening of 
bipolar ionized regions (see Sects.~\ref{ss:evolstr}). 
The vertical line denotes a sudden change of this ratio between 
days 3.395 and 3.759, reflecting a change in the physical 
parameters of the warm WDP (see Sect.~\ref{sss:ourshocks}). 
          }
\label{fig:fsfn}
\end{figure}

In our case, the gradual decrease of ${\dot{M}}_{\rm WD}$ (see 
Fig.~\ref{fig:dotM}) suggests the gradual opening of bipolar, 
less dense ionized regions, already a few days after 
the outburst onset. 
This effect is indicated by a rapid decrease in the luminosity 
$L_{\rm WD}^{\rm warm}$ after $t_{\rm max}$ (see Table~\ref{tab:par}, 
Fig.~\ref{fig:lrtcont}\,b). 
The flux of the warm WDP, $F_{\rm WD}^{\rm warm}(\lambda)$, was 
rapidly declining relative to the flux of the nebular continuum, 
$F_{\rm N}(\lambda)$, especially in the blue part of the spectrum 
(see Figs.~\ref{fig:sedopt} and \ref{fig:seduvir}). For example, 
their ratio, $F_{\rm WD}^{\rm warm}/F_{\rm N}$ at 
$\lambda = 4,700$\,\AA, decreased from $\sim$30 on day 0.130 
to $\sim$0.7 on day 3.759 after $t_{\rm max}$ 
(see Fig.~\ref{fig:fsfn}). 

As the optically thin/thick interface between fast and slow 
(compressed) ejecta gradually opens, at some point, the angle 
of its opening will coincide with the line of sight (i.e., it will 
be equal to the orbital inclination, $i$). From this time on, we 
can see more into the central parts of the ejecta through 
the optically thinner medium, allowing us to directly observe 
the hot WDP. Such a situation probably occurred around day~4. 
This view is supported by the following results. 
\begin{enumerate}
\item
After this time, when the ionizing source (i.e., the hot WDP) rises 
above the optically thick horizon of the ejecta, the absorption 
in the direction towards the observer decreased significantly. 
This is supported by, 
(i) 
a rapid disappearance of the high-velocity absorption component of 
the hydrogen lines on day 3--4, which is created in the accumulated 
wind from the giant at the front of the expanding warm WDP by 
absorbing its light (see Fig.~\ref{fig:hahb}, 
Sect~\ref{sss:ourshocks}), 
(ii) 
a significant decrease in the flux absorbed by the narrow 
component from this time (see Fig.~\ref{fig:ha}\,e), and 
balancing the RV of the \hb\ and \ha\ line as the 
absorbing column rapidly decreased 
(see Sect.~\ref{sss:hanarrow}, Fig.~\ref{fig:ha}\,h), 
(iii) 
modeling of the X-ray NICER (0.5--10\,keV) spectra showed 
a decrease in the hydrogen column density ($N_{\rm H}$) on 
the line of sight by a factor of $\sim$3.5 from $\sim$day~1 
to $\sim$day~4--5 (see Fig.~6 of \cite{2023ApJ...955...37O} 
and Fig.~6\,d of \cite{2024ApJ...960..125I}), and by 
(iv) 
a gradual increase in the hard X-ray flux measured by 
{\it Swift}-XRT and NICER to a maximum around day 5 (see Fig.~2 
and 3 of \cite{2022ApJ...935...44C} and 
\cite{2023ApJ...955...37O}), which also could be related to 
the gradual opening of the ionized shocked polar regions of 
the ejecta (see Sect.~\ref{sss:ourshocks}). 
\item
The first $IUE$ spectrum, made on day 6.6, directly 
measured the stellar component of radiation from the hot WDP, which 
dominated the far-UV (see Fig.~\ref{fig:seduvir}\,d), confirming 
that the opening of the optically thin ($\approx$ionized) polar 
regions was larger than $i$ within our interpretation. 
\item
Between day 3.395 and 3.759, the warm WDP cooled by $\sim$2,300\,K, 
corresponding to an almost doubling of its effective radius 
(Table~\ref{tab:par}). Its emission shifted to the red part of 
the spectrum, while the nebular continuum began to dominate in 
the blue, leading to a flattening of the optical spectrum 
(see Figs.~\ref{fig:sedopt}, \ref{fig:seduvir}, \ref{fig:fsfn}, 
\ref{fig:cont}, and \ref{fig:uves}). 
We interpret this significant temperature drop as a consequence 
of the rise of a cold, dense shell behind the internal shocks 
above the horizon of the compressed equatorial outflow, which 
thus began to contribute more to the thermal radiation from 
the weakening warm WDP (see Sect.~\ref{sss:ourshocks}). 
\end{enumerate}
%
%
%
\subsection{Density enhancement in the orbital plane}
\label{ss:deop}
After the continuum flattened, the warm WDP was evident in 
the SED models as an extra component until the end of our 
observations (the solid blue line in Figs.~\ref{fig:sedopt} and 
\ref{fig:seduvir}). However, its luminosity and temperature 
gradually decreased, so that its presence was sometimes not 
necessary for SED modeling (day 31.349 and 33.361; 
Table~\ref{tab:par}, Fig.~\ref{fig:seduvir}\,h). 
According to our interpretation of the ejecta evolution 
(Sect.~\ref{ss:evolstr}), the continued expansion of the ejecta 
and the decrease in the mass-loss rate led to a dilution of the 
ejecta. Its optically densest part became geometrically thinner, 
forming a disk-like structure surrounding the WD in the equatorial 
(orbital) plane. 
According to our SED models, it can persist within the ejecta for 
a longer time until it becomes optically thin in the continuum 
and fully ionized (see Fig.~\ref{fig:seduvir}) -- the warm WDP 
disappears. 
However, a higher density in the equatorial region compared to 
the polar region persists. Moreover, the focusing of the stellar 
wind from the giant towards the orbital plane increases the density 
here, ahead of the ejecta\footnote{Assuming the ejecta moves at 
a velocity of $\lesssim$1\,800\kms\ in the orbital plane 
\citep[][]{2007A&A...464..119C}, the front of the ejecta reaches 
a distance of $\lesssim$31\,AU on day 30.}. 
This effect has been theoretically demonstrated in several studies 
\citep[e.g.,][ for RS~Oph]{2016MNRAS.457..822B} as well as 
confirmed directly from observations 
\citep[e.g.,][]{2016A&A...588A..83S}. 
In our study, this effect is supported by narrow components in 
the spectral lines whose profile is not of the P-Cygni type 
(see Sects.~\ref{sss:hanarrow} and \ref{ss:environment}). 
%

The presence of the density enhancement on the orbital plane was 
directly indicated in 5\,GHz radio imaging made by the European 
VLBI Network on day 34, when the wind from the giant, centered 
in the orbital plane, obscured the receding lobe of the bipolar 
ejecta \citep[][]{2022A&A...666L...6M}. 
By analyzing the high-resolution imaging of the evolving bipolar 
outflows and assuming symmetry of both the approaching and 
receding lobes of the ejecta, \cite{2024A&A...692A.107L} 
estimated the density at/around the orbital plane as a function 
of radial distance from the binary, which is responsible for 
the flux attenuation between the lobes. 
%
After the 2006 explosion, by analyzing the Keck Interferometer 
Nuller observations on $\sim$day~3.8, \cite{2008ApJ...677.1253B} 
also discussed the results in terms of a model that includes an 
increase in density in the orbital plane. 

Finally, we note that the detection of increased density in 
the orbital plane from radio observations represents further 
independent indication of the focusing of the giant's wind 
towards the orbital plane, which makes the wind-mass-transfer 
in S-type symbiotic stars very efficient 
\citep[see][ and references therein]{2023A&A...680A..60S}. 
%
%
\subsection{Surrounding wind environment}
\label{ss:environment}
According to \cite{2025CoSka..55c..47M}, the narrow emission 
component comes from the pre-existing RG wind, which may 
be ionized by the UV-flash photons generated approximately 1 day 
after the TNR onset, until the formation of expanding ejecta 
absorbs them \citep[][]{2016ApJ...830...40K}\footnote{We add 
that the hot WDP, which appears at early stages of the outburst 
and gives rise to strong nebular emission even before the optical 
maximum, can also ionize the surroundings of the binary 
(Sect.~\ref{sss:parhot}).}. 
Accordingly, at this time ($\sim$ time $t_0$), the maximum part 
of the giant wind is ionized. As a result, the first observations 
will always indicate a maximum flux of the narrow emission 
component. Its subsequent gradual decrease 
(Fig.~\ref{fig:ha}\,e) is the result of ongoing recombinations 
in the ionized wind and the reduction in the volume of 
the undisturbed part of the wind due to the expansion 
of the ejecta. In the closest parts to the binary, up to distances 
of a few AU with densities of several times 10$^{10}$ to 
10$^{8}$\cmt\ \citep[][]{2016MNRAS.457..822B}, recombinations 
will occur on time scales of minutes to hours. 
In these parts, on the line of sight between the forming warm WDP 
and the observer, the value of $N_{\rm H}$ will therefore increase, 
which causes an increase in the absorption of light emitted by 
the underlying warm WDP -- the flux absorbed by the narrow 
component increases. 
However, the expansion of the ejecta gradually shortens the 
column of undisturbed wind from the giant above the warm WDP, 
which slows down the growth of $N_{\rm H}$. The resulting slight 
increase in absorption by the narrow component during the first 
days (see Fig.~\ref{fig:ha}\,e) is therefore the result 
of the rivalry of these two processes -- the neutralization of 
the initially ionized wind from time $t_0$ and the shortening 
of its column by the expansion of the ejecta. 
According to our interpretation of the evolution of the ejecta 
structure (Sect.~\ref{ss:evolstr}), the value of $N_{\rm H}$ 
significantly reduces after around day 5, when the hot WDP 
starts ionizing the material on the line of sight. 
As a result, the flux absorbed in the undisturbed wind 
decreases significantly on a timescale of several days 
(Fig.~\ref{fig:ha}\,e). 

The evolution of the width of both narrow emission and absorption 
components follows the evolution of their fluxes. This is because 
the latter depends on the size of the column of emitting or 
absorbing material, as described above. Accordingly, in later 
stages of evolution, the column of undisturbed wind is further 
away from the binary, where theoretical models expect a smaller 
dispersion of RVs, and vice versa. 
Also, their very low fluxes correspond to smaller FWZI, since 
their wings are too weak to be measured accurately. 

Finally, the RVs of the narrow emission lines 
(Fig.~\ref{fig:ha}\,h) show interesting behavior. 
After subtracting the systemic velocity, the narrow emission is 
located between 24 and 35\kms, around the terminal velocity of 
the RG wind. 
This means that the narrow emission component is emitted by the 
giant's wind, which is mostly moving away from the observer, 
which is consistent with the non-P~Cyg profile of narrow 
components (Sect.~\ref{sss:hanarrow}). 
The non-spherically symmetric distribution of the giant's wind 
with a reduced density in the polar sides is also suggested by 
the hydrodynamic modeling of the wind-mass-transfer in symbiotic 
binaries \citep[e.g.,][]{2016MNRAS.457..822B,2020MNRAS.493.2606B} 
and the finding from observations that the giant's wind is focused 
towards the orbital plane at the expense of the polar regions 
\citep[e.g.,][]{2016A&A...588A..83S,
                2022A&A...666L...6M}. 
%
%
\subsection{Shocks in the RS~Oph ejecta}
\label{ss:shocks}
\subsubsection{A sketch of shocks in nova outflows}
\label{sss:shocks}
The detection of gamma rays in the early stages of nova evolution 
indicates the presence of shock waves inside the nova ejecta 
\citep[e.g.,][]{2013A&A...551A..37M}. 
According to recent studies 
\citep[e.g.,][]{2014MNRAS.442..713M,
                2015MNRAS.450.2739M,
                2017NatAs...1..697L,
                2018A&A...612A..38M,
                2021ARA&A..59..391C} 
aimed at explaining the nature of non-thermal emission (especially 
gamma rays) measured during the explosion of some CNe, we summarize 
the principle of shock waves formation in nova outflows as follows. 

The shock model assumes the existence of an initial, relatively 
slow equatorially-focused outflow that is followed by a fast, more 
spherically symmetric outflow/wind. The fast wind then penetrates 
the previous, slower, and denser flow, generating an outwardly 
propagating forward-reverse shock structure, the so-called 
internal shocks. 
The shocks can accelerate a fraction of the charged particles 
to relativistic velocities that emit gamma rays through hadronic 
and leptonic interactions in a dense cold layer downstream of 
the shocks. The latter is the result of a rapid cooling of the hot 
shocked plasma by converting the rest of its energy into X-rays 
and UV emission. The cool gas collects behind the forward and 
reverse shocks, creating the so-called 'cold central shell'. 
According to the shock dynamics, the cold central shell will 
quickly reach an approximate steady state with a constant 
velocity, identical to the velocity of shocks 
\citep[see Eqs.~(11) and (12) of][]{2017NatAs...1..697L}. 
The harder thermal emission from the shocks is then absorbed and 
reprocessed by the dense slow ejecta ahead of the shocks into 
optical emission \citep[see Fig.~1 of][]{2015MNRAS.450.2739M}. 

The observed correlation between optical and $\gamma$-ray LCs 
around the optical maximum until several weeks after it 
\citep[see Fig.~2 of][]{2022ApJ...935...44C} supports the above 
scenario and indicates that a significant fraction of the optical 
light comes from reprocessed shock emission. 
In the following section, we probe this property within our model. 

\subsubsection{Shocks in the wind-compression model of RS~Oph}
\label{sss:ourshocks}
Figure~\ref{fig:sketch} shows a sketch of the geometric and 
ionization structure of the RS~Oph ejecta as inferred from our 
SED modeling (Sect.~\ref{ss:sed}) and its interpretation 
(see Sects.~\ref{ss:ionejecta} to \ref{ss:deop}). The location
of the shocks and the cold central shell in the ejecta is 
outlined according to Fig.~1 of \cite{2015MNRAS.450.2739M}. 
The bipolar density and velocity distribution in the mass outflow, 
which is essential for the formation of shocks in the ejecta 
during nuclear outbursts on the WD surface, is fairly well 
demonstrated by observations 
\citep[e.g.,][ for CNe]{2014Natur.515..234S,
                        2014Natur.514..339C,
                        2013ApJ...768...49R}. 
Also, in less violent Z~And-type outbursts in SySts, the ejected 
material expands more slowly in the equatorial region due to its 
compression here than material with reduced density in the polar 
regions. The former is indicated by the emergence of P~Cyg 
profiles, while the latter is characterized by the broad emission 
wings of hydrogen lines during the outbursts; best documented for 
systems with high orbital inclination 
\citep[see][]{2006A&A...453..279S,2006A&A...457.1003S}. 
However, the development of the bipolar structure from time $t_0$ 
has not been recorded yet. 

Within our model, the process of wind compression into the orbital 
plane (i.e., the transition from spherical geometry at time $t_0$ 
to bipolar) occurs as fast as the wind expands. Already after 14 
hours (day -0.366), the nebular emission was represented by bright 
hydrogen emissions, and on day 0.13 it was possible to estimate 
the amount of its continuum, which on day 0.617 clearly dominated 
the short-wave part of the spectrum (see Figs.~\ref{fig:sedopt}\,d 
and \ref{fig:fsfn} for $\lambda$3,600\,\AA). 
The compressed equatorial part of the outflow/wind is cooled by 
recombinations and bremsstrahlung due to increasing 
density, and slowed down by increasing frictional forces. 
The continued fast wind emitted from the WD surface then 
penetrates the previously emitted, but already compressed and 
slowed wind, creating conditions for the formation of internal 
shocks. 

{\sf Luminosity generated by shocks.} 
According to the ejecta structure (Fig.~\ref{fig:sketch}) and 
the principle of shocks formation in it (Sect.~\ref{sss:shocks}), 
the luminosity of the warm WDP, $L_{\rm WD}^{\rm warm}$, in our 
model, should be comparable to the luminosity of the internal 
shocks. 
Assuming that the velocity of the slow outflow is close to 
the velocity of the propagating shock, the total shock power, 
$L_{\rm sh}$, is dominated by the reverse shock, i.e. 
\citep[see Eq.~(15) of][]{2017NatAs...1..697L}, 
\begin{equation}
  L_{\rm sh} = \frac{9}{32}f_{\Omega}
               \frac{f^{1/2}\dot M_{\rm WD}}{(1 - f_{\Omega})}
               \frac{(v_{\rm f} - v_{\rm sh})^3}{v_{\rm f}} ,
\label{eq:lshock}
\end{equation}
where $f_{\Omega}$ is the fractional solid angle of the slow 
equatorially-focused outflow (i.e., $f = 1-f_{\Omega}$ in 
Eq.~(\ref{eq:dotM})), $f^{1/2}\dot M_{\rm WD}/(1 - f_{\Omega})$ 
is the spherical equivalent of the mass-loss rate, $v_{\rm f}$ 
is the velocity of the wind, and $v_{\rm sh}$ is the velocity 
of the propagating internal shock, assumed to be close to 
the expanding velocity of the cold central shell behind 
the shocks. 

\begin{figure}[t]
\begin{center}
\resizebox{\hsize}{!}{\includegraphics[angle=-90]{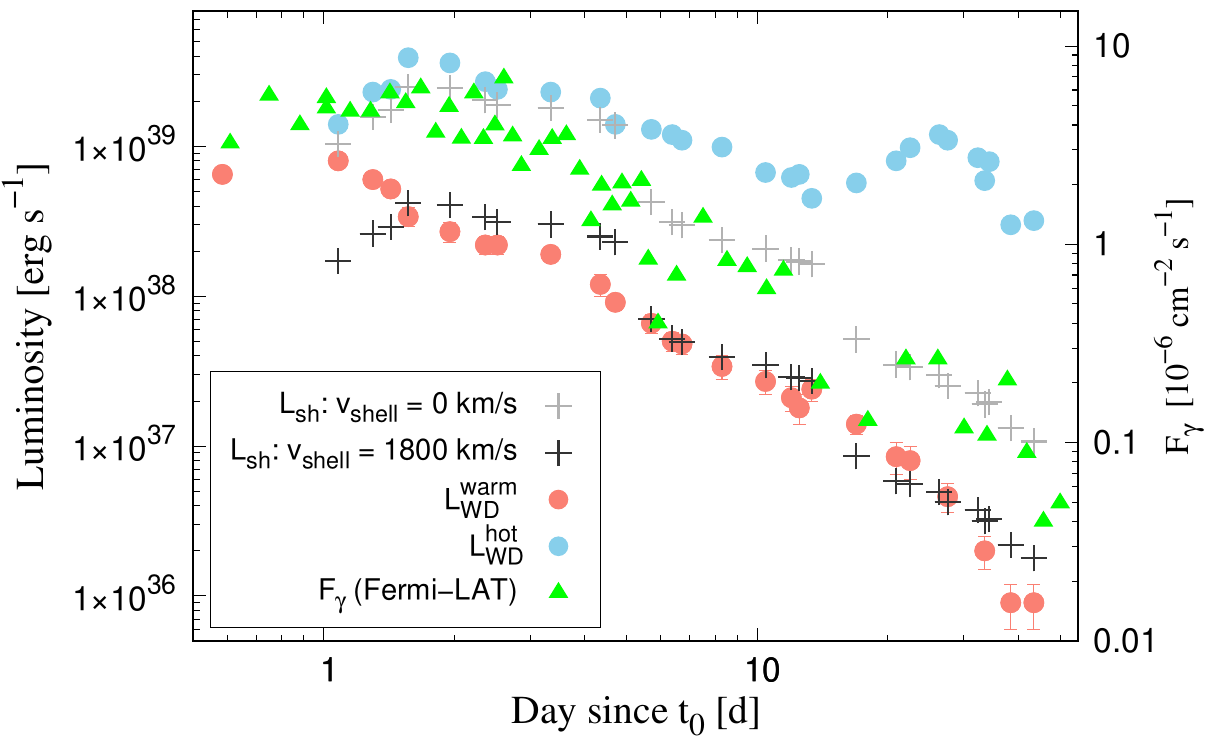}}
\end{center}
\caption{
Comparison of the warm WDP luminosity ($L_{\rm WD}^{\rm warm}$, 
Table~\ref{tab:par}), 
the luminosity generated by internal 
shocks in the slow equatorial outflow ($L_{\rm sh}$, 
Eq.~(\ref{eq:lshock})), and the Fermi--LAT $\gamma$-ray photon 
fluxes \citep[$F_{\gamma}$, from Fig.~2 of][]{2022ApJ...935...44C} 
as a function of the nova age. The comparability of 
$L_{\rm WD}^{\rm warm}$ and $L_{\rm sh}$, and their correlation 
with $F_{\gamma}$ confirms that internal shocks generate 
a significant part of the luminosity of the warm WDP during 
its presence in the spectrum. 
          }
\label{fig:lsh}
\end{figure}

In our model, the optically thin polar part of the ejecta is 
gradually opening, which narrows the residual compressed part of 
the slow equatorial wind; the parameter $f_{\Omega}$ decreases. 
On $\sim$day~5 after $t_0$, the opening angle of the optically 
thin/thick interface between the fast and slow (compressed) 
ejecta was approximately equal to $i = 52{\degr}$ 
(see Sect.~\ref{ss:evolstr}), which corresponds to 
$f_{\Omega}=0.62$. 
In estimating $L_{\rm sh}$, we adopted this value until day 4.7. 
Then, until day 13.4, we used $f_{\Omega}=0.34$, and for 
the rest of the observations (day 16.9 to 43.5 after $t_0$) 
we chose $f_{\Omega}=0.17$. 
Further, we used the values of $\dot M_{\rm WD}$ from 
Table~\ref{tab:par}, which correspond to $f=1$, 
$v_{\rm f} = 4,000$\kms\ based on the terminal velocities of 
the broad \ha\ wings (see Fig.~\ref{fig:ha}\,d), and for 
$v_{\rm sh}$, according to \cite{2007A&A...464..119C}, we chose 
two limiting values, 0 and 1\,800\kms. 
Figure~\ref{fig:lsh} shows a good agreement between 
the values of $L_{\rm WD}^{\rm warm}$ and $L_{\rm sh}$ for 
$v_{\rm sh} = 1\,800$\kms. 
However, it is likely that the actual values of $L_{\rm sh}$ are 
smaller than the values of $L_{\rm WD}^{\rm warm}$ determined 
directly from the spectrum, because the efficiency of generating 
shocks is expected to be higher near the equator due to 
the largest velocity difference, $v_{\rm f}-v_{\rm sh}$, than 
at higher latitudes of the slow outflow under the wind compression 
model. 

The comparability of $L_{\rm WD}^{\rm warm}$ and $L_{\rm sh}$ 
values throughout the entire period, until the dispersion of 
the ejecta at the beginning of the SSS phase (Fig.~\ref{fig:lsh}), 
suggests that the slow equatorial outflow had been continuously 
replenished by the compression of the initially 
spherically-symmetric fast wind from the WD during this period. 
This result suggests that the wind compression model can create 
a bipolar structure of the outflow shortly after the ejection of 
the optically thick envelope at time $t_0$, which can persist 
for a longer time. Its density and velocity contrasts, suitable 
for shock formation, are limited by the value of $\dot M_{\rm WD}$. 
When the $\dot M_{\rm WD}$ drops significantly, the optically 
thick equatorial disk structure disappears\footnote{In the same 
way, the enhanced value of $\dot M_{\rm WD}$ during Z~And-type 
outbursts in SySts is responsible for the formation of 
an optically thick disk-like structure in the orbital plane, 
which disappears as the $\dot M_{\rm WD}$ decreases, and 
the system enters a quiescent phase 
\citep[see Sect.~3.1 of][]{2012A&A...548A..21C}.}, 
and shock formation and, hence, $\gamma$-rays are also 
significantly attenuated or disappear, creating conditions 
for observing SSS emission from the burning WD. 
%

{\sf The cold central shell.}
The high cadence of our spectra allowed us to capture a sudden 
drop in the temperature of the warm WDP by $\sim$2,300\,K 
from day 3.395 to day 3.759 after $t_{\rm max}$, but at 
a comparable luminosity and a much larger effective radius 
(see Table~\ref{tab:par}). The warm WDP emission shifted to longer 
wavelengths -- the $F_{\rm WD}^{\rm warm}/F_{\rm N}$ ratio 
suddenly decreased in the blue part of the spectrum 
($\lambda$\,3,600 -- 4,700\,\AA), but increased at 
$\lambda$\,7,300\,\AA\ (see Fig.~\ref{fig:fsfn}). 
Within the internal shocks model in the compressed wind
(Fig.~\ref{fig:sketch}), this effect could be interpreted 
as a result of narrowing the slower 
dense outflow to the equator, below the line of sight 
(i.e., $f_{\Omega}<0.62$), when the radiation from the warm WDP 
weakens as it is viewed from the side. In contrast, radiation 
from the cold, dense shell behind the internal shocks becomes 
visible when viewing the slow outflow from above at later times 
($>$day~4--5 after $t_0$) and can thus contribute to the warm 
thermal radiation. 
This explanation is supported by the long-lasting presence of 
the very cold radiation in the spectrum 
($T_{\rm BB}\sim 5,200 - 3,600$\,K; the blue line in 
Figs.~\ref{fig:sedopt} and \ref{fig:seduvir}), whose
contribution decreases along with the Fermi-LAT $\gamma$-ray 
radiation. 
Also, the {\it Swift}-XRT emission (2--10\,keV) from the hot 
base of the internal reverse shock should be better observed 
at $f_{\Omega}<0.62$ because it suffers from less absorption, 
which would explain the shift of its maximum to days 5--6 after 
$t_0$ \citep[see Fig.~2 of][]{2022ApJ...935...44C}. 

{\sf External shocks.} 
In the case of symbiotic binaries, the ejecta penetrates 
the surrounding relatively stable and thin stellar wind from 
the giant, generating external shocks at the expense of its 
kinetic energy. 
The high-velocity absorptions observed in the \ha\ profile during 
the first 3--4 days (see Fig.~\ref{fig:hahb}) 
probably originate in the cold absorbing shell accumulated behind 
the external shocks (Sect.~\ref{sss:habroad}), which represents 
their manifestation in the optical spectrum at a very early stage. 
The evolution of the broad \ha\ component indicates a relatively 
small deceleration of the ejecta in the line of sight due to 
external shocks -- the broad wings are still present, although very 
weak (see Fig.~\ref{fig:brandnar}), with only a slight decrease of its 
terminal velocity to about 4,000\kms\ (Fig.~\ref{fig:ha}\,d). 
The small deceleration of the ejecta indicated by the \ha\ line 
profile is likely a consequence of the significant dilution of 
the wind from the giant in the polar regions. 

\section{Summary}
\label{s:summ}
In this work, we focused on the analysis of the optical continuum 
of the recurrent symbiotic nova RS~Oph during its recent 2021 
explosion, from 9 hours before the brightness maximum to 
the SSS phase ($\sim$day~42). Using the method of 
the multi-wavelength SED modeling, we determined the ionization 
structure of the ejecta 
and its physical parameters as a function of the nova age 
(Figs.~\ref{fig:sketch}, \ref{fig:sedopt}, \ref{fig:seduvir}, 
\ref{fig:lrtcont}, \ref{fig:dotM}, Table~\ref{tab:par}). 
Additional properties of the ejecta and the surrounding environment 
were obtained by analyzing the evolution of the \ha\ line using 
medium to very high resolution spectroscopy (Sects.~\ref{ss:ha}, 
and \ref{ss:environment}). 
The main results of our analysis can be summarized as follows. 
\begin{enumerate}
\item
On day $-$0.366 (i.e., 9 hours before $t_{\rm max}$ or 14 hours 
after $t_0$), the warm WD pseudophotosphere 
(warm WDP, see Sect.~\ref{ss:sed}) resembled an A-type star with 
$T_{\rm eff}$ of $\sim$11,000\,K, inflated to 
$\sim$114\,$(d/1.6\,{\rm kpc})$\ro, generating luminosity of 
$L_{\rm WD}^{\rm warm}\sim 7\times 10^{38}\,(d/1.6\,{\rm kpc})^2$\es. 
On day $+$0.130 (3 hours after $t_{\rm max}$), the $T_{\rm eff}$ of 
the warm WDP dropped to $\sim$7,250\,K and its effective radius 
$R_{\rm WD}^{\rm eff}$ (see Sect~\ref{sss:parwarm}) reached 
a maximum of $\sim$290\,$(d/1.6\,{\rm kpc})$\ro. The estimate of 
the nebular continuum emission measure, 
$EM\gtrsim10^{62}\,(d/1.6\,{\rm kpc})^2$\cmt, limited 
the luminosity of the hot WD pseudophotosphere 
$L_{\rm WD}^{\rm hot}$ (hot WDP, see Sect.~\ref{ss:sed}) 
to $\gtrsim 1.4\times 10^{39}\,(d/1.6\,{\rm kpc})^2$\es, and 
the mass-loss rate $\dot M_{\rm WD}$ to 
$\gtrsim$2.3$\times 10^{-4}$\myr. 
During the next 12 hours (day 0.617), $EM$ and $L_{\rm WD}^{\rm hot}$ 
increased by a factor of $>$3, reached their maximum, and the nebular 
continuum began to rival the radiation from the warm WDP 
(see Fig.~\ref{fig:seduvir}\,a). 
The corresponding maximum value of $\dot M_{\rm WD}$ was 
$\sim$5$\times 10^{-4}$\myr. Integrating $\dot M_{\rm WD}$ 
from day 0.13 to day 42.6 corresponds to an average value of 
the ejected ionized mass of 1.2$\times 10^{-5} $\mo\ 
(Sects.~\ref{ss:mdot}, \ref{ss:mass}, and Fig.~\ref{fig:dotM}). 
\item
Disentangling the spectrum into the stellar and nebular components 
of radiation from the very beginning of the explosion revealed 
the biconical structure of the ejecta: A flared density-enhanced 
disk-like structure around the equator, whose outer rim (i.e., 
the warm WDP) obscures a strong ionizing source (i.e., the hot WDP) 
in the line of sight, whose radiation ionizes the rarefied polar 
regions giving rise to the strong nebular emission
(Sect.~\ref{ss:ionejecta}, Fig.~\ref{fig:sketch}). 
We explain the bipolar structure of the ejecta and its evolution 
by rotation of the accreting WD, which leads to a compression of 
the outflowing wind towards the equatorial plane 
(see Sects.~\ref{ss:cause}, and \ref{ss:evolstr}). 
\item
On $\sim$day~4 (or $\sim$day~5 after $t_0$), the continuum suddenly 
flattened (Fig.~\ref{fig:cont}), the $T_{\rm BB}$ of the warm WDP 
dropped by $\sim$2,300\,K with subsequent gradual decline of 
its luminosity and cooling 
(Table~\ref{tab:par}, Figs.~\ref{fig:seduvir} and \ref{fig:lrtcont}\,d). 
Simultaneously, a significant decrease in absorption towards 
the observer was indicated, and the hot WDP became directly 
observable. 
According to the ejecta evolution (see Sect.~\ref{ss:evolstr}), 
these effects can be caused by narrowing the optically thick 
equatorial outflow below the line of sight (i.e., 
the fractional solid angle $f_{\Omega} < 0.62$, 
see Eq.~(\ref{eq:lshock})). 
Accordingly, the sudden change in $T_{\rm BB}$ can be interpreted 
by the direct visibility of the cold dense shell behind the internal 
shocks (see Fig.~\ref{fig:sketch}) after $\sim$day~4. 
The comparability of $L_{\rm WD}^{\rm warm}$ with the luminosity of 
the internal shocks $L_{\rm sh}$ (Eq.~(\ref{eq:lshock})) and their 
correlation with the Fermi-LAT $\gamma$-ray fluxes until 
$\sim$day~42, support this interpretation 
(see Fig.~\ref{fig:lsh}, Sect.~\ref{sss:ourshocks}). 
\item
Very high resolution UVES spectra ($R\gtrsim10^5$) showed that 
the profile of the narrow line components is not of the P~Cyg 
type (see Fig.~\ref{fig:brandnar}). 
This fact, and the position of the narrow emission component 
near the terminal velocity of the RG wind are consistent 
with the non-spherically symmetric distribution of the giant's 
wind with a reduced density in the polar regions 
(see Sects.~\ref{sss:hanarrow}, \ref{ss:deop}, and 
\ref{ss:environment}). 
\end{enumerate}
The wind compression model (Sect.~\ref{ss:cause}, 
Fig.~\ref{fig:sketch}) represents a promising mechanism shaping 
the bipolar structure of the ejecta from very early stages, with 
sufficient density and velocity contrasts to generate internal 
shocks. Our UV/optical/near-IR continuum analysis (Sect.~\ref{ss:sed}) 
confirmed the presence of strong internal shocks that form within 
the compressed, slower equatorial outflow, the power of which 
is comparable with the warm WDP luminosity from the maximum 
until our last observations on day~42 
(see Sect.~\ref{sss:ourshocks}, and Fig.~\ref{fig:lsh}). 
We conclude that the disentangling method allows us to specify 
the efficiency and evolution of $\gamma$-ray emission more precisely, 
and to estimate the structure of mass ejection in novae. 
%
%
\begin{acknowledgments}
The authors thank the anonymous referee for useful comments. 
This work was supported by a grant of the Slovak Academy 
of Sciences, VEGA No. 2/0003/25, and by the Slovak Research 
and Development Agency under the contracts No. APVV-20-0148, 
and APVV-24-0160. The research of MW was partially supported 
by the project Cooperatio -- Physics of Charles University 
in Prague. 
The amateur spectra presented in this paper were obtained 
within the {\it Astronomical Ring for Access to Spectroscopy 
(ARAS)}, an initiative promoting cooperation between professional 
and amateur astronomers in the field of spectroscopy. 
The authors thank ARAS observers for their contributions made 
within the ARAS program coordinated by Francois Teyssier. 
This paper is in part based on data from Paranal Observatory, 
ESO, Chile. The observations have been taken under a target 
opportunity program 105.20B6.001, 105.20B6.002, PI Paolo Molaro. 
We obtained the spectra from the ESO Science Archive Facility 
with DOI: https://doi.org/10.18727/archive/50. 
This publication also uses ultraviolet spectroscopy made with 
the {\it International Ultraviolet Explorer}, obtained from 
the MAST data archive at the Space Telescope Science Institute, 
which is operated by the Association of Universities for 
Research in Astronomy, Inc., under NASA contract NAS\,5-26555. 
We also acknowledge the variable-star observations from the AAVSO 
International Database contributed by observers worldwide and used 
in this research. 
\end{acknowledgments}

\begin{contribution}
AS was responsible for data analysis and interpretation, and for 
writing the initial draft of the manuscript. MV, FT, MF, M\v{S}, 
and MW contributed to the acquisition and processing of 
spectroscopic observations (see Sect.~\ref{ss:spec}). 
SS contributed to the acquisition and processing of the photometric 
observations (see Sect.~\ref{ss:phot}, and Appendix~\ref{app:phot}). 
\end{contribution}

\facilities{IUE, AAVSO, ARAS, Perek:2m, UT2 Kueyen:8m} 
%

\appendix

%
\section{Multicolor optical photometry}
\label{app:phot}
In this appendix, we present the table with our multicolor 
$UBVR_{\rm C}I_{\rm C}$ photometry of RS~Oph during its 2021 
outburst (see Sect.~\ref{ss:phot}, Table~\ref{tab:ubvri}). 
%
\begin{table*}[h!]
\caption{$UBVR_{\rm C}I_{\rm C}$ photometry of RS~Oph obtained 
at the Star\'a Lesn\'a Observatory during its 2021 outburst.}
\begin{center}
\begin{tabular}{cccccccccccc}
\hline
\hline
\noalign{\smallskip}
HJD        &     Date           & 
\multicolumn{2}{c}{$U$}         &
\multicolumn{2}{c}{$B$}         &
\multicolumn{2}{c}{$V$}         & 
\multicolumn{2}{c}{$R_{\rm C}$} &
\multicolumn{2}{c}{$I_{\rm C}$} \\
2\,459... & mm-dd.ddd   &  mag & error & mag & error & mag & error 
          &  mag & error & mag & error  \\
\noalign{\smallskip}
\hline
\noalign{\smallskip}
464.296 & 09-06.796 &   9.284 &   0.070 &   9.644 &   0.063 &   9.164 &   0.058 &   7.350 &   0.088 &   7.458 &   0.073 \\
467.256 & 09-09.756 &   9.564 &   0.056 &   9.926 &   0.048 &   9.379 &   0.043 &   7.588 &   0.079 &   7.667 &   0.063 \\ 
476.265 & 09-18.765 &   9.989 &   0.053 &  10.310 &   0.045 &   9.743 &   0.039 &   8.020 &   0.075 &   8.057 &   0.060 \\ 
490.225 & 10-02.725 &  10.442 &   0.049 &  10.664 &   0.044 &  10.090 &   0.037 &   8.484 &   0.069 &   8.388 &   0.059 \\ 
493.218 & 10-05.718 &  10.499 &   0.043 &  10.693 &   0.037 &  10.073 &   0.028 &   8.534 &   0.064 &   8.393 &   0.055 \\ 
496.216 & 10-08.716 &  10.551 &   0.036 &  10.786 &   0.029 &  10.124 &   0.019 &         &         &         &         \\
500.206 & 10-12.706 &  10.696 &   0.034 &  10.870 &   0.029 &  10.201 &   0.018 &   8.692 &   0.058 &   8.527 &   0.050 \\ 
503.209 & 10-15.709 &  10.753 &   0.036 &  10.968 &   0.029 &  10.259 &   0.017 &   8.795 &   0.056 &   8.591 &   0.048 \\ 
512.193 & 10-24.693 &  10.939 &   0.042 &  11.106 &   0.038 &  10.485 &   0.028 &   9.051 &   0.059 &   8.784 &   0.054 \\ 
514.191 & 10-26.691 &  11.063 &   0.039 &  11.201 &   0.035 &  10.531 &   0.022 &   9.079 &   0.056 &   8.812 &   0.051 \\ 
517.182 & 10-29.682 &  11.154 &   0.038 &  11.351 &   0.033 &  10.685 &   0.021 &   9.237 &   0.057 &   8.950 &   0.052 \\ 
519.179 & 10-31.679 &  11.238 &   0.043 &  11.396 &   0.039 &  10.772 &   0.029 &   9.331 &   0.061 &   9.057 &   0.056 \\ 
524.181 & 11-05.681 &  11.633 &   0.046 &  11.711 &   0.041 &  11.001 &   0.023 &   9.637 &   0.058 &   9.291 &   0.056 \\
\noalign{\smallskip}
\hline
\end{tabular}
\end{center}
\label{tab:ubvri}
\end{table*}
%
%
%
\section{Spectra calibration}
\label{app:calib}
To calibrate the spectra in relative units (mostly the 
low-resolution spectra in Table~\ref{tab:low}), it is necessary 
to know the absolute flux at least at one wavelength. For this 
purpose, we used multicolor photometry acquired simultaneously 
with the spectrum. Due to the presence of emission lines in 
the spectrum, we first eliminated their influence on the true 
continuum. According to the method of
\cite{2007NewA...12..597S}, we determined corrections for 
emission lines, $\Delta m = m_{\rm obs} - m_{\rm cont}$, 
where $m_{\rm obs}$ is the observed magnitude, and 
$m_{\rm cont}$ is the magnitude of the true continuum. 
Their values are listed in Table~\ref{tab:dml}. 
Then we converted the emission-line-free $UBV$ and 
$R_{\rm C}I_{\rm C}$ magnitudes to fluxes according to 
the calibration of \cite{1982asph.book.....H} and 
\cite{1979PASP...91..589B}, respectively. 
In most cases, we scaled the spectra to the dereddened and
emission-line-free fluxes obtained from the $V$ magnitudes. 
Fluxes from other photometric filters served as a verification 
of the continuum profile. A systematic flux shift in the $B$ 
filter (up to 10--15\% above the continuum) was probably 
caused by the blending of emission lines in this region. 

\begin{figure}[h!]
\begin{center}
\resizebox{10cm}{!}{\includegraphics[angle=-90]{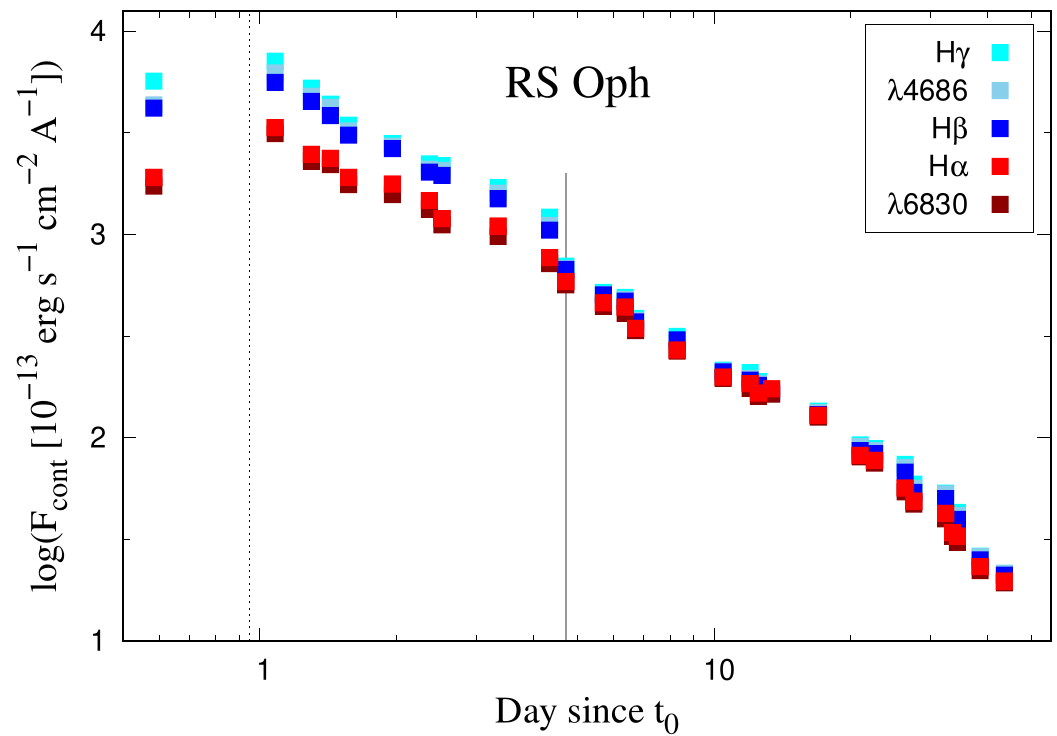}}
\end{center}
\caption{
Evolution of the continuum fluxes, $F_{\rm cont.}$, at/around 
selected emission lines (see keys) given by SED models. 
The vertical dotted line indicates the time of the optical 
maximum (see Table~\ref{tab:maxima}), and the gray line at 
day 4.7 denotes the time of spectrum flattening, when the continuum 
fluxes in the blue part (at \hg\ and \hb) became nearly equal 
to those in the red part (at \ha\ and $\lambda$6,830). 
}
\label{fig:cont}
\end{figure}

Having the low-resolution spectra calibrated, we determined 
absolute fluxes in the continuum at the wavelengths of selected 
emission lines along the outburst (see Fig~\ref{fig:cont}). 
Using these calibration curves, we can calibrate the corresponding 
regions in the medium-resolution spectra by interpolating adjacent 
continuum fluxes to the time of their acquisition. 
%
%
\begin{table}[h!]
\caption{Corrections of the $U$, $B$, $V$, and $R_{\rm C}$ 
         magnitudes for emission lines measured on the spectra 
         in Tables~\ref{tab:low} and \ref{tab:uves}. 
}
\begin{center}
\begin{tabular}{rccccc}
\hline
\hline
\noalign{\smallskip}
Age$^{\rm a}$~
& Sp. range 
& $\Delta U$ 
& $\Delta B$ 
& $\Delta V$ 
& $\Delta R_{\rm C}$ \\
(days) & (nm) & (mag) & (mag) & (mag) & (mag) \\
\hline
\noalign{\smallskip}
-0.366 & 390 - 724 &   --    &   --    &  -0.012 &  -0.005 \\
 0.130 & 370 - 860 &   --    &  -0.016 &  -0.002 &  -0.026 \\
 0.348 & 390 - 750 &   --    &  -0.077 &  -0.016 &  -0.031 \\
 0.478 & 371 - 739 &   --    &  -0.082 &  -0.023 &  -0.042 \\
 0.617 & 304 - 946 &  -0.005 &  -0.120 &  -0.084 &  -0.069 \\
 1.006 & 366 - 880 &  -0.005 &  -0.110 &  -0.057 &  -0.103 \\
 1.409 & 364 - 785 &  -0.006 &  -0.332 &  -0.107 &  -0.154 \\
 1.567 & 304 - 946 &  -0.019 &  -0.220 &  -0.096 &  -0.195 \\
 2.391 & 330 - 855 &  -0.029 &  -0.247 &  -0.118 &  -0.292 \\
 3.395 & 330 - 855 &  -0.049 &  -0.272 &  -0.116 &  -0.439 \\
 3.759 & 304 - 946 &  -0.040 &  -0.371 &  -0.142 &  -0.495 \\
 4.751 & 304 - 946 &  -0.065 &  -0.405 &  -0.211 &  -0.590 \\
 5.416 & 330 - 855 &  -0.064 &  -0.452 &  -0.186 &  -0.545 \\
 5.759 & 304 - 946 &  -0.074 &  -0.529 &  -0.300 &  -0.673 \\
 7.343 & 385 - 750 &   --    &  -0.489 &  -0.277 &  -0.746 \\
 9.510 & 304 - 946 &  -0.085 &  -0.614 &  -0.397 &  -0.915 \\
11.048 & 360 - 875 &  -0.039 &  -0.487 &  -0.320 &  -0.866 \\
11.559 & 304 - 946 &  -0.110 &  -0.585 &  -0.332 &  -0.904 \\
12.421 & 330 - 855 &  -0.077 &  -0.559 &  -0.331 &  -0.877 \\
15.998 & 355 - 874 &  -0.068 &  -0.610 &  -0.355 &  -0.661 \\
19.989 & 357 - 880 &  -0.088 &  -0.758 &  -0.429 &  -0.720 \\
21.583 & 304 - 946 &  -0.189 &  -0.772 &  -0.346 &  -0.596 \\
25.356 & 330 - 855 &  -0.142 &  -0.854 &  -0.426 &  -1.144 \\
26.559 & 304 - 946 &  -0.217 &  -0.849 &  -0.444 &  -1.171 \\
31.349 & 330 - 855 &  -0.261 &  -0.669 &  -0.428 &  -1.145 \\
32.549 & 304 - 946 &  -0.267 &  -0.907 &  -0.473 &  -1.225 \\
33.361 & 370 - 750 &   --    &  -0.650 &  -0.394 &  -1.072 \\
37.532 & 304 - 946 &  -0.284 &  -0.889 &  -0.528 &  -1.225 \\
42.566 & 304 - 946 &  -0.320 &  -0.930 &  -0.566 &  -1.350 \\
\hline
\end{tabular}
\end{center}
{\bf Notes.} 
$^a$ As in Table~\ref{tab:par}. 
%
\label{tab:dml}
\end{table}
%
%
%
%
%
\section{Evolution of the optical continuum on the UVES spectra}
\label{app:cont}
\begin{figure*}[h!]
\begin{center}
\resizebox{\hsize}{!}{\includegraphics[angle=-90]{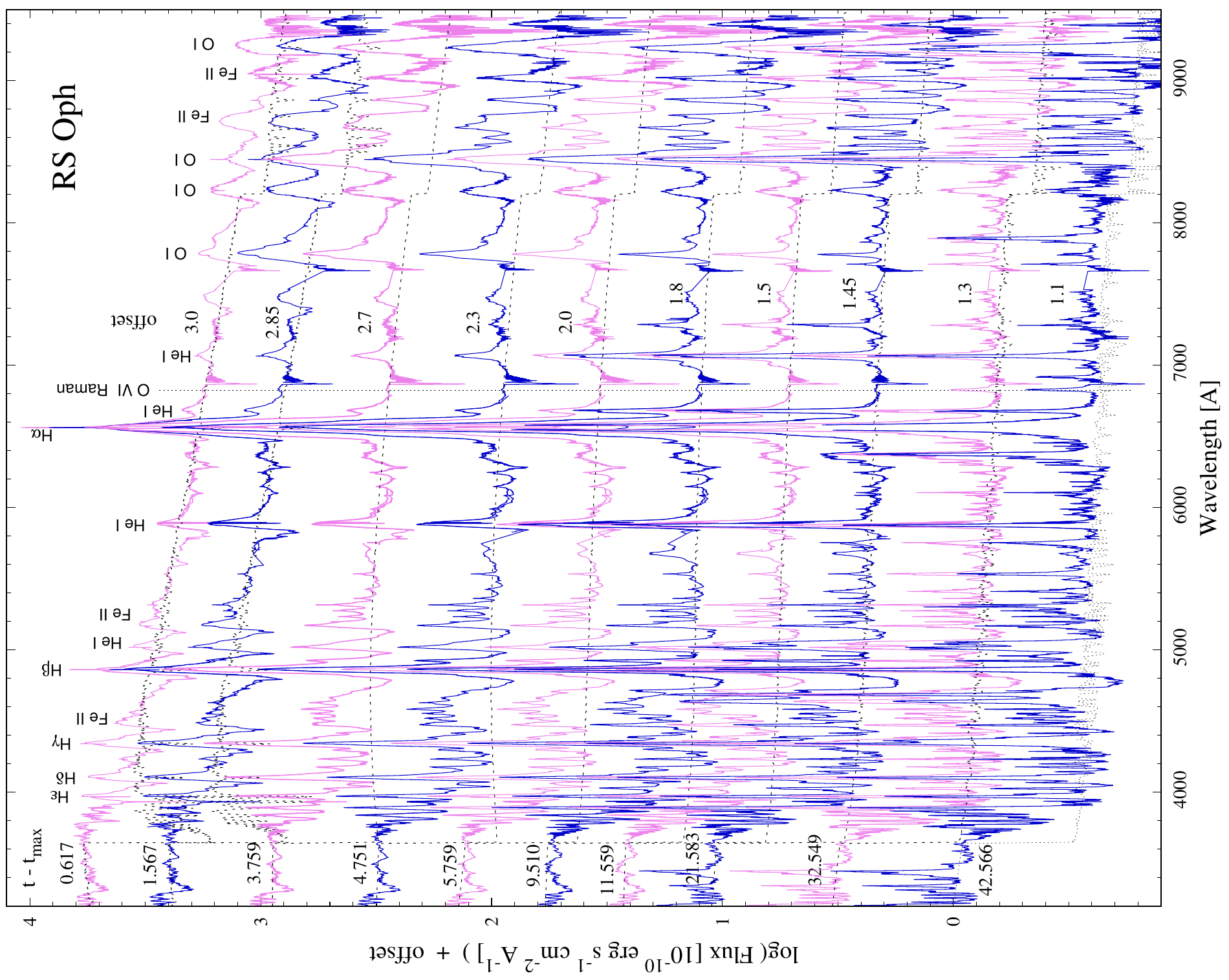}}
\end{center}
\caption{
The evolution of the RS~Oph optical spectrum measured by the UVES 
spectrograph from day 0.617 to day 42.566. A strong nebular 
continuum is indicated by the Balmer jump in emission. It 
dominates the spectrum from day 3.7, when the overall continuum 
flattens (see also Fig.~\ref{fig:cont}). 
Dotted lines represent our SED models (Eq.~(\ref{eq:fl1}) or 
(\ref{eq:fl2}) for parameters in Table~\ref{tab:par}). 
}
\label{fig:uves}
\end{figure*}

In this appendix, we present a figure showing the evolution of 
the RS~Oph continuum from just after $t_{\rm max}$ (day 0.617) 
to our last observation (day 42.566). In this example, we used 
high-resolution UVES spectra smoothed with a 1\,\AA\ filter 
which show the range from 3,200 to 9,500\,\AA\ 
(see Figure~\ref{fig:uves}). Our SED models in the figure are 
given by the parameters in Table~\ref{tab:par}, determined by 
Eq.~(\ref{eq:fl1}) or (\ref{eq:fl2}). 
%
%
%
%
\section{Evolution of the \ha\ and \hb\ line profiles}
\label{app:hahb}
\begin{figure*}[h!]
\begin{center}
\resizebox{17cm}{!}{\includegraphics[angle=-90]{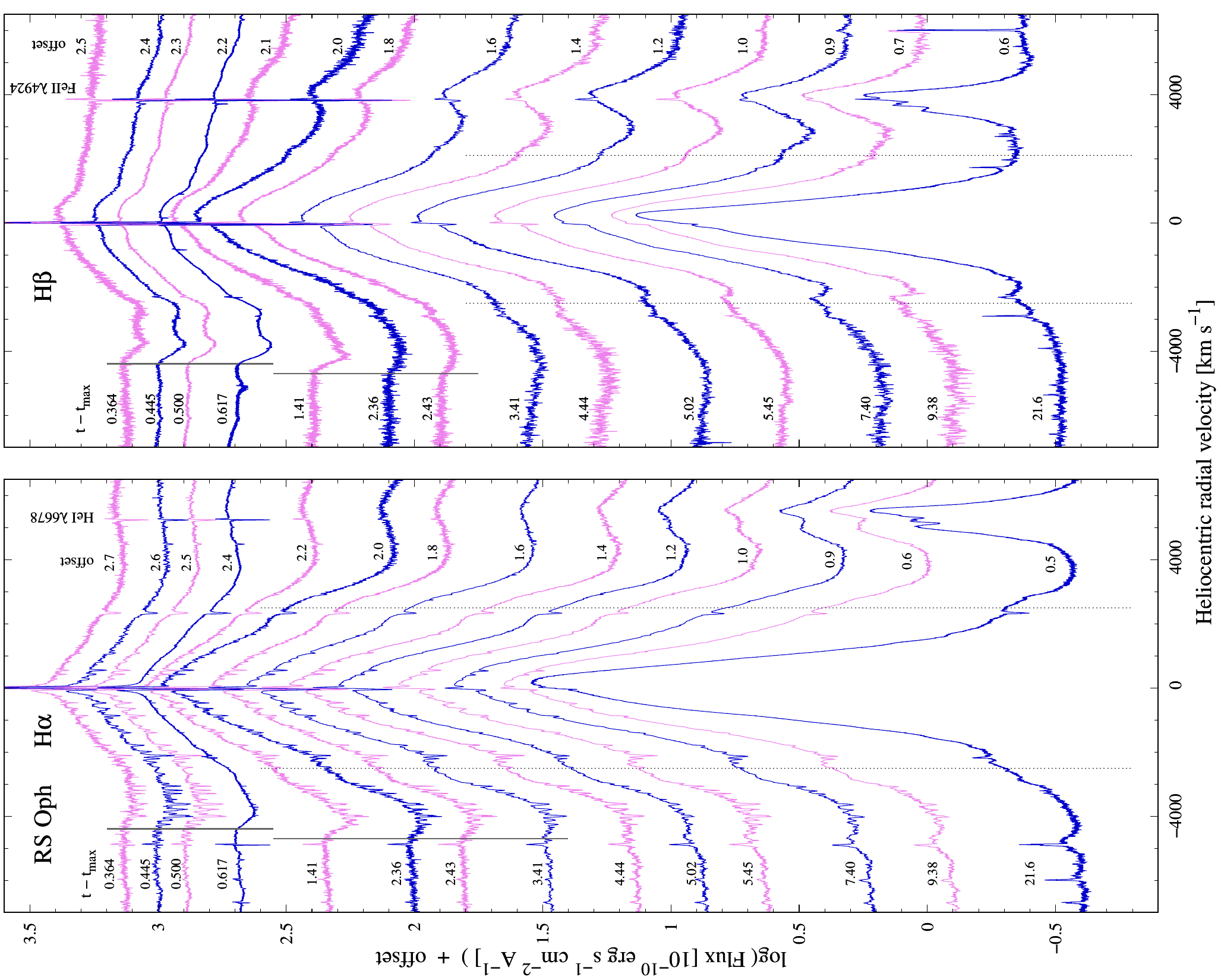}}
\end{center}
\caption{
Evolution of the \ha\ (left) and \hb\ (right) line profiles 
during the first hours and days of the RS~Oph explosion. 
Vertical gray lines denote the terminal velocity of the broad 
absorption component, while the dotted lines denote emission 
bumps located around $\pm$2,500\kms. 
}
\label{fig:hahb}
\end{figure*}

In this appendix, we present a figure showing the evolution 
of the \ha\ and \hb\ line profiles during the first hours and 
days of the RS~Oph explosion in 2021 (see Figure~\ref{fig:hahb}). 
%
%
%
%
%
\section{Tables of spectroscopic observations}
\label{app:spectra}
In this appendix, we present tables of spectroscopic 
observations used in this study (see Tables~\ref{tab:low}, 
\ref{tab:uves}, and \ref{tab:medium}). 
%
%
\begin{table}
\caption{Log of low-resolution spectroscopic 
           observations$^{\dag}$}
\begin{center}
\begin{tabular}{rcccc}
\hline
\hline
\noalign{\smallskip}
Age$^{a}$~~~~&   JD         & Date$^{b}$ & Range & Obs.$^{c}$ \\
JD-JD$_{\rm max}$& 2,459...    & mm-dd.ddd  &  [nm] &            \\
\noalign{\smallskip}
\hline
\noalign{\smallskip}
 -0.366~ & 435.594 & 08-09.094 & 390 - 724 & (i)   \\ 
  0.130~ & 436.090 & 08-09.590 & 370 - 860 & (ii)  \\
  0.348~ & 436.308 & 08-09.808 & 390 - 750 & (iii) \\ 
  0.478~ & 436.438 & 08-09.938 & 371 - 739 & (iv)  \\
  1.006~ & 436.966 & 08-10.466 & 366 - 880 & (ii)  \\
  1.361~ & 437.321 & 08-10.821 & 390 - 750 & (iii) \\ 
  1.431~ & 437.392 & 08-10.891 & 364 - 500 & (v)   \\ 
  1.435~ & 437.395 & 08-10.895 & 375 - 785 & (vi)  \\ 
  2.375~ & 438.335 & 08-11.835 & 330 - 570 & (vii) \\ 
  2.406~ & 438.366 & 08-11.866 & 570 - 855 & (vii) \\ 
  3.388~ & 439.348 & 08-12.848 & 330 - 570 & (vii) \\ 
  3.401~ & 439.361 & 08-12.861 & 570 - 855 & (vii) \\ 
  5.404~ & 441.364 & 08-14.864 & 330 - 570 & (vii) \\ 
  5.428~ & 441.388 & 08-14.888 & 570 - 855 & (vii) \\ 
  7.343~ & 443.303 & 08-16.803 & 385 - 750 & (iii) \\ 
 11.048~ & 447.008 & 08-20.508 & 360 - 875 & (ii)  \\
 12.414~ & 448.374 & 08-21.874 & 330 - 570 & (vii) \\ 
 12.427~ & 448.387 & 08-21.887 & 570 - 855 & (vii) \\ 
 15.998~ & 451.958 & 08-25.458 & 355 - 874 & (ii)  \\
 19.989~ & 455.949 & 08-29.449 & 357 - 880 & (ii)  \\
 25.343~ & 461.303 & 09-03.803 & 330 - 570 & (vii) \\ 
 25.368~ & 461.328 & 09-03.828 & 570 - 855 & (vii) \\ 
 31.333~ & 467.293 & 09-09.793 & 330 - 570 & (vii) \\ 
 31.365~ & 467.325 & 09-09.825 & 570 - 855 & (vii) \\ 
 33.361~ & 469.321 & 09-11.821 & 370 - 750 & (viii)\\ 
\noalign{\smallskip}
\hline
\end{tabular}
\end{center}
{\bf Notes.}\\
  $^{\dag}$ Resolving power from 500 to 1\,300 (Table~\ref{tab:spec}), 
  $^{(a)}$ as in Table~\ref{tab:par}, 
  $^{(b)}$ month and day of 2021 -- the mid of exposure in UT, 
  $^{(c)}$ label of the observatory (Table~\ref{tab:spec}). 
\label{tab:low}
\end{table}
%
%
%
\begin{table}
\footnotesize
\caption{Log of high-resolution UVES spectra$^{\dag}$}
\begin{center}
\begin{tabular}{rcccr}
\hline
\hline
\noalign{\smallskip}
Age$^{a}$~~~~&   JD      & Date$^{b}$ & Grism$^{c}$ & $T_{\rm exp}$ \\
JD-JD$_{\rm max}$& 2,459... & mm-dd.ddd  &             &     (s)       \\
\noalign{\smallskip}
\hline
\noalign{\smallskip}
  0.614~ & 436.574 & 08-10.074 &  CD1 &  60  \\
  0.614~ & 436.574 & 08-10.074 &  CD3 &  40  \\
  0.619~ & 436.579 & 08-10.079 &  CD2 &  15  \\
  0.619~ & 436.579 & 08-10.079 &  CD4 &  15  \\
\noalign{\smallskip}
  1.561~ & 437.521 & 08-11.021 &  CD1 &  30  \\
  1.561~ & 437.521 & 08-11.021 &  CD3 &  10  \\
  1.565~ & 437.525 & 08-11.025 &  CD2 &  10  \\
  1.565~ & 437.525 & 08-11.025 &  CD4 &  ~4  \\
  1.574~ & 437.534 & 08-11.034 &  CD4 &  ~2  \\
\noalign{\smallskip}
  3.755~ & 439.715 & 08-13.215 &  CD3 & 180  \\
  3.755~ & 439.715 & 08-13.215 &  CD1 & 360  \\
  3.762~ & 439.722 & 08-13.222 &  CD4 &  60  \\
  3.762~ & 439.722 & 08-13.222 &  CD2 &  60  \\
\noalign{\smallskip}
  4.748~ & 440.708 & 08-14.208 &  CD3 & 180  \\
  4.748~ & 440.708 & 08-14.208 &  CD1 & 360  \\
  4.755~ & 440.715 & 08-14.215 &  CD4 &  60  \\
  4.755~ & 440.715 & 08-14.215 &  CD2 &  60  \\
\noalign{\smallskip}
  5.755~ & 441.715 & 08-15.215 &  CD3 & 360  \\
  5.755~ & 441.715 & 08-15.215 &  CD1 & 480  \\
  5.763~ & 441.723 & 08-15.223 &  CD4 & 120  \\
  5.763~ & 441.723 & 08-15.223 &  CD2 & 120  \\
\noalign{\smallskip}
  9.506~ & 445.466 & 08-18.966 &  CD3 & 360  \\
  9.506~ & 445.466 & 08-18.966 &  CD1 & 480  \\
  9.513~ & 445.473 & 08-18.973 &  CD4 & 120  \\
  9.513~ & 445.473 & 08-18.973 &  CD2 & 120  \\
\noalign{\smallskip}
 11.551~ & 447.511 & 08-21.011 &  CD3 & 360  \\
 11.551~ & 447.511 & 08-21.011 &  CD1 & 480  \\
 11.567~ & 447.527 & 08-21.027 &  CD4 & 120  \\
 11.567~ & 447.527 & 08-21.027 &  CD2 & 120  \\
\noalign{\smallskip}
 21.579~ & 457.539 & 08-31.039 &  CD1 & 480  \\
 21.587~ & 447.547 & 08-31.047 &  CD2 & 120  \\
 21.578~ & 447.538 & 08-31.038 &  CD3 & 360  \\
 21.587~ & 457.547 & 08-31.047 &  CD4 & 120  \\
\noalign{\smallskip}
 26.554~ & 462.514 & 09-05.014 &  CD1 & 480  \\
 26.564~ & 462.524 & 09-05.024 &  CD2 & 120  \\
 26.554~ & 462.514 & 09-05.014 &  CD3 & 360  \\
 26.563~ & 462.523 & 09-05.023 &  CD4 & 120  \\
\noalign{\smallskip}
 32.530~ & 468.490 & 09-10.990 &  CD1 & 600  \\
 32.556~ & 468.516 & 09-11.016 &  CD2 & 180  \\
 32.543~ & 468.503 & 09-11.003 &  CD3 & 240  \\
 32.559~ & 468.519 & 09-11.019 &  CD4 &  15  \\
 32.556~ & 468.516 & 09-11.016 &  CD4 & 180  \\
\noalign{\smallskip}
 37.535~ & 473.495 & 09-15.995 &  CD1 & 600  \\
 37.529~ & 473.489 & 09-15.989 &  CD2 & 180  \\
 37.535~ & 473.495 & 09-15.995 &  CD3 & 540  \\
 37.532~ & 473.492 & 09-15.992 &  CD4 &   5  \\
 37.529~ & 473.489 & 09-15.989 &  CD4 & 180  \\
\noalign{\smallskip}
 42.558~ & 478.518 & 09-21.018 &  CD1 & 600  \\
 42.570~ & 478.530 & 09-21.030 &  CD2 & 180  \\
 42.558~ & 478.518 & 09-21.018 &  CD3 & 540  \\
 42.573~ & 478.533 & 09-21.033 &  CD4 &   5  \\
 42.570~ & 478.530 & 09-21.030 &  CD4 & 180  \\
\hline
\end{tabular}
\end{center}
{\bf Notes.} 
  $^{\dag}$ Resolving power from 58,640 to 107,200,
  $^{(a)}$ as in Table~\ref{tab:spec}, 
  $^{(b)}$ month and day of 2021 -- the start of exposure in UT, 
  $^{(c)}$ corresponding spectral ranges are 304--392, 373--500,
           473--684 and 566--946\,nm for grism CD1, CD2, CD3 and 
           CD4. 
%
\label{tab:uves}
\normalsize
\end{table}
%
%
%
\begin{table}
\footnotesize
\caption{Log of medium-resolution spectroscopic
           observations$^{\dag}$}
\begin{center}
\begin{tabular}{rcccc}
\hline
\hline
\noalign{\smallskip}
Age$^{a}$~~~~&   JD         & Date$^{b}$ & Range & Obs.$^{c}$ \\
JD-JD$_{\rm max}$& 2,459...    & mm-dd.ddd  &  [nm] &       \\
\noalign{\smallskip}
\hline
\noalign{\smallskip}
 0.364 & 436.324 & 08-09.824 & 420-736 & (ix)   \\ 
 0.378 & 436.338 & 08-09.838 & 420-736 & (ix)   \\ 
 0.389 & 436.349 & 08-09.849 & 641-688 & (Ondrejov)   \\ 
 0.392 & 436.352 & 08-09.852 & 420-736 & (ix)   \\ 
 0.406 & 436.366 & 08-09.866 & 420-736 & (ix)   \\ 
 0.412 & 436.372 & 08-09.872 & 374-975 & (x)  \\ 
 0.420 & 436.380 & 08-09.880 & 420-746 & (ix)   \\ 
 0.422 & 436.382 & 08-09.882 & 392-760 & (xi) \\ 
 0.422 & 436.382 & 08-09.882 & 641-688 & (Ondrejov) \\ 
 0.434 & 436.394 & 08-09.894 & 420-736 & (ix)   \\ 
 0.444 & 436.404 & 08-09.904 & 400-761 & (xii)  \\ 
 0.445 & 436.405 & 08-09.905 & 392-760 & (xi) \\ 
 0.472 & 436.432 & 08-09.932 & 400-761 & (xii)  \\ 
 0.481 & 436.441 & 08-09.941 & 394-894 & (xiii)   \\ 
 0.494 & 436.454 & 08-09.954 & 400-761 & (xii)  \\ 
 0.500 & 436.460 & 08-09.960 & 392-760 & (xi) \\ 
 0.513 & 436.473 & 08-09.973 & 394-894 & (xiii)   \\ 
 0.546 & 436.506 & 08-10.006 & 394-894 & (xiii)   \\ 
 1.364 & 437.324 & 08-10.824 & 420-736 & (ix)   \\ 
 1.378 & 437.338 & 08-10.838 & 420-746 & (ix)   \\ 
 1.390 & 437.350 & 08-10.850 & 392-752 & (xii)  \\ 
 1.392 & 437.352 & 08-10.852 & 420-736 & (ix)   \\ 
 1.406 & 437.366 & 08-10.866 & 420-736 & (ix)   \\ 
 1.420 & 437.380 & 08-10.880 & 420-746 & (ix)   \\ 
 1.434 & 437.394 & 08-10.894 & 420-736 & (ix)   \\ 
 1.460 & 437.420 & 08-10.920 & 392-752 & (xii)  \\ 
 2.059 & 438.019 & 08-11.519 & 650-674 & (xiv)  \\ 
 2.360 & 438.320 & 08-11.820 & 420-746 & (ix)   \\ 
 2.375 & 438.335 & 08-11.835 & 420-746 & (ix)   \\ 
 2.389 & 438.349 & 08-11.849 & 420-746 & (ix)   \\ 
 2.403 & 438.363 & 08-11.863 & 420-746 & (ix)   \\ 
 2.409 & 438.369 & 08-11.869 & 430-752 & (xii)  \\ 
 2.418 & 438.378 & 08-11.878 & 420-746 & (ix)   \\ 
 2.432 & 438.392 & 08-11.892 & 420-736 & (ix)   \\ 
 2.485 & 438.445 & 08-11.945 & 648-664 & (xv) \\ 
 3.386 & 439.346 & 08-12.846 & 641-688 & (Ondrejov) \\ 
 3.414 & 439.374 & 08-12.874 & 387-894 & (xiii)   \\ 
 3.449 & 439.409 & 08-12.909 & 647-698 & (xvi)\\ 
 3.455 & 439.415 & 08-12.915 & 648-664 & (xv) \\ 
 3.488 & 439.448 & 08-12.948 & 392-760 & (xi) \\ 
 4.363 & 440.323 & 08-13.823 & 420-736 & (ix)   \\ 
 4.369 & 440.329 & 08-13.829 & 626-674 & (Ondrejov) \\ 
 4.392 & 440.352 & 08-13.852 & 420-736 & (ix)   \\ 
 4.406 & 440.366 & 08-13.866 & 420-736 & (ix)   \\ 
 4.420 & 440.380 & 08-13.880 & 420-746 & (ix)   \\ 
 4.424 & 440.384 & 08-13.884 & 398-758 & (xi) \\ 
 4.434 & 440.394 & 08-13.894 & 420-746 & (ix)   \\ 
 4.439 & 440.399 & 08-13.899 & 400-890 & (xiii)   \\ 
 4.493 & 440.453 & 08-13.912 & 640-667 & (vii) \\ 
 5.018 & 440.978 & 08-14.478 & 400-741 & (xvii)  \\ 
 5.343 & 441.303 & 08-14.803 & 626-674 & (Ondrejov)   \\ 
 5.378 & 441.338 & 08-14.838 & 400-761 & (xii)  \\ 
 5.409 & 441.369 & 08-14.869 & 619-688 & (xviii)   \\ 
 5.423 & 441.383 & 08-14.883 & 420-736 & (ix)   \\ 
 5.425 & 441.385 & 08-14.885 & 400-761 & (xii)  \\ 
 5.452 & 441.412 & 08-14.912 & 392-759 & (xi) \\ 
 6.444 & 442.404 & 08-15.904 & 378-948 & (x)  \\ 
 7.368 & 443.328 & 08-16.828 & 648-664 & (xvi)\\ 
\hline 
\end{tabular}
\end{center}
\end{table}
\addtocounter{table}{-1}
\begin{table}[t!]
\footnotesize
\caption{continued}
\begin{center}
\begin{tabular}{rcccc}
\hline
\hline
\noalign{\smallskip}
Age$^{a}$~~~~&   JD         & Date$^{b}$ & Range & Obs.$^{c}$ \\
JD-JD$_{\rm max}$   & 2\,459...    & mm-dd.ddd  &  [nm] &    \\
\noalign{\smallskip}
\hline
\noalign{\smallskip}
 7.371 & 443.331 & 08-16.831 & 430-735 & (ix)   \\ 
 7.397 & 443.357 & 08-16.857 & 394-894 & (xiii)   \\ 
 8.388 & 444.348 & 08-17.848 & 395-890 & (xiii)   \\ 
 8.402 & 444.362 & 08-17.862 & 643-677 & (xvi)\\ 
 9.384 & 445.344 & 08-18.844 & 387-894 & (xiii)   \\ 
 9.724 & 445.684 & 08-19.184 & 648-664 & (xix)  \\ 
10.400 & 446.360 & 08-19.860 & 420-890 & (xiii)   \\ 
10.473 & 446.433 & 08-19.933 & 648-664 & (xv) \\ 
11.395 & 447.355 & 08-20.855 & 473-735 & (ix)   \\ 
11.431 & 447.391 & 08-20.891 & 387-894 & (xiii)   \\ 
11.440 & 447.400 & 08-20.900 & 648-669 & (xv) \\ 
11.711 & 447.671 & 08-21.171 & 648-664 & (xix)  \\ 
12.355 & 448.315 & 08-21.815 & 641-688 & (Ondrejov)   \\ 
12.366 & 448.326 & 08-21.826 & 641-688 & (Ondrejov)   \\ 
12.456 & 448.416 & 08-21.916 & 387-894 & (xiii)   \\ 
13.147 & 449.107 & 08-22.607 & 400-740 & (xvii)  \\ 
13.467 & 449.427 & 08-22.927 & 387-894 & (xiii)   \\ 
14.402 & 450.362 & 08-23.862 & 420-890 & (xiii)   \\ 
15.419 & 451.379 & 08-24.879 & 392-759 & (xi) \\ 
15.456 & 451.416 & 08-24.916 & 648-663 & (xv) \\ 
16.374 & 452.334 & 08-25.834 & 475-735 & (ix)   \\ 
16.439 & 452.399 & 08-25.899 & 392-759 & (xi) \\ 
16.448 & 452.408 & 08-25.908 & 648-663 & (xv) \\ 
17.384 & 453.344 & 08-26.844 & 387-894 & (xiii)   \\ 
18.431 & 454.391 & 08-27.891 & 648-663 & (xv) \\ 
18.452 & 454.412 & 08-27.912 & 392-759 & (xi) \\ 
19.356 & 455.316 & 08-28.816 & 475-735 & (ix)   \\ 
19.376 & 455.336 & 08-28.836 & 381-894 & (xiii)   \\ 
20.373 & 456.333 & 08-29.833 & 381-894 & (xiii)   \\ 
22.085 & 458.045 & 08-31.545 & 650-674 & (xiv)  \\ 
22.420 & 458.380 & 08-31.880 & 648-663 & (xv) \\ 
23.347 & 459.307 & 09-01.807 & 475-735 & (ix)   \\ 
23.381 & 459.341 & 09-01.841 & 392-759 & (xi) \\ 
24.317 & 460.277 & 09-02.777 & 641-688 & (Ondrejov) \\ 
24.386 & 460.346 & 09-02.846 & 392-759 & (xi) \\ 
24.422 & 460.382 & 09-02.882 & 648-663 & (xv) \\ 
25.417 & 461.377 & 09-03.877 & 381-894 & (xiii)   \\ 
26.378 & 462.338 & 09-04.838 & 381-894 & (xiii)   \\ 
27.375 & 463.335 & 09-05.835 & 392-892 & (xiii)   \\ 
27.407 & 463.367 & 09-05.867 & 392-759 & (xi) \\ 
28.394 & 464.354 & 09-06.854 & 392-759 & (xi) \\ 
29.345 & 465.305 & 09-07.805 & 644-666 & (xx) \\ 
30.391 & 466.351 & 09-08.851 & 387-893 & (xiii)   \\ 
31.336 & 467.296 & 09-09.796 & 641-688 & (Ondrejov) \\ 
31.397 & 467.357 & 09-09.857 & 644-666 & (xx) \\ 
32.389 & 468.349 & 09-10.849 & 381-886 & (xiii)   \\ 
34.379 & 470.339 & 09-12.839 & 402-894 & (xiii)   \\ 
35.331 & 471.291 & 09-13.791 & 644-666 & (xx)  \\ 
35.995 & 471.955 & 09-14.455 & 400-741 & (xvii) \\ 
38.328 & 474.288 & 09-16.788 & 644-666 & (xx)  \\ 
38.335 & 474.295 & 09-16.795 & 648-664 & (xxi) \\ 
41.071 & 477.031 & 09-19.531 & 400-741 & (xvii)  \\ 
41.390 & 477.350 & 09-19.850 & 410-890 & (xiii)   \\ 
44.040 & 480.000 & 09-22.500 & 400-741 & (xvii)  \\ 
47.311 & 483.271 & 09-25.771 & 641-688 & (Ondrejov)   \\ 
\hline
\end{tabular}
\end{center}
{\bf Notes.}
  $^{\dag}$ Resolving power from 8,500 to 20,000 
            (Table~\ref{tab:spec}), 
  $^{(a)}$ as in Table~\ref{tab:par}, 
  $^{(b)}$ as in Table~\ref{tab:low}, 
  $^{(c)}$ label of the observatory (Table~\ref{tab:spec}, 
           Sect.~\ref{ss:spec}). 
\label{tab:medium}
\end{table}
%
%
%
\section{Parameters of the \ha\ line}
\label{app:linepar}
In this appendix, we present a table of \ha\ line 
parameters (see Table~\ref{tab:hapar}).
%
%
\begin{table*}[h!]
\footnotesize
\caption{Parameters of the \ha\ line profile} 
\begin{center}
\begin{tabular}{cccccccrcccc}
\hline
\hline
\noalign{\smallskip}
Age$^{a}$                   & 
Date$^{b}$                  &
$F_{\alpha}^{c}$            & 
$F_{\rm cont.}^{d}$         & 
FWHM$^{e}$                  &
$v_{\infty}^{f}$            &
$RV_{\rm abs}^{g}$          &
$RV_{\rm em}^{g}$           &
$F_{\rm abs}^{h}$          &
$F_{\rm em}^{h}$           &
$\dot M_{\rm WD}^{i}$       &
Obs.$^{j}$                  \\
\noalign{\smallskip}
\hline
\noalign{\smallskip}
 0.364 & 0809.824 & 15.74 & 2.598 & 55.40 & -4,448 & -74.95 & -11.91 &0.94 &12.27 & 1.57 & (ix)  \\
 0.378 & 0809.838 & 16.66 & 2.570 & 57.20 & -4,424 & -75.44 & -11.94 &0.88 &12.89 & 1.62 & (ix)  \\
 0.389 & 0809.849 & 15.79 & 2.548 & 55.64 & -4,738 & -73.96 & -11.37 &0.71 &11.95 & 1.69 &(Ondrejov)\\
 0.392 & 0809.852 & 15.98 & 2.542 & 57.35 & -4,351 & -75.01 & -12.43 &0.88 &12.25 & 1.56 & (ix)  \\
 0.406 & 0809.866 & 15.47 & 2.514 & 57.20 & -4,430 & -76.42 & -13.38 &0.91 &12.52 & 1.57 & (ix)  \\
 0.412 & 0809.872 & 16.16 & 2.502 & 57.05 & -4,595 & -75.97 & -13.39 &0.84 &12.41 & 1.66 & (x) \\
 0.420 & 0809.880 & 15.52 & 2.486 & 57.70 & -4,358 & -75.08 & -12.96 &0.82 &12.17 & 1.55 & (ix)  \\
 0.422 & 0809.882 & 15.49 & 2.482 & 58.60 & -4,221 & -80.11 & -13.87 &0.66 &12.24 & 1.50 & (xi)\\
 0.422 & 0809.882 & 15.68 & 2.482 & 59.10 & -4,736 & -74.41 & -12.29 &0.90 &12.39 & 1.55 &(Ondrejov)\\
 0.434 & 0809.894 & 15.63 & 2.458 & 57.55 & -4,219 & -76.02 & -13.44 &0.93 &12.39 & 1.51 & (ix)  \\
 0.444 & 0809.904 & 16.92 & 2.438 & 60.00 & -4,370 & -74.67 & -11.18 &1.10 &12.06 & 1.63 & (xii) \\
 0.445 & 0809.905 & 15.27 & 2.436 & 59.10 & -4,398 & -78.34 & -13.47 &0.93 &11.58 & 1.55 & (xi)\\   
 0.472 & 0809.932 & 17.48 & 2.382 & 58.55 & -4,395 & -74.29 & -12.16 &1.18 &11.37 & 1.67 & (xii) \\ 
 0.481 & 0809.941 & 18.50 & 2.367 & 59.70 & -4,444 & -73.50 &  -9.55 &1.12 &12.09 & 1.74 & (xiii)  \\ 
 0.494 & 0809.954 & 18.45 & 2.352 & 59.45 & -4,383 & -75.24 & -12.66 &1.23 &11.70 & 1.74 & (xii) \\ 
 0.500 & 0809.960 & 17.44 & 2.345 & 59.95 & -4,397 & -76.62 & -12.21 &0.83 &11.57 & 1.70 & (xi)\\ 
 0.513 & 0809.973 & 19.74 & 2.330 & 59.78 & -4,434 & -72.59 &  -9.10 &1.14 &12.67 & 1.85 & (xiii)  \\   
 0.546 & 0810.006 & 20.94 & 2.293 & 59.65 & -4,484 & -72.94 &  -9.87 &1.05 &11.23 & 1.98 & (xiii)  \\  
 0.617 & 0810.076 & 18.83 & 1.919 & 59.63 & -4,497 & -74.24 & -12.57 &2.02 &10.97 & 1.99 &(Paranal)\\
 1.364 & 0810.824 & 34.45 & 1.470 & 58.28 & -4,592 & -72.52 & -11.08 &1.89 & 8.37 & 2.52 & (ix)  \\
 1.378 & 0810.838 & 34.77 & 1.470 & 58.10 & -4,619 & -72.09 & -11.79 &1.94 & 8.00 & 2.53 & (ix)  \\
 1.390 & 0810.850 & 34.10 & 1.470 & 56.10 & -4,657 & -72.12 & -11.37 &2.86 & 8.35& 2.52 & (xii) \\
 1.392 & 0810.852 & 35.10 & 1.470 & 58.40 & -4,630 & -73.04 & -12.28 &1.99 & 8.01& 2.55 & (ix)  \\
 1.406 & 0810.866 & 34.72 & 1.470 & 58.25 & -4,665 & -73.31 & -11.41 &2.04 & 7.82& 2.54 & (ix)  \\
 1.420 & 0810.880 & 35.18 & 1.470 & 58.30 & -4,618 & -73.56 & -12.35 &2.00 & 8.18& 2.53 & (ix)  \\
 1.434 & 0810.894 & 36.25 & 1.469 & 58.20 & -4,589 & -74.51 & -12.84 &2.12 & 8.11& 2.55 & (ix)  \\  
 1.460 & 0810.920 & 34.31 & 1.459 & 56.25 & -4,610 & -72.28 & -12.90 &2.72 & 7.66& 2.48 & (xii) \\
 1.567 & 0811.027 & 34.15 & 1.200 & 58.38 & -4,621 & -73.95 & -12.74 &2.86 & 8.57& 2.45 &(Paranal)\\
 2.059 & 0811.519 &  --   &   --  &  --   &   --  & -71.60 &  -9.01 &3.34 & 6.20& --   & (xiv)  \\
 2.360 & 0811.820 & 51.34 & 1.118 & 53.70 & -4,588 & -71.90 & -10.68 &3.82 & 6.21& 2.92 & (ix)   \\
 2.375 & 0811.835 & 50.95 & 1.112 & 53.70 & -4,681 & -71.02 & -11.18 &3.42 & 6.62& 2.96 & (ix)   \\
 2.389 & 0811.849 & 53.87 & 1.107 & 53.85 & -4,728 & -71.97 & -10.30 &3.47 & 6.77& 3.08 & (ix)   \\
 2.403 & 0811.863 & 51.45 & 1.103 & 54.05 & -4,747 & -72.00 & -11.24 &3.44 & 6.16& 3.02 & (ix)   \\
 2.409 & 0811.869 & 54.02 & 1.101 & 55.35 & -4,724 & -74.30 & -12.17 &4.48 & 6.17& 3.08 & (xii)  \\
 2.418 & 0811.878 & 51.37 & 1.099 & 54.05 & -4,720 & -71.12 & -10.82 &3.38 & 6.47& 3.00 & (ix)   \\
 2.432 & 0811.892 & 51.84 & 1.095 & 53.65 & -4,704 & -70.70 &  -9.94 &3.32 & 6.44& 3.00 & (ix)   \\
 2.485 & 0811.945 &   --  &  --   &   --  &   --  & -73.10 & -11.89 &4.68 & 6.15& --   & (xv) \\
 3.386 & 0812.846 & 56.51 & 0.815 & 50.33 & -4,781 & -70.76 &  -9.09 &2.90 & 3.94& 3.12 & (Ondrejov)\\
 3.414 & 0812.874 & 58.07 & 0.809 & 49.10 & -4,547 & -73.70 & -10.20 &2.97 & 4.65& 2.99 & (xiii)   \\
 3.449 & 0812.909 &   --  &  --   &   --  &   --  & -73.78 & -10.74 &4.55 & 4.71& --   & (xvi)\\
 3.455 & 0812.915 &   --  &  --   &   --  &   --  & -73.33 &  -9.83 &4.50 & 3.69& --   & (xv) \\
 3.488 & 0812.948 & 56.69 & 0.796 & 48.65 & -4,555 & -71.12 &  -9.45 &2.91 & 3.62& 2.90 & (xi) \\
 3.759 & 0813.219 & 45.22 & 0.587 & 46.80 & -4,599 & -74.56 & -10.15 &3.69 & --  & 2.41 &(Paranal) \\
 4.363 & 0813.823 & 58.43 & 0.636 & 43.40 & -4,416 & -70.67 &  -9.46 &2.16 & 3.01& 2.61 & (ix)   \\
 4.369 & 0813.829 & 57.90 & 0.635 & 45.03 & -4,795 & -69.85 &  -8.18 &2.01 & 3.07& 2.58 & (Ondrejov)\\
 4.392 & 0813.852 & 55.24 & 0.631 & 44.65 & -4,331 & -71.65 &  -8.61 &2.09 & 3.06& 2.49 & (ix)   \\
 4.406 & 0813.866 & 59.04 & 0.628 & 45.05 & -4,352 & -69.86 &  -9.10 &1.79 & 2.66& 2.58 & (ix)   \\
 4.420 & 0813.880 & 59.43 & 0.626 & 45.05 & -4,371 & -70.34 &  -8.68 &1.97 & 3.04& 2.60 & (ix)   \\
 4.424 & 0813.884 & 59.27 & 0.625 & 44.78 & -4,382 & -69.90 &  -6.19 &2.30 & 2.41& 2.60 & (xi) \\
 4.434 & 0813.894 & 57.09 & 0.623 & 44.75 & -4,342 & -69.47 &  -9.63 &2.08 & 2.86& 2.53 & (ix)   \\
 4.439 & 0813.899 & 57.91 & 0.622 & 43.45 & -4,400 &   --   &    --  &1.87 & 2.95& 2.58 & (xiii)   \\
 4.493 & 0813.953 & 59.71 & 0.612 & 44.05 & -4,450 & -70.57 &  -8.90 &3.19 & 3.32& 2.65 & (vii) \\
 4.751 & 0814.211 & 44.77 & 0.461 & 42.78 & -4,598 & -72.55 & -10.88 &2.36 & 3.41& 2.36 &(Paranal) \\
 5.018 & 0814.478 & 57.67 & 0.516 & 42.54 & -4,355 & -70.67 & -10.37 &2.03 & 2.72& 2.38 & (xvii)  \\
\hline
\end{tabular}
\end{center}
\end{table*}
\addtocounter{table}{-1}
\begin{table*}[t!]
\footnotesize
\caption{continued}
\begin{center}
\begin{tabular}{cccccccrcccc}
\hline
\hline
\noalign{\smallskip}
Age$^{a}$                   & 
Date$^{b}$                  &
$F_{\alpha}^{c}$            & 
$F_{\rm cont.}^{d}$         & 
FWHM$^{e}$                  &
$v_{\infty}^{f}$            &
$RV_{\rm abs}^{g}$          &
$RV_{\rm em}^{g}$           &
$F_{\rm abs}^{h}$          &
$F_{\rm em}^{h}$           &

$\dot M_{\rm WD}^{i}$       &
Obs.$^{j}$                  \\
\noalign{\smallskip}
\hline
\noalign{\smallskip}
 5.343 & 0814.803 & 52.50 & 0.457 & 40.88  & -4,300 & -67.56 &  -8.63 &0.73 & 2.80& 2.09 & (Ondrejov)\\
 5.378 & 0814.838 & 53.57 & 0.451 & 41.00  & -4,348 & -67.79 &  -8.41 &0.98 & 2.97& 2.06 & (xii)  \\
 5.409 & 0814.869 & 55.15 & 0.445 & 40.65  & -4,262 &    --  &    --  &0.24 & 1.44& 2.03 & (xviii) \\
 5.423 & 0814.883 & 52.62 & 0.444 & 41.45  & -4,150 & -68.36 & -10.80 &1.21 & 2.42& 1.92 & (ix)   \\
 5.425 & 0814.885 & 55.60 & 0.444 & 41.05  & -4,343 & -69.27 &  -8.97 &0.92 & 2.57& 2.06 & (xii)  \\
 5.452 & 0814.912 & 51.07 & 0.441  & 40.55 & -4,349 & -70.70 &  -8.58 &0.82 & 1.83& 1.97 & (xi) \\
 5.759 & 0815.219 & 40.15 & 0.346  & 40.37 & -4,577 & -71.02 & -11.63 &1.34 & 2.49& 1.74 &(Paranal) \\  
 6.444 & 0815.904 & 49.39 & 0.351  & 38.45 & -4,260 & -72.33 & -12.49 &0.41 & 2.17& 1.83 & (x) \\
 7.368 & 0816.828 &   --  &   --   &  --   &   --  & -70.61 & -10.76 &0.16 & 1.35& --   & (xvi)\\
 7.371 & 0816.831 & 39.29 & 0.269  & 36.15 & -4,172 & -77.00 & -11.91 &0.28 & 1.40& 1.64 & (ix)   \\
 7.397 & 0816.857 & 40.26 & 0.269  & 37.35 & -4,165 & -71.22 &  -4.52 &0.38 & 1.21& 1.66 & (xiii) \\
 8.388 & 0817.848 & 40.88 & 0.246  & 33.65 & -4,332 & -71.67 &  -5.44 &0.31 & 1.07& 1.70 & (xiii) \\
 8.402 & 0817.862 &   --  &   --   &   --  &   --  & -71.40 & -11.56 &0.15 & 1.41& --   & (xvi)\\
 9.384 & 0818.844 & 40.28 & 0.223  & 32.35 & -4,175 &   --   &    --  &0.17 & 0.81& 1.59 & (xiii)   \\
 9.510 & 0818.970 & 35.29 & 0.199  & 31.86 & -4,563 & -71.71 &  -9.82 &0.60 & 0.67& 1.63 &(Paranal) \\
 9.724 & 0819.184 &   --  &   --   &   --  &   --  & -71.50 & -10.73 &0.10 & 0.77& --   & (xix)  \\
10.400 & 0819.860 & 36.69 & 0.200  & 30.15 & -4,063 & -72.59 &  -5.44 &0.12 & 0.75& 1.42 &(xiii)   \\
10.473 & 0819.933 &   --  &   --   &   --  &   --  & -73.43 & -10.85 &0.10 & 0.80& --   & (xv) \\
11.395 & 0820.855 & 30.94 & 0.182  & 29.20 & -4,064 & -73.06 &  -9.56 &0.06 & 0.46& 1.19 & (ix) \\
11.431 & 0820.891 & 35.38 & 0.182  & 28.90 & -4,100 & -70.30 &    --  &0.03 & 0.53& 1.28 & (xiii) \\
11.440 & 0820.900 &   --  &   --   &   --  &   --  & -72.70 & -10.57 & --  & --  & --   & (xv) \\
11.559 & 0821.019 & 30.33 & 0.166  & 28.44 & -4,484 & -72.11 &  -9.07 &0.36 & 0.41& 1.26 &(Paranal) \\
11.711 & 0821.171 &   --  &   --   &   --  &   --  & -69.69 &  -6.65 &0.11 & 0.54& --   & (xix)  \\
12.355 & 0821.815 & 34.67 & 0.175  & 26.32 & -5,425 & -73.04 &  -7.26 &0.03 & 0.35& 1.50 &(Ondrejov)\\
12.366 & 0821.826 & 33.87 & 0.174  & 26.32 & -5,316 & -75.78 &  -7.72 &0.03 & 0.36& 1.49 &(Ondrejov)\\
12.456 & 0821.916 & 36.53 & 0.174  & 26.25 & -4,101 & -71.22 &  -3.61 &0.04 & 0.51& 1.48 & (xiii)   \\
13.147 & 0822.607 & 35.06 & 0.165  & 25.98 & -4,275 & -77.20 &  -5.48 &0.16 & 0.41& 1.48 & (xvii)  \\
13.467 & 0822.927 & 34.12 & 0.161  & 25.55 & -4,064 & -70.48 & -10.64 &0.00 & 0.46& 1.36 & (xiii)   \\
14.402 & 0823.862 & 34.20 & 0.149  & 24.00 & -4,075 & -70.30 &    --  &0.01 & 0.55& 1.32 & (xiii)   \\
15.419 & 0824.879 & 26.97 & 0.136  & 23.20 & -3,976 &   --   &  -7.78 &0.01 & 0.31& 1.09 & (xi) \\
15.456 & 0824.916 &   --  &  --    &  --   &   --  & -73.64 & -10.14 &0.06 & 0.24& --   & (xv) \\
16.374 & 0825.834 & 28.17 & 0.125  & 22.45 & -4,063 & -79.61 &  -9.27 &0.03 & 0.41& 1.08 & (ix)   \\
16.439 & 0825.899 & 28.13 & 0.124  & 22.40 & -3,933 &   --   &  -3.92 &0.00 & 0.36& 1.05 & (xi) \\
16.448 & 0825.908 &  --   &  --    &  --   &   --  & -72.92 & -10.10 &0.04 & 0.42&  --  & (xv) \\
17.384 & 0826.844 & 30.34 & 0.113  & 21.75 & -3,951 &   --   &   --   & --  & 0.32& 1.05 & (xiii)   \\
18.431 & 0827.891 &   --  &   --   &   --  &   --  & -72.83 & -10.25 &0.04 & 0.38&   -- & (xv) \\
18.452 & 0827.912 & 22.35 & 0.100  & 21.25 & -3,937 &   --   &  -8.46 &0.00 & 0.23& 0.855& (xi) \\
19.356 & 0828.816 & 21.39 & 0.0895 & 20.90 & -3,888 &   --   &  -7.54 &0.00 & 0.23& 0.789& (ix)   \\
19.376 & 0828.836 & 24.01 & 0.0892 & 20.80 & -4,138 &   --   &  -7.12 &0.00 & 0.27& 0.890& (xiii)   \\
20.373 & 0829.833 & 26.94 & 0.0802 & 20.05 & -4,315 &   --   & -10.50 &0.01 & 0.35& 0.932& (xiii)   \\
21.583 & 0831.043 & 23.47 & 0.0775 & 19.38 & -4,345 & -71.06 &  -8.48 &0.12 & 0.29& 0.851&(Paranal)\\
22.085 & 0831.545 &   --  &    --  &    -- &   --  & -71.62 &  -7.21 &0.02 & 0.32&  --  & (xiv)  \\
22.420 & 0831.880 &   --  &    --  &    -- &   --  & -72.16 & -11.40 &0.03 & 0.28&  --  & (xv) \\
23.347 & 0901.807 & 20.00 & 0.0660 & 17.95 & -4,061 &   --   &  -2.27 &0.05 & 0.21& 0.704& (ix)   \\
23.381 & 0901.841 & 20.33 & 0.0658 & 17.05 & -3,933 &   --   &   --   & --  & --  & 0.687& (xi) \\
24.317 & 0802.777 & 15.97 & 0.0614 & 18.24 & -4,752 &   --   &  -6.35 & --  & 0.08& 0.624&(Ondrejov)\\
24.386 & 0902.846 & 16.81 & 0.0611 & 18.10 & -3,898 &   --   &  -4.79 & --  & 0.26& 0.616& (xi) \\
24.422 & 0902.882 &   --  &    --  &   --  &   --  & -73.39 & -12.64 &0.02 & 0.27&  --  & (xv) \\
25.417 & 0903.877 & 20.44 & 0.0563 & 16.45 & -4,074 &   --   &  -9.11 & --  & 0.36& 0.702& (xiii)   \\
26.378 & 0904.838 & 21.13 & 0.0541 & 16.25 & -4,386 &   --   &  -5.52 &0.00 & 0.30& 0.715& (xiii)   \\
26.559 & 0905.019 & 15.57 & 0.0487 & 17.42 & -4,272 & -70.90 &  -8.12 &0.06 & 0.15& 0.599&(Paranal) \\
27.375 & 0905.835 & 18.63 & 0.0518 & 15.90 & -4,301 &   --   &  -4.72 & --  & 0.22& 0.641& (xiii)   \\
27.407 & 0905.867 & 16.10 & 0.0517 & 17.10 & -4,136 &   --   &  -8.91 & --  & 0.20& 0.573& (xi) \\
28.394 & 0906.854 & 14.99 & 0.0494 & 15.50 & -4,068 &   --   &  -4.43 & --  & 0.15& 0.533& (xi) \\
29.345 & 0907.805 & 14.57 & 0.0472 & 15.55 & -3,956 &   --   &   --   & --  & --  & 0.503& (xx) \\
30.391 & 0908.851 & 17.17 & 0.0447 & 14.50 & -4,071 &   --   &  -5.09 & --  & --  & 0.549& (xiii)   \\
31.336 & 0909.796 & 13.67 & 0.0425 & 14.54 & -4,589 &   --   &  -4.98 & --  & 0.02& 0.520&(Ondrejov)\\
\hline 
\end{tabular}
\end{center}
\end{table*}
\addtocounter{table}{-1}
\begin{table*}[t!]
\footnotesize
\caption{continued}
\begin{center}
\begin{tabular}{cccccccrcccc}
\hline
\hline
\noalign{\smallskip}
Age$^{a}$                   & 
Date$^{b}$                  &
$F_{\alpha}^{c}$            & 
$F_{\rm cont.}^{d}$         & 
FWHM$^{e}$                  &
$v_{\infty}^{f}$            &
$RV_{\rm abs}^{g}$          &
$RV_{\rm em}^{g}$           &
$F_{\rm abs}^{h}$          &
$F_{\rm em}^{h}$           &
$\dot M_{\rm WD}^{i}$       &
Obs.$^{j}$                  \\
\noalign{\smallskip}
\hline
\noalign{\smallskip}
31.397 & 0909.857 & 14.59 & 0.0423 & 14.33 & -3,929 &   --   &   --   & --  & --  & 0.481& (xx) \\
32.389 & 0910.849 & 12.96 & 0.0386 & 14.30 & -4,064 &   --   &  -2.08 & --  & --  & 0.446& (xiii)   \\
32.549 & 0911.009 & 11.55 & 0.0341 & 15.18 & -4,299 & -71.02 &  -4.33 &0.08 & 0.12& 0.441&(Paranal) \\
34.379 & 0912.839 & 11.48 & 0.0341 & 13.85 & -4,040 &   --   &   --   & --  & --  & 0.417& (xiii)   \\
35.331 & 0913.791 & 10.07 & 0.0334 & 14.19 & -4,131 &   --   &   --   & --  & --  & 0.382& (xx) \\
35.995 & 0914.455 & 10.34 & 0.0329 & 13.95 & -4,131 &   --   &   --   & --  & --  & 0.372& (xvii)  \\
37.532 & 0915.992 & 07.24 & 0.0232 & 13.58 & -4,306 & -72.20 &   --   &0.02 & --  & 0.292&(Paranal) \\
38.328 & 0916.788 & 09.17 & 0.0311 & 12.57 & -3,766&   --   &   --   & -- & -- & 0.288& (xx)     \\
38.335 & 0916.795 & 09.68 & 0.0311 & 12.89 & -3,857&   --   &   --   & -- & -- & 0.303& (xxi)    \\
41.071 & 0919.531 & 09.79 & 0.0291 & 12.41 & -4,040&   --   &   --   & -- & -- & 0.315& (xvii)   \\
41.390 & 0919.850 & 10.70 & 0.0289 & 12.25 & -4,086&   --   &   --   & -- & -- & 0.328& (xiii)   \\
42.566 & 0921.026 & 08.17 & 0.0197 & 12.18 & -4,040&   --   &   --   & -- & -- & 0.283&(Paranal) \\
44.040 & 0922.500 & 08.23 & 0.0272 & 11.90 & -4,040&   --   &   --   & -- & -- & --   & (xvii)   \\
\hline
\end{tabular}
\end{center}
{\bf Notes.}
$^{(a)}$ As in Table~\ref{tab:par}, 
$^{(b)}$ as in Table~\ref{tab:low}, 
$^{(c)}$ $F_{\alpha}$ -- flux of the broad component 
         in 10$^{-9}$\ecs, 
$^{(d)}$ $F_{\rm cont.}$ -- flux of the local continuum in 
         10$^{-10}$\ecsa, 
$^{(e)}$ FWHM -- Full Width at Half Maximum in \AA, 
$^{(f)}$ $v_{\infty}$ -- terminal velocity from the blue wing 
         in \kms, 
$^{(g)}$ $RV_{\rm abs}$ and $RV_{\rm em}$ -- radial velocities 
         of narrow absorption and emission components in \kms, 
$^{(h)}$ $F_{\rm abs}$ and $F_{\rm em}$ -- their fluxes 
         in 10$^{-10}$\ecs, 
$^{(i)}$ $\dot M_{\rm WD}$ -- mass-loss rate in $10^{-4}$\myr, 
$^{(j)}$ label of the observatory (Table~\ref{tab:spec}, 
         Sect.~\ref{ss:spec}). 
The parameter uncertainties are described in 
Sects.~\ref{sss:habroad}, \ref{sss:hanarrow}, and \ref{ss:mdot}. 
\label{tab:hapar}
\end{table*}

\clearpage

\bibliography{rs2021.bib}{}

\bibliographystyle{aasjournal}

\end{document}